\RequirePackage{luatex85}
\documentclass[12pt]{article}
\usepackage{a4wide}
\usepackage[centertags]{amsmath}
\usepackage{amssymb}
\usepackage[sort&compress,numbers]{natbib}
\usepackage{ifpdf}
\usepackage{setspace}
\usepackage{amsfonts}
\usepackage{color}
\usepackage{threeparttable}
\usepackage{cancel}
\usepackage{tikz}
\usepackage{ulem}
\usetikzlibrary{decorations.pathmorphing}
\usepackage{feynmf}
\usepackage[compat=1.1.0]{tikz-feynman}
\tikzfeynmanset{compat=1.1.0}
\usepackage{verbatim}
\usepackage{xcolor}
\usepackage{bbm}
\usepackage{tikz}
\usepackage{pgfplots}
\pgfplotsset{compat=1.16}
\usepackage{tikz}
\usepackage{amsmath}
\usepackage{pgfplots}
\usepackage{float}

\usepackage{lscape}

\usepackage{blindtext, rotating}

\usepackage{yfonts}

\usetikzlibrary{decorations.markings}
\ifpdf
\usepackage[colorlinks=true,
linkcolor=black,
citecolor=black,
urlcolor=blue,
filecolor=blue,
pdfstartview=FitV,
pdftitle={},
pdfauthor={},
pdfsubject={Hydrodynamics},
pdfkeywords={hydrodynamics, AdS/CFT},
pdfpagemode=None,
bookmarksopen=true
]{hyperref}
\else
\usepackage[hypertex]{hyperref}
\fi
\usepackage{rotating}
\usepackage{rotating}
\makeatletter
\@addtoreset{equation}{section}

\newcommand{\qn}{{\textswab{q}}}
\newcommand{\wn}{{\textswab{w}}}

\makeatletter
\renewcommand\section{\@startsection {section}{1}{\z@}%
	{-3.5ex \@plus -1ex \@minus -.2ex}
	{2.3ex \@plus.2ex}%
	{\normalfont\large\bfseries}}
\renewcommand\subsection{\@startsection{subsection}{2}{\z@}%
	{-3.25ex\@plus -1ex \@minus -.2ex}%
	{1.5ex \@plus .2ex}%
	{\normalfont\bfseries}}

\def\sec#1{Sec.\ \ref{#1}}

\def\App#1{Appendix \ref{#1}}

\usepackage{graphicx,epstopdf}   
\usepackage{sidecap}
\usepackage{caption}
\usepackage{subcaption}

\title{
	Three-point functions from a Schwinger-Keldysh effective action, resummed in derivatives}
\author{Navid Abbasi$^{a,b,c}$\footnote{abbasi@lzu.edu.cn},  \ Dirk H.\ Rischke$^{b,d}$\footnote{drischke@itp.uni-frankfurt.de}\\[2mm]
	\small{\textit{$^{a}$School of Nuclear Science and Technology, Lanzhou University,}}\\
	\small{\textit	{ 
			222 South Tianshui Road, Lanzhou 730000, China }} \\[2mm]
	\small{\textit{$^{b}$Institut für Theoretische Physik, Johann Wolfgang Goethe–Universität,}}\\
	\small{\textit	{ 
			Max-von-Laue-Str.\ 1, D-60438 Frankfurt am Main, Germany}} \\[2mm]
	\small{\textit{$^{c}$ExtreMe Matter Institute EMMI,
			GSI Helmholtzzentrum für Schwerionenforschung,}}\\
	\small{\textit	{ 
			Planckstrasse 1,
			D-64291 Darmstadt, Germany}} \\[2mm]
	\small{\textit{$^{d}$Helmholtz Research Academy Hesse for FAIR,}}\\
	\small{\textit	{ 
			Max-von-Laue-Str.\ 12, D-60438 Frankfurt am Main, Germany}} \\
}

\begin{document}
	
	\setlength{\baselineskip}{16pt}
	\begin{titlepage}
		\maketitle
		
		\vspace{-36pt}
		
		\begin{abstract}
			The search for the conjectured QCD critical point in heavy-ion collisions  requires to account for far-from equilibrium effects as well as fluctuations, and in particular non-Gaussian fluctuations, in the modeling of the dynamics of the hot and dense matter created in such collisions. 
			In order to study far-from equilibrium effects as well as fluctuations, in this work we construct a Schwinger-Keldysh effective field theory (EFT) for the diffusion of the density to all orders in derivatives. 
			The dissipation in the free part of our EFT follows the Boltzmann equation in the relaxation-time approximation (RTA). 
			The interaction part of the EFT is constructed based on the self-interaction of the density field. 
			We analytically find the quadratic and cubic parts of the KMS-invariant EFT in  closed form, resummed in  derivatives. 
			We then explicitly compute the symmetrized three-point function at tree level, and investigate its analytical structure in detail. 
			We also analytically calculate the branch-point singularity that appears in the structure of the two-point response function due to loop effects.
			Our results are important for future studies of the real-time dynamics of the correlation functions and the possible relation to thermalization when the system is far from equilibrium.
		\end{abstract}
		\thispagestyle{empty}
		\setcounter{page}{0}
	\end{titlepage}

	\renewcommand{\baselinestretch}{1}  
	\tableofcontents
	\renewcommand{\baselinestretch}{1.2}  
	\section{Introduction}
	\label{intro}
	Quantum chromodynamics (QCD) is the fundamental theory of the strong interaction.
	Heavy-ion collisions probe QCD matter at extremes of temperature and density, with the goal to clarify the structure of the QCD phase diagram and identify possible phase transitions.
	One such transition is related to the dynamical breaking of the chiral symmetry of QCD in the vacuum and its restoration at high temperatures and densities.
	It has been conjectured that there exists a second-order critical endpoint of a line of first-order transitions at low temperatures and intermediate densities \cite{An:2021wof,Du:2024wjm} (see also Ref.~\cite{Stephanov:2024xkn} for a review of recent developments).
	The search for this endpoint in heavy-ion collisions is one of the major goals of the Beam Energy Scan (BES) experiment at RHIC, as well as experiments at other accelerator facilities. 
	Varying the collision energy, different regions of the phase diagram can be explored, allowing to measure excitation functions of various observables.
	It has been predicted that the critical endpoint manifests itself in the non-monotonic behavior of the excitation function of particle-multiplicity cumulants.
	Among them, non-Gaussian particle-multiplicity cumulants are considered to be the quantities which are most sensitive to the existence of the critical endpoint \cite{Stephanov:2008qz}.

	In the last few years, many effective dynamical models have been proposed to simulate non-Gaussian particle-multiplicity cumulants, see for instance Refs.~\cite{An:2020vri,Chattopadhyay:2024jlh,Tang:2023zvj,An:2022tfk,Nahrgang:2018afz}. 
	However, at present there are no simulation data obtained within microscopic models \footnote{Recently, Ref.~\cite{Pantelidou:2022ftm} calculated the three-point functions of the conserved charge in a holographic setup. }. 
	In this work, we aim to take a step towards this problem. 
	The microscopic model we will use is the classical dynamical theory of massless relativistic particles, which obey the Boltzmann equation in the relaxation-time approximation (RTA). 
	This can serve as a simple  microscopic model to describe the thermalization of large-$N$ gauge-theory systems \cite{Romatschke:2015gic} \footnote{There are also recent works about Boltzmann equation with a collision operator in scalar field theory. }.

	Hydrodynamic correlation functions for certain densities can be computed in the framework of (stochastic) hydrodynamics \cite{Martin,Kovtun:2012rj,Akamatsu:2016llw,Chao:2020kcf} and effective field theory (EFT) \cite{Chen-Lin:2018kfl,Jain:2020zhu,Delacretaz:2020nit} (see Ref.~\cite{Basar:2024srd} for a review of recent developments). 
	In addition to these effective models, analytical results for hydrodynamic correlation functions are known for at least three systems with specific microscopic dynamics: holographic matter \cite{Son:2002sd}, the kinetic system discussed in the previous paragraph \cite{Romatschke:2015gic,Bajec:2024jez} (see also Refs.~\cite{Kurkela:2017xis,Dash:2023ppc,Brants:2024wrx,Hu:2024tnn}
 for an RTA with momentum-dependent relaxation time), and the SYK model \cite{Maldacena:2016hyu,Sachdev:2023try}. 
	However, most of these results are related to Gaussian (two-point) correlation functions. For \textit{non-Gaussian correlation functions}, things become more complicated. 
	In the context of EFT, these can be calculated by incorporating interactions into the effective Lagrangian. 
	Recently, this has been used to compute three-point correlation functions of the density in simple diffusive systems in the framework of Schwinger-Keldysh EFT \cite{Delacretaz:2023ypv} (see also Ref.~\cite{Sogabe:2021svv}). 
	For holographic systems, explicit results have been found in Ref.~\cite{Pantelidou:2022ftm} (see also Ref.~\cite{Jana:2020vyx})\footnote{See Refs.~\cite{Saremi:2011nh,Grozdanov:2016fkt} for holographic calculations in the small-frequency/momentum limit. 
		In the same limit, see Ref.~\cite{Moore:2010bu} for a calculation of three-point correlation functions of the stress tensor in a weakly coupled real scalar field theory.
		For the calculation of three-point thermal correlators in  $CFT_2$ in momentum space, see Ref.~\cite{Becker:2014jla}, and at large operator dimension, see Ref.~\cite{Rodriguez-Gomez:2021mkk}.}.
	The aim of this work is to compute the three-point correlation functions for another well-known microscopic system: a kinetic system of massless particles in RTA.

	In this work we do not simply perform a pure kinetic-theory calculation. 
	The idea is to first construct a Schwinger-Keldysh effective action, whose classical equation of motion is the linear equation for charge diffusion in kinetic theory in RTA. 
	This will be done in \sec{SK_EFT}. 
	As a first step in this study, we propose a new method to find this equation to infinite order in the derivative expansion (\sec{sec_diff}). 
	Having found this equation, we then construct the associated Schwinger-Keldysh effective action to quadratic order in dynamic fields. 
	This is not our final result, it just gives us the two-point correlation function. 
	In agreement with what was found by Romatschke in Ref.~\cite{Romatschke:2015gic}, for momenta $k$ smaller than a certain value $k^*$, the correlation function has a simple pole and  branch cut between $-i/\tau-k<\omega<-i/\tau+k$ in the lower half of the complex-frequency plane. 
	For $k>k^*$ the pole disappears.

	The calculation of the two-point function in a diffusive system is well known in the literature, both in kinetic theory \cite{Romatschke:2015gic} and in the EFT setup \cite{Kovtun:2012rj,Chen-Lin:2018kfl}. 
	However, there are two points which distinguishes our work from the  existing literature:
	\begin{enumerate}
		\item Compared to other well-known Schwinger-Keldysh EFTs associated with diffusive systems with finite-order derivatives \cite{Chen-Lin:2018kfl}, our action is expressed to infinite order in derivatives. 
		More importantly, we resum the derivative expansion and show that the action has a simple closed form.
		\item Studies of kinetic theory existing in the literature focus on computing either response functions \cite{Romatschke:2015gic,Bajec:2024jez} or correlation functions \cite{landau1981kinetic,SoaresRocha:2024afv}. 
		However, in the framework of Schwinger-Keldysh EFT, we will be able to systematically derive both response and correlation functions. 
		We will show that the physical correlation function associated with our effective action is exactly related to the response function found by Romatschke in Ref.~\cite{Romatschke:2015gic} via the fluctuation-dissipation theorem.
	\end{enumerate}
	The main part of our work will be done in \sec{SK_int} and is devoted to the study of nonlinear effects in the system. 
	Motivated by the need to study non-Gaussian correlation functions, as discussed earlier, our main goal is to investigate the symmetrized three-point function. 
	The leading contribution comes from the cubic action. 
	For this, we construct the KMS-invariant cubic action, to infinite order in derivatives, and present it in a closed form.\footnote{In \protect \sec{SK_EFT}, we briefly review the Schwinger-Keldysh EFT.} 
	We do this systematically by including the self-interaction of the charge density.

	Although the analytic structure of the vertices looks very rich in our Lagrangian  (due to the resummation of derivatives), we  find that the analytic structure of the tree-level correlator in momentum space is completely fixed in terms of the analytic structure of the external legs.\footnote{This is similar to the results associated with a scalar boundary operator dual to a scalar field with cubic interactions in the bulk of AdS \cite{Pantelidou:2022ftm}. }  
	It turns out that as long as any external momentum is below the certain value $k^*$, the corresponding leg will only produce a simple diffusion pole in the lower half-plane.
	We will also find that the branch-cut discontinuity of the correlator is the union of the branch cuts associated with the external legs. 
	We will show all of this by explicitly calculating the symmetric three-point function. 
	
	In the limit that the coupling constant is small, the effect of quartic and higher-order interactions on the three-point function, i.e., loop effects, can be neglected.
	However, in the same limit the two-point function is corrected by the loop effect originating from the cubic interaction \cite{Kovtun:2012rj,Chen-Lin:2018kfl,Jain:2020zhu,Abbasi:2021fcz,Abbasi:2022aao} (see also Ref.~\cite{Grozdanov:2024fle}). In particular, it is known that a branch-point singularity emerges in the structure of the response function due to the loop effect. 
	By performing an explicit calculation we find the location of the branch point in a closed formula. 
	At leading order in derivatives, our formula reduces to the famous $ \omega_{\text{b.p.}}=-\frac{i}{2}D k^2$ result \cite{Chen-Lin:2018kfl,Delacretaz:2020nit}. 
	For the entire range of $k<k^*$, this branch point continues to persist. 
	As a result, we find that there are two branch cuts in the structure of the two-point response function; one which is the specific feature of weakly coupled kinetic systems $-i/\tau-k<\omega<-i/\tau+k$, and one which emerges due to the interactions $- i \infty<\omega < \omega_{\text{b.p.}}$.
	
	In \sec{kin_theory} we discuss how to apply the Martin-Siggia-Rose (MSR) formalism to the Boltzmann equation and find the effective action. 
	Our result is consistent with our Schwinger-Keldysh EFT, as it should be when the noise field is taken to have a Gaussian distribution. 
	Finally, in \sec{Review} we review five concrete results of our work. 
	We then briefly discuss some applications of our results.

	\section{Density fluctuations from kinetic theory}
	\label{sec_diff}
	The \textcolor{black}{relativistic} Boltzmann equation for the single-particle distribution function $f(x,\boldsymbol{p})$ in phase space in the absence of external sources and in RTA is given by
	\textcolor{black}{
	\begin{equation}\label{Boltzmann_rel}
		p^{\mu}\partial_{\mu} f(x,\boldsymbol{p})=\frac{p^{\alpha}u_{\alpha}}{\tau}\big[f(x,\boldsymbol{p})-f^{(0)}(x,\boldsymbol{p})\big]\;,
	\end{equation}
}
	where 
	$\tau$ is the (constant) relaxation time. \textcolor{black}{The particles are assumed to be on-shell, i.e., $p^0 = \epsilon(\boldsymbol{p})$,  with $\epsilon(\boldsymbol{p})$ being the dispersion relation of particles. 
	In this work, we consider a relativistic system of massless particles, so $\epsilon(\boldsymbol{p})=|\boldsymbol{p}| \equiv p$. 
    In addition, we assume that the system is in the high-temperature, low-density limit, $T\gg \mu$, where quantum effects are negligible.
    This is true except for momenta $p \ll \mu$, which, however, corresponds only to a tiny fraction of the occupied states, since typical momenta are of order $p \sim T$. 
    Since later on we will integrate over momentum space, these states are further suppressed by the integration measure $d^3p \sim dp\, p^2$. 
    Thus, the equilibrium distribution function $f^{(0)}$ follows the Maxwell-Boltzmann distribution.
    For the sake of simplicity, we also do not consider antiparticles, although there is in principle no obstacle to include them. 
    In RTA, the Boltzmann equation for antiparticles would simply decouple from that for particles.
				In the rest frame of the system, where 
		 $u^{\mu}=(1,0,0,0)$, the local-equilibrium distribution function  is given by}
	\begin{equation} \label{eq:eq_dist}
		f^{(0)}(x,\boldsymbol{p})=\exp\bigg[-\frac{p-\mu(x)}{T}\bigg]\;.
	\end{equation}
	Note that in this work we keep the background temperature $T$ constant 	\textcolor{black}{and $\mu(x)$ is a dynamical variable in our setup. The kinetic equation in the absence of external sources then  takes the following form in the rest frame of the system,}
	\begin{equation}\label{Boltzmann}
		\partial_{t} f(x,\boldsymbol{p})+\boldsymbol{v \cdot \nabla} f(x,\boldsymbol{p})=-\frac{f(x,\boldsymbol{p})-f^{(0)}(x,\boldsymbol{p})}{\tau}\;.
	\end{equation}
    \textcolor{black}{Here, $\boldsymbol{v}=\partial \epsilon/\partial \boldsymbol{p}$. Since the particles are assumed to be massless,
	$v \equiv |\boldsymbol{v}| = 1$, as we work in natural units, i.e., the velocity of light is $c=1$.}
 Directly integrating Eq.~\eqref{Boltzmann} over momentum space with measure $\int_{\boldsymbol{p}}\equiv\int\frac{d^3p}{(2\pi)^3}$ just gives the equation of charge conservation:
	\begin{equation}\label{conservation}
		\partial_t J^0 + \boldsymbol{\nabla}\cdot \boldsymbol{J}=0\;,
	\end{equation}
	where $J^0$ is the charge density, $\boldsymbol{J}$ is the charge current,
	\begin{eqnarray}\label{J_0}
		J^0(x)&=&\int_{\boldsymbol{p}} f(x, \boldsymbol{p})\;,\\\label{J_i}
		\boldsymbol{J}(x)&=&\int_{\boldsymbol{p}} \boldsymbol{v}f(x, \boldsymbol{p})\;,
	\end{eqnarray}
	We have used the Landau matching condition \cite{Romatschke:2017ejr}
	\begin{equation}\label{matching}
		\int_{\boldsymbol{p}}\big[f(x,\boldsymbol{p})-f^{(0)}(x,\boldsymbol{p})\big]=0\;,
	\end{equation}
\textcolor{black}{	to ensure the conservation of particle number. This makes the momentum-space integral over} the right-hand side of Eq.~\eqref{Boltzmann} vanish.
	Now, in order to compute $J^0$ and $\boldsymbol{J}$, we need to solve the Boltzmann equation in RTA. 
	We do this in a derivative expansion.
	
	\subsection{How to derive the diffusion  equation (to large orders in derivatives) from kinetic theory?}
	\label{resum}
	In the long-wavelength limit, the microscopic scale represented by $\tau$ is much smaller than the scale over which $f(x,\boldsymbol{p})$ varies.
	In this case, the ``scaled" Boltzmann equation \cite{Cercignani:2002}
	
	\begin{equation}\label{Boltzamnn_1}
		\varepsilon\,\textbf{D} f=-\frac{f-f^{(0)}}{\tau}\;,
	\end{equation}
	with 
	\begin{equation}\label{f_n_general}
		\mathbf{D}=\,\partial_t+\boldsymbol{v \cdot \nabla}\;,
	\end{equation}
	can be solved \textcolor{black}{exactly as follows:\footnote{\textcolor{black}{We thank one of the referees for pointing this out. In the first version of this paper, we had derived this equation differently.}}
	\begin{equation}\label{exact}
f=\frac{1}{1+\varepsilon \,\tau\, \textbf{D}}\,f^{(0)}\;.
	\end{equation}
	This solution can be perturbatively expanded around} the local-equilibrium distribution function  (\ref{eq:eq_dist}),
	\begin{equation}\label{Chapman_Enskog}
		f=\,f^{(0)}+\,\varepsilon f^{(1)}+\,\varepsilon^2f^{(2)}+\ldots\;,
	\end{equation}
	where $\varepsilon$ is an auxiliary parameter counting the number of derivatives, which will be set to 1 at the end of the calculation, such that at order $\ell$ the distribution function is given by $f\equiv \sum_{m=0}^{\textcolor{black}{\ell}} f^{(m)}$.  
	Note that for the sake of brevity  we have suppressed the arguments of $f(x,\boldsymbol{p})$.
	
Substituting the exact solution \eqref{exact} into the Boltzmann equation, we arrive at
	\begin{equation}\label{Boltzamnn_2}
	\textcolor{black}{\frac{\textbf{D}}{1+\varepsilon \,\tau\, \textbf{D}}f^{(0)}=\,}
		-\frac{f-f^{(0)}}{\tau}\;.
	\end{equation}
	To summarize, we have converted the original Boltzmann equation \eqref{Boltzmann} into Eq.~\eqref{Boltzamnn_2}. 
	Although the right-hand side is the same, the left-hand side has changed significantly. 
	The price we have paid to get rid of the full out-of-equilibrium distribution function $f$ is to include \textcolor{black}{infinitely many} additional derivatives in the \textcolor{black}{resummed} form given on the left-hand side of Eq.~\eqref{Boltzamnn_2}. 
	But the bonus is that the derivatives just act on the local-equilibrium distribution function $f^{(0)}$.
	
	Integrating Eq.~\eqref{Boltzamnn_2} over  momentum space, and applying Eq.~\eqref{matching}, gives the conservation equation in a derivative expansion. 
We find that the integrated kinetic equation \textcolor{black}{expanded in derivatives} at order $\ell$ is given by:
	\begin{equation}\label{higher_diffusion}\boxed{
			\int_{\boldsymbol{p}}\,\mathbf{D}\big[1-\tau \mathbf{D}+(-\tau \mathbf{D})^2+\ldots+(-\tau \,\mathbf{D})^{\textcolor{black}{\ell}}\big] f^{(0)}=\,0\;.}
	\end{equation}
	Interestingly, it is easy to see that the matching condition \eqref{matching} at order $\ell$ is nothing but the conservation equation at order $\ell-1$ in the $\varepsilon$-expansion. 
	For instance at $\ell=2$ the matching condition is $\int_{\boldsymbol{p}}(f^{(1)}+f^{(2)})=0$, which by virtue of    Eq.~\eqref{exact} is written as $\int_{\boldsymbol{p}}\textbf{D}(1-\tau \textbf{D})f^{(0)}=0$.
	
		\textcolor{black}{
			Note that the unknown function in Eq.~\eqref{higher_diffusion} is $\mu(x)$, which is contained in the expression for $f^{(0)}(x,\boldsymbol{p})$. In the rest frame of the system, $f^{(0)}(x,\boldsymbol{p})$ is an isotropic function of $\boldsymbol{p}$. Therefore, the integration over $\boldsymbol{p}$ can be performed independently of how $\partial_t$ and $\partial_i$ (within the $\textbf{D}$ operator) act on $\mu(x)$.    
            This suggests introducing the function $n(x) \equiv \int_{\boldsymbol{p}}f^{(0)}(x,\boldsymbol{p})$. Then by performing the momentum-space integral, we obtain the main result of this section,}
	\begin{equation}\label{diffusion_all_order}
		\left[\tilde{\partial}_t+\left(-\tilde{\partial}_t^2-\frac{1}{3}\tilde{\boldsymbol{\nabla}}^2 \right)+\left(\tilde{\partial}_t\tilde{\boldsymbol{\nabla}}^2+\tilde{\partial}_t^3 \right)+\left(-\tilde{\partial}_t^4-2\tilde{\partial}_t^2\tilde{\boldsymbol{\nabla}}^2-\frac{1}{5}\tilde{\boldsymbol{\nabla}}^4\right)+\ldots\right]n(x)=\,0\;,
	\end{equation}
	where $\tilde{\partial}_t=\,\tau \partial_t,\,\tilde{\boldsymbol{\nabla}}=\,\tau \boldsymbol{\nabla}\,$.
	This equation is in fact the \textit{diffusion equation for the density $n(x)$ including all orders in derivatives \textcolor{black}{in the rest frame of the system}}.
	Note also that this equation is in fact equivalent to \textcolor{black}{the rest-frame representation of} Eq.~\eqref{conservation}.
	
	\subsection{Derivative-counting scheme}
	For a (non-relativistic) Markovian process, the distance that a particle travels within a time interval $\Delta t$ scales as $|\Delta \textbf{x}| \sim \sqrt{\Delta t}$, which motivates the following 
	power counting for the time and space derivatives
	\begin{equation}\label{scaling}
		\partial_t\sim \boldsymbol{\nabla}^2\sim \epsilon^2\;,
	\end{equation}
	where $\epsilon$ should not be confused with $\varepsilon$ in Eq.~\eqref{Chapman_Enskog}. 
	Applying Eq.~\eqref{scaling} to Eq.~\eqref{diffusion_all_order}, we group terms of the same order in $\epsilon$ and obtain:
	\begin{equation}\label{diff_eq_higher_orders}
		\left[\left(\tilde{\partial}_t-\frac{1}{3}\tilde{\boldsymbol{\nabla}}^2 \right)+\left(-\tilde{\partial}_t^2+\tilde{\partial}_t\tilde{\boldsymbol{\nabla}}^2-\frac{1}{5}\tilde{\boldsymbol{\nabla}}^4 \right)+\left(\tilde{\partial}_t^3-2\tilde{\partial}_t^2\tilde{\boldsymbol{\nabla}}^2+\tilde{\partial}_t\tilde{\boldsymbol{\nabla}}^4-\frac{1}{7}\tilde{\boldsymbol{\nabla}}^6 \right)+\ldots\right]n(x)=0\;.
	\end{equation}
	Restricting ourselves to the terms of leading order in $\epsilon$ (the terms in the first parentheses), we get the famous Fick's law of diffusion, i.e., the $\mathcal{O}(\epsilon^2)$ diffusion equation. 
	The other parentheses represent higher-order corrections to the leading-order diffusion equation, i.e., $\mathcal{O}(\epsilon^4)$, $\mathcal{O}(\epsilon^6)$, etc. 
	Again, this equation is equivalent to Eq.~\eqref{conservation}. 
	Now we can explicitly represent $\text{J}^0(x)$ and $\text{J}^i(x)$ in terms of $n(x)$.
	
	\subsection{Density and current}
	\label{sec2.3}
	Substituting Eq.~\eqref{exact} into Eqs.~\eqref{J_0} and \eqref{J_i} \textcolor{black}{and expanding in derivatives}, the constitutive relations now read as follows (we temporarily omit the argument $x$)
	\begin{displaymath}
		\begin{split}
			J^0=&\int_{\boldsymbol{p}}\big(1-\tau \mathbf{D}+\ldots\big) f^{(0)} \\
			=&\,  n-\tau \dot{n}+\left(\frac{\tau^2}{3}\nabla^2n+\tau^2 \ddot{n}\right)-\tau^3 \left(\nabla^2\dot{n}+\dddot{n}\right)+\tau^4\left(\frac{1}{5}\nabla^4n+\ldots\right)+\ldots\;, \\		
            J^i=&\int_{\boldsymbol{p}} v^i\big(1-\tau \mathbf{D}+\ldots\big) f^{(0)}\,=\,-\frac{\tau}{3}\nabla_i n+\frac{2\tau^2}{3} \nabla_i\dot{n}-\tau^3\left(\frac{1}{5}\nabla_i\nabla^2n+ \nabla_i\ddot{n}\right)+\ldots\;. 
		\end{split}
	\end{displaymath}
	Here we grouped the terms according to the number of derivatives, i.e., following the $\varepsilon$-expansion (where $\partial_t \sim \boldsymbol{\nabla} \sim \varepsilon$). 
	However, we can group them also according to the $\epsilon$-expansion (where $\partial_t \sim \boldsymbol{\nabla}^2 \sim \epsilon^2$):
	\begin{equation}\label{Current_re_group}
		\begin{split}
			J^0=&\,n-\tau\left(\dot{n}-\frac{\tau}{3}\nabla^2n\right)+\tau^2\left( \ddot{n}-\tau \nabla^2\dot{n}+\frac{\tau^2}{5}\nabla^4n\right)+\ldots\;, \\
			J^i=&\,-\frac{\tau}{3}\nabla_i n+\tau^2\left(\frac{2}{3} \nabla_i\dot{n}-\frac{\tau}{5}\nabla_i\nabla^2n\right)-\tau^3\left(\nabla_i\ddot{n}-\frac{4\tau}{5}\nabla_i\nabla^2\dot{n}+\frac{\tau^2}{7}\nabla_i \nabla^4n\right)+\ldots\;.
		\end{split}
	\end{equation}
	At first glance, it seems that $J^{0}(x)\ne n(x)$, while the Landau matching condition \eqref{matching} requires $J^{0}(x)= n(x)$, order by order in the $\varepsilon$ (or $\epsilon$)-expansion. 
	As we discussed below Eq.~\eqref{higher_diffusion}, the Landau matching condition \textcolor{black}{\eqref{matching}} at any order is exactly the conservation equation at one lower order of the expansion. 
	Considering this, one can show that Eq.~\eqref{Current_re_group} reduces to $J^{0}(x)=n(x)$, at any order in the expansion. 
	In Table~\ref{Table}, we have shown this for the first few orders in the $\epsilon$-expansion.
	\begin{table}[!htb]
		\label{table one}
		\begin{center}
			\begin{tabular}{|c|c|c|c|c|}
				\hline
				\hline
				&	 & &   $J^{0}$ upon applying\\
				$\ell$&	 $J^0$ to $\mathcal{O}(\epsilon^{\ell})$ & Eq.~\eqref{conservation} &  ``Eq.~\eqref{conservation} \\
				&	 & with $J^0$ to $\mathcal{O}(\epsilon^{\ell})$ &   with $J^0$ to $\mathcal{O}(\epsilon^{\ell-2})$"\\
				\hline
				\hline
				&&&\\
				0&$n$	& $\dot{n} -\frac{\tau}{3}\nabla^2 n=0$ &  $n$\\
				&&&\\
				\hline
				&&&\\
				2 & $n-\tau\big(\dot{n}-\frac{\tau}{3}\nabla^2n\big)$	& $\dot{n}-\tau\big(\ddot{n}-\frac{\tau}{3}\nabla^2\dot{n}\big)$ & $n$  \\ 
				&&   $-\frac{\tau}{3}\nabla^2 n+\tau^2\big(\frac{2}{3} \nabla^2\dot{n}-\frac{\tau}{5}\nabla^4n\big)=0$   &\\
				\hline
				&&&\\
				4 & $n-\tau\big(\dot{n}-\frac{\tau}{3}\nabla^2n\big)$	& $\cdots$  & $n$  \\ 
				&  $+\tau^2\big( \ddot{n}-\tau \nabla^2\dot{n}+\frac{\tau^2}{5}\nabla^4n\big)$ &   &   \\
				\hline
				&&&\\					
				$	\vdots$&&& $n$\\
				\hline
				\hline
			\end{tabular}
		\end{center}
		\caption{Illustration of $J^{0}(x)=n(x)$, order by order in the derivative expansion.  
			We have shown this for the first few orders in derivatives. 
			For higher orders the expressions are more complicated. 
			The dots in the $\ell=4$ row are not important for finding  $J^0$ to this order in the third column.  }
		\label{Table}
	\end{table}
	
	When written to all orders in derivatives, one can then simply drop all derivative terms in $J^{0}$. 
	Similarly, we can apply lower-order equations of motion to $J^i$ and carefully simplify it. 
	Interestingly, we find that the diffusion equation takes the following familiar form (see \App{diff_familiar} for the details)
	\begin{equation}\label{diffusion_changed}
		\partial_t n+\,
		\nabla_i\bigg[-\frac{\tau}{3}\nabla_i n+\frac{\tau^3}{45}\nabla_i \nabla^2 n-\frac{2\tau^5}{945}\nabla_i \nabla^4 n+\ldots\bigg]=0\;.
	\end{equation}
	The two equations \eqref{diff_eq_higher_orders} and \eqref{diffusion_changed} are obviously equivalent in the derivative expansion. 
	In what follows, we continue to work with Eq.~\eqref{diff_eq_higher_orders}. 
	We find this equation more useful, since by using it we will be able to express our results in resummed form.

	\subsection{Large-order transport coefficients}
	By Fourier-transforming the diffusion equation \eqref{diff_eq_higher_orders}, we can find the diffusion mode perturbatively in a momentum expansion. 
	We define the Fourier transform as 
	\begin{equation}\label{Fourier}
		Q(\omega, \textbf{k})=\int_{-\infty}^{+\infty} dt \int d^3 x \,e^{i \omega t - i \textbf{k}\cdot \textbf{x}}Q(t,\textbf{x})
	\end{equation}
	for a quantity $Q(t,\textbf{x})$.
	The diffusion mode is then formally written as 
	\begin{equation}\label{diffusion_mode}
		\omega(k^2)=\,-i\,\mathcal{D}(k^2)\,k^2\;,
	\end{equation}
	where the diffusion constant is found to be given by the following expansion
	\begin{equation}\label{higher_transport_coeff}
		\mathcal{D}(k^2)=\,\frac{\tau}{3}\left[1+\frac{1}{15}(\tau k)^2+\frac{2}{315}(\tau k)^4+\frac{1}{1575}(\tau k)^6+\frac{2}{31185}(\tau k)^8+\ldots\right]\;,
	\end{equation}
	which is also obviously consistent with Eq.~\eqref{diffusion_changed}.
	A simple ratio test reveals that the series is convergent for $k<k_c$,\footnote{See Ref.~\cite{Heller:2020hnq} for a discussion on the convergence of the derivative expansion in sound and shear channels associated with the same theory. 
		See also Ref.~\cite{Gangadharan:2024ovs}.}
	where $k \equiv |\textbf{k}|$  and \textcolor{black}{$k_c\approx 3.14/\tau\,$, as we will see below.}

\textcolor{black}{Note that this calculation is inspired by earlier work on the convergence
 of the derivative expansion,  initiated in Refs.~\cite{Withers:2018srf,Grozdanov:2019kge}, and extended to the M\"uller-Israel-Stewart theory in Ref.~\cite{Heller:2020uuy}. The approach we take here for diffusion is mathematically equivalent to the analysis performed for the shear channel in Refs.~\cite{Grozdanov:2019kge,Heller:2020uuy}.}
	\subsection{Resummation of the diffusion equation}
	\label{SK_quadratic}
	Since Eq.~\eqref{higher_diffusion} is in the form of a geometric series, it can be resummed as follows 
	\begin{equation}\label{Resummed}
		\boxed{
			\int_{\boldsymbol{p}}\,\frac{\mathbf{D}}{1+\tau \textbf{D}} f^{(0)}(x,\boldsymbol{p})=\,0}\,\,\,\,\,\,\,\text{or}\,\,\,\,\,\,\,\,\,\,\,\boxed{
			\int_{\Omega}\,\frac{\mathbf{D}}{1+\tau \textbf{D}} n(x)=\,0} \;.
	\end{equation}
	Here $f^{(0)}$ only depends on $|\boldsymbol{p}|=p$, cf.~Eq.~\eqref{eq:eq_dist}. 
	We have also defined the integration over solid angle, $\int_{\Omega}=\frac{1}{4\pi}\int_0^{2\pi}d\phi \int_{-1}^1 d(\cos \theta) $. 
	The  $\Omega$-dependence of the $\boldsymbol{p}$-integrand is just through the $\cos \theta$ in $\textbf{D}=\partial_t+\textcolor{black}{\boldsymbol{v \cdot \nabla}}$, where \textcolor{black}{$\boldsymbol{v}=(\sin \theta \cos \phi, \sin \theta \sin \phi, \cos \theta)$}.
	This is why the two equations in Eq.~\eqref{Resummed} can be equivalent (\textcolor{black}{see the discussion in the second paragraph below Eq.~\eqref{higher_diffusion}}). 

    \textcolor{black}{In order to perform the $\Omega$-integral in Eq.~\eqref{Resummed}, we expand the derivative operator:}
	\begin{equation}\label{EoM_Fourier}
		\begin{split}
          \color{black}  4\pi\left[\tilde{\partial}_t+\left(-\tilde{\partial}_t^2-\frac{1}{3}\tilde{\boldsymbol{\nabla}}^2 \right)+\left(\tilde{\partial}_t\tilde{\boldsymbol{\nabla}}^2+\tilde{\partial}_t^3 \right)+\left(-\tilde{\partial}_t^4-2\tilde{\partial}_t^2\tilde{\boldsymbol{\nabla}}^2-\frac{1}{5}\tilde{\boldsymbol{\nabla}}^4\right)+\cdots\right]\frac{n(x)}{4\pi}=&\,0
		\end{split}
	\end{equation}
        \color{black}The special form of the derivatives appearing in the above equation suggest to rewrite the above equation in a resummed form.  Each pair of parentheses in this equation contains the terms at a specific order in derivatives, that is a polynomial of $\tilde{\partial}_t$ and $\tilde{\boldsymbol{\nabla}}$. Interestingly, the form of these polynomials is reminiscent of $(\tilde{\partial}_t+\tilde{\nabla})^n-(\tilde{\partial}_t-\tilde{\nabla})^n$ for $n=2,3,4,\ldots$. Let us define
\begin{equation}
L_{\pm}\equiv \,\tilde{\partial}_t\pm\tilde{\nabla}\,,\,\,\,\,\,\,\,\,\,\,\tilde{\nabla}^2\equiv \tilde{\boldsymbol{\nabla}}^2\;.
\end{equation}
Then the consecutive terms in Eq.~\eqref{EoM_Fourier} can be written as 
\begin{equation}\label{}
\begin{split}
\tilde{\partial}_t=\,&\,\frac{1}{2\tilde{\nabla}}\,\frac{L_{+}^2-L_{-}^2}{2}\;,\\
-\tilde{\partial}_t^2-\frac{1}{3}\tilde{\boldsymbol{\nabla}}^2 =\,&\,\frac{1}{2 \tilde{\nabla}}\,\frac{L_{+}^3-L_{-}^3}{3}\;,\\
\tilde{\partial}_t\tilde{\boldsymbol{\nabla}}^2+\tilde{\partial}_t^3 =\,&\,\frac{1}{2 \tilde{\nabla}}\,\frac{L_{+}^4-L_{-}^4}{4}\;,\\
-\tilde{\partial}_t^4-2\tilde{\partial}_t^2\tilde{\boldsymbol{\nabla}}^2-\frac{1}{5}\tilde{\boldsymbol{\nabla}}^4=\,&\,\frac{1}{2 \tilde{\nabla}}\,\frac{L_{+}^5-L_{-}^5}{5}\;.
\end{split}
\end{equation}
Then Eq.~\eqref{EoM_Fourier} can be rewritten as 
\begin{equation}\label{}
\left(-1+\frac{1}{2 \tilde{\nabla}}\sum_{n=1}\frac{( L_{+})^n-(L_{-})^n}{n}\right)n=\,0\;.
\end{equation}
The summation is nothing but $\ln(1+  L_{+})-\ln(1+  L_{-})$; thus one writes
\begin{equation}\label{confirming_linear_response}
			\Longrightarrow\,\,\,\,\,\,\,\,\,\,\,\, \left(1-\frac{1}{2\tau \nabla}\ln \frac{\frac{1}{\tau}+\partial_t+\nabla}{\frac{1}{\tau}+\partial_t-\nabla}\right)n(x)\,=\,0\;.
\end{equation}
 This equation is indeed the resummed version of Eqs.~\eqref{diffusion_all_order} and \eqref{diff_eq_higher_orders}.
      \color{black} 	
	
	Note that this equation can be also derived from Romatschke's work \cite{Romatschke:2015gic}, simply by integrating his Eq.~(10) and setting the external electric field to zero.
	Thus as it is expected, by transforming $n(t, \textbf{x})$ to Fourier space, the eigenmode of the above equation is found to be exactly the same as the diffusion pole found from the linear-response analysis in Ref.~\cite{Romatschke:2015gic}\footnote{In the context of $1+1$-dimensional QFTs, it was recently shown that when the system is near a conformal field theory (CFT) with a large central charge, the constitutive relations in the hydrodynamic limit can be found to all orders in derivatives, leading to resummed dispersion relations \cite{Davison:2024msq}.}
	\begin{equation}\label{mode_non_pertrub}
		\omega_{\text{p}}=-\frac{i}{\tau}+\, i k \cot\big(\tau k \big)\;.
	\end{equation}
	This identically reproduces Eqs.~\eqref{diffusion_mode} and \eqref{higher_transport_coeff} when expanding in $k$ \textcolor{black}{(see Ref.~\cite{Bajec:2024jez} for representing Eq.~\eqref{higher_transport_coeff}  in terms of the
Bernoulli numbers)}. 
	However,  Eq.~\eqref{mode_non_pertrub} applies for the whole range of momenta, including $k>k_c$. 
	Thus this is nothing but the analytic continuation of Eq.~\eqref{diffusion_mode} to the entire complex-momentum plane.  
	This also indicates that the radius of convergence is determined by the  pole singularity of the $\cot (\tau k)$ function, the nearest to $k=0$, namely \textcolor{black}{$k_{p}=\frac{\pi}{\tau}$}, which agrees with $k_c$ found earlier.

	In summary, we regard Eq.~\eqref{Resummed} as a local derivative expansion (see Eq.~\eqref{diffusion_all_order}) with a finite radius of convergence in $k$-space.
	The operator $(1+\tau \textbf{D})^{-1}$ is used, although it is clearly non-local, only to express the derivative expansion in resummed form. 
	Integrating over solid angle yields the logarithm, see Eq.~\eqref{confirming_linear_response}.
	In Fourier space, this logarithm must be expanded in $\omega$ and $k$, giving a polynomial in $\omega$ and $k$, which is sometimes referred to as the derivative expansion in frequency/momentum space. This expansion does not converge 
	for all values of $\omega$ and $k$. However, we always treat the logarithm as a whole; in 
	this way, we achieve an analytical continuation of the derivative expansion from its convergence domain to the entire complex-frequency and complex-momentum plane.
	
	In the following sections, we follow the same logic: we express the results, namely the Fourier-space correlation function, in a form involving the same logarithm.
	However, we always consider that this must be expanded in terms of $\omega$ and $k$.

	Before ending this section, let us note an important feature  of the resummed equations given by Eq.~\eqref{Resummed}. 
	The left-hand side box is a general result for kinetic theory in RTA. 
	The equilibrium distribution function $f^{(0)}$ could be associated with any arbitrary microscopic system. 
	The right-hand side box, however, is specifically related to a system of massless particles described by the Maxwell-Boltzmann distribution in equilibrium. 
	We will continue to work with this system in the following.

	\section{Kinetic-theory correlation functions from Schwinger-Keldysh EFT}
	\label{SK_EFT}
	So far we discussed how to systematically derive the diffusion equation from kinetic theory in RTA.
	Equation \eqref{diffusion_all_order} (or equivalently Eq.~\eqref{Resummed}) reflects the dissipative feature of the system. 
	However, in order to consistently describe the state of the system, we must take into account the effect of fluctuations around equilibrium. 
	The latter is equivalent to investigating the \textit{correlation functions} of the theory. 
	We will not explore this in the kinetic-theory framework. 
	Instead, we will construct a Schwinger-Keldysh EFT to describe the fluctuations of a density field whose dissipation follows from Eq.~\eqref{Resummed}.
	The latter is sufficient for constructing the non-interacting classical effective action  in the Schwinger-Keldysh framework.\footnote{In this work we will only study statistical fluctuations. Quantum fluctuations will not be included.} 
	The fluctuation part is fixed by imposing the Kubo-Martin-Schwinger (KMS) constraint. 
	The whole interacting part in our EFT will then be constructed just based on general EFT arguments.

	
	\subsection{Review of the Schwinger-Keldysh EFT}
	\label{sec3.1}
	In this subsection, we recall some relevant aspects of Schwinger-Keldysh EFT developed in Ref.~\cite{Crossley:2015evo} (see Ref.~\cite{Liu:2018kfw} for a review of the topic \footnote{See also Refs.~\cite{Grozdanov:2013dba,Glorioso:2016gsa,Glorioso:2017fpd,Glorioso:2017lcn,Jensen:2018hse,Jensen:2018hhx,Jensen:2017kzi,Haehl:2018lcu,Kovtun:2014hpa}.}).  
	In the Schwinger-Keldysh framework, the low-energy, long-wavelength dynamics is described in terms of two fields living on the two legs of a Closed Time-Path (CTP) contour: $\phi_{1,2}$ or equivalently $\phi=\frac{1}{2}(\phi_1+\phi_2)$ and $\phi_a=\phi_1-\phi_2$. 
	The CTP path integral in the presence of external sources can then be written as
	\begin{equation}\label{gen_func}
		Z[A_{\mu,1},A_{\mu,2}]=\,\int \mathcal{D}\phi_1\mathcal{D}\phi_2 e^{i S_{\text{EFT}}[\phi_{1,2};A_{\mu,12}]}\;,
	\end{equation}
	with $S_{\text{EFT}}=\int dt d^{3} x \mathcal{L} \,\equiv\int_{x}\mathcal{L}$. 
	For a diffusive system, the classical Lagrangian in the $r-a$ basis is quadratic in $\phi_a$ \cite{Crossley:2015evo}:
	\begin{equation}\label{SK_general}
		\mathcal{L}=\,H B_{a, 0}+ G_i B_{a,i}+ i B_{a,0}M_1B_{a,0}+ i B_{a,i}M_2B_{a,i} + i B_{a,0}M_{3,i}B_{a,i} \;,
	\end{equation}
	where $H$ and $G_i$ are functionals of $B_{r,t}$ and $\partial_tB_{r,i}$, with $B_{r,\mu}\equiv \partial_{\mu}\phi+{A}_{r,\mu}$. 
	In addition, $B_{a,\mu}=\partial_{\mu}\phi_a+A_{a,\mu} $. 
	Note that $A_{r,a}$ are the external sources coupled to $\phi$ and $\phi_a$. 
	
	A few comments (see Ref.~\cite{Crossley:2015evo} for details):
	\begin{itemize}
		\item Each term in $\mathcal{L}$ must have at least one factor of $B_a$.
		\item Varying the classical action with respect to $\phi_a$, and setting the external sources to zero and also $\phi_a=0$ (as the boundary condition), one finds:
		\begin{equation}\label{EoM_EFT}
			\partial_t H+\nabla_i G_i=0\;.
		\end{equation}
		This conservation equation is in fact the \textit{classical}  conservation equation for the current density.
		Since our EFT is supposed to describe  diffusion, we take Eq.~\eqref{EoM_EFT} to be the diffusion equation. 
		Depending on whether both $H$ and $G_i$ are linear functions of the density or not, Eq.~\eqref{EoM_EFT} would correspond to the linear or the non-linear diffusion equation, respectively.
		\item  $H$ and $G_i$ represent the dissipative aspects of diffusion.
		Once they are given, $M_1$, $M_2$, and $M_{3,i}$ are fixed via imposing the dynamical KMS symmetry. 
		The latter determines the fluctuation aspects of diffusion in the system. 
		For the free (non-interacting) Lagrangian, $M_1$, $M_2$, and $M_{3,i}$  are just constructed from derivative operators.  
		\item The Schwinger-Keldysh EFT can  include the effect of both statistical and quantum fluctuations.
		By limiting the KMS condition to the classical limit, we will only include the effect of statistical fluctuations in this work. 
		\item Correlation functions are given by:
		\begin{equation}\label{G_general}
			G^{\mu_1\cdots\mu_n}_{\alpha_a, \cdots, \alpha_n}(t_1,t_2,\cdots,t_n)=\frac{1}{i^{n_r}}\frac{\delta^n \ln Z}{\delta A_{\bar{\alpha}_1\mu_1}(t_1)\cdots \delta A_{\bar{\alpha}_n\mu_n}(t_n)}\bigg|_{A_{a,r}=0}\;,
		\end{equation}
		where $\alpha=\{r,a\}$ while $\bar{\alpha}=\{a,r\}$. In this work, we will be eventually interested in calculating $G_{rrr}$ defined as
		\begin{equation}\label{def_Grrr}
			G_{rrr}(t_2,\textbf{x}_2;t_1,\textbf{x}_1)\equiv	G^{000}_{rrr}(t_2,\textbf{x}_2;t_1,\textbf{x}_1)=\frac{1}{i^3}\frac{\delta^3 \ln Z}{\delta A_{a0}(t_2,\textbf{x}_2)A_{a0}(t_1,\textbf{x}_1)A_{a0}(0,\textbf{0})}\bigg|_{A_{a,r}=0}\,.
		\end{equation}
	\end{itemize}

	\subsection{Resummed quadratic Schwinger-Keldysh action and free correlators}
	\label{SK}
	Following the discussion in the beginning of the section and also in the previous subsection, we incorporate the \textit{linear} diffusion equation from our kinetic-theory calculation into the effective Lagrangian. 
	Equation \eqref{Resummed} is indeed what assumes the role of Eq.~\eqref{EoM_EFT}. 
	Identifying $\mu\equiv \dot{\phi}$,\footnote{This is indeed reasonable, because $B_{r0}=A_{r0}+\dot{\phi}$.} \textcolor{black}{near equilibrium we have $\mu(t,\textbf{x})=\mu_0+\delta \mu(t,\textbf{x})$. 
    Equation~\eqref{Resummed}, however, is in terms of $n(x)$.  
    Similarly, we have $n(t,\textbf{x})=n_0+\delta n(t,\textbf{x})$. 
    Then   $\delta n$ can be related to $\delta \mu$ via $	\delta n(\mu)=\,n(\mu(t,\textbf{x}))-n(\mu_0)\equiv \,\chi\,\delta \mu$}, where $\chi$ is the static charge-number susceptibility (see Eq.~\eqref{tau_chi} and the explanations below it for a careful definition). \textcolor{black}{For the sake of brevity,  we drop the $\delta$'s and simply write $n$ and $\mu$. Consequently,} the free Lagrangian takes the following form:
	\begin{equation}\label{L_0}
		\mathcal{L}^{\text{free}}\textcolor{black}{[n, \phi_a]}	=\,	-\int_{\Omega}\left( \textbf{D}\frac{1}{1+\tau \textbf{D}}n \right)\,\phi_a+\mathcal{O}(\phi_a^2)\;.
	\end{equation}
	In order to be consistent with the general form of the Schwinger-Keldysh Lagrangian \eqref{SK_general} (see also Ref.~\cite{Crossley:2015evo}) and also for our later requirements, we need to integrate by parts the terms in Eq.~\eqref{L_0}. 
	As usual, this integration by parts is done on the level of the effective action.
	We find the following resummed effective quadratic Lagrangian (see \App{KMS_L_2} for details)
	\begin{equation}\label{L_2}\boxed{
			\mathcal{L}^{\text{free}}\textcolor{black}{[n, \phi_a]}	= \int_{\Omega}\bigg[(\textbf{D}\phi_a)\frac{1}{1+ \tau\textbf{D}} n  +iT\sigma\,\, (\textbf{D}\phi_a)\,\bigg(\frac{1}{1+ \tau \textbf{D}}\bigg)_{\Theta}\textbf{D} \phi_a\bigg]} \;,
	\end{equation}
	where $\sigma\equiv \chi \tau$ and we have also defined
	\begin{equation}
		(A)_{\Theta}\equiv\frac{1}{2}(1+\Theta)A\;,
	\end{equation}
	for any quantity $A$.
	Here, $\Theta \equiv \mathcal{PT}$ is the discrete parity and time-reversal transformation, acting on $A$.
	The quantity $(A)_\Theta$ is thus invariant under this transformation, i.e., it is symmetric under $\mathcal{PT}$.
	Note that the Lagrangian \eqref{L_2} looks  non-local.  
	However, as was discussed below Eq.~\eqref{mode_non_pertrub}, in practice we treat the Lagrangian as a derivative expansion.
	Since in momentum space, the expansion can be always analytically continued to the entire momentum plane (except for the singular points), we choose to express the Lagrangian by this seemingly non-local form.\footnote{In holography, the linear constitutive relations associated with diffusion have been found in Ref.~\cite{Bu:2015ame} 
		The latter is equivalent to having the quadratic (free) effective action for the boundary theory; see also \cite{Bu:2020jfo}. 
		In this work, however, we will go beyond this and construct the cubic action corresponding to the (leading) nonlinear constitutive relations to all order in derivatives. }
	
	Writing the effective action in the form  
	\begin{equation}\label{S_parameterization}
		S^{\text{free}}_{EFT}=\frac{i}{2}\int_{\omega,\textbf{k}}\Phi^{a}_{\omega,\textbf{k}}P_{ab}(\omega,\textbf{k})\Phi^{b}_{-\omega,-\textbf{k}}\;,
	\end{equation}
	where $P$ is a positive definite matrix, the two-point function reads
	\begin{equation}\label{correlators}
		\langle\Phi_{a} \Phi_{b} \rangle_{\omega, \textbf{k}} \equiv \langle\Phi_{a}(\omega, \textbf{k}) \Phi_{b}(-\omega, -\textbf{k}) \rangle=\,(P^{-1})_{ab}(\omega,\textbf{k})\,.
	\end{equation}
	Introducing the notation $p\equiv(\omega, \textbf{k})$,  $S_{EFT}^{\text{free}}=\int_{x}\mathcal{L}^{\text{free}}$ takes the following form  in Fourier space
	\begin{equation}\label{S_EFT}
		S^{\text{free}}_{EFT}=\int_{\omega, \textbf{k}} \,\int_{\Omega}\, \bigg[-\frac{D_p}{1+ \tau D_p} n_{p}\,\phi_{a,-p}+  i T\sigma  D_p D^*_{p}\,
		\text{Re}\Big(\frac{1}{1+\tau D_p}\Big)\phi_{a,p}\,\phi_{a,-p}\bigg]\;,
	\end{equation}
	where $D_{p}\equiv D_{\omega,\textbf{k}}=\,- i \omega + i k\,\cos \theta$. 
	Using Eq.~\eqref{correlators}, we find
	\begin{eqnarray}\label{n_phi_prop}
		\langle n\phi_{a}\rangle_p&=&\frac{- i\,\tau}{1-L_p}\;,\\\label{phi_n_prop}
		\,\,\,\,\,\,\,\,\,\,\langle \phi_an\rangle_p&=&\frac{- i\,\tau}{1-L^*_{p}}\;, \\\label{n_n_prop}
		\langle n n \rangle_p&=&\frac{T \chi \tau\, \Big(2-L_p-L^*_{p}\Big)}{(1-L_p)(1-L^*_{p})}=\,i
		T\chi\,\Big(\langle n \phi_{a} \rangle_p+\langle \phi_a n \rangle_p\Big)\;,
	\end{eqnarray}
	where the last equality on the right-hand side of Eq.~\eqref{n_n_prop} is the manifestation of the fluctuation-dissipation theorem, and where 
	\begin{equation}\label{L_p}
		L_p=\frac{1}{2 i \tau k}\ln\left(\frac{\frac{i}{\tau}+\omega-k}{\frac{i}{\tau}+\omega+k}\right)\;.
	\end{equation}
	See Appendix \ref{integrals} for details.
	
	Let us find the small-frequency/momentum limit of the above result. 
	Assuming that $\omega\sim k^2\sim \epsilon^2$, to leading order in $\epsilon$ we find the familiar hydrodynamic correlators \cite{Chen-Lin:2018kfl,Jain:2020zhu,Michailidis:2023mkd}
	\begin{eqnarray}\label{n_phi_prop_hydro}
		\langle n \phi_{a}\rangle_p=\,	\langle \phi_a n \rangle_{-p}&=&\frac{1}{\omega + i\frac{\tau}{3} k^2}\;,\\\label{n_n_prop_hydro}
		\langle n n \rangle_p&=&\frac{2 T \chi \frac{\tau}{3}k^2}{\omega^2 + \big(\frac{\tau}{3}\big)^2 k^4}\;.
	\end{eqnarray}
	To obtain Eqs.~\eqref{n_phi_prop_hydro}, we expand the denominators in Eqs.~\eqref{n_phi_prop} and \eqref{phi_n_prop} and keep the terms to second, i.e., leading order in $\epsilon^2$. 
	The expression in Eq.~\eqref{n_n_prop_hydro} was found applying the right-hand side of Eq.~\eqref{n_n_prop}.

	\subsection{Physical correlators}
	\label{sec3.3}
	According to Eq.~\eqref{G_general}, physical correlators can be computed by performing derivatives with respect to the sources.
	For this reason, we need to rewrite the effective Lagrangian in the presence of such sources. 
	Since we want to find the $G_{rr}$ correlator, we only need to turn on the $a0$ sources. 
	Following the standard techniques developed in Ref.~\cite{Crossley:2015evo}, we find (see Appendix \ref{KMS_L_2} for details; in particular Eqs.~\eqref{gauging} and \eqref{gauging_L_2})
	\begin{equation}\label{L_2_sousrced}
		\begin{split}
			\mathcal{L}^{\text{free}}_{\text{source}}(x)	=	& \; A_{a0}\frac{1}{1+ \tau\textbf{D}} n  +i T \sigma A_{a0}\bigg(\frac{1}{1+\tau \textbf{D}}\bigg)_{\Theta}\textbf{D} \phi_a \\
			& +i T \sigma (\textbf{D} \phi_a)\bigg(\frac{1}{1+\tau \textbf{D}}\bigg)_{\Theta} A_{a0}+ i T \sigma A_{a0}\bigg(\frac{1}{1+  \tau\textbf{D}}\bigg)_{\Theta} A_{a0}  \;.
		\end{split}
	\end{equation}
	%
	Now we are ready to read off the physical correlators. 
	We choose to derive $G_{rr}(p)=G^S_{J^0 J^0}(p,-p)$ defined as
	\begin{equation}\label{G_nn_s}
		\begin{split}
			G^{S}_{J^0J^0}(p,-p)&=\,\frac{1}{i^2}\frac{\delta^2 \ln Z}{\delta A_{a0}(-p)\delta A_{a0}(p)}\\
			&=\,\Big\langle \int_{\Omega}\Big[\frac{1}{1+\tau D_{p}}n_{p}+ 2 i T \sigma \text{Re}\left(\frac{1}{1+ \tau D_{p}}\right)D_{p}\phi_{a,p}\Big]\\
			&\,\,\,\,\,\,\,\,\,\,\,\,\times\int_{\Omega}\Big[\frac{1}{1+\tau D^*_{p}}n_{-p}+ 2 i T \sigma \text{Re}\left(\frac{1}{1+ \tau D^*_{p}}\right)D^*_{p}\phi_{a,-p}\Big] \Big\rangle\\
			& \quad +\,2 T \sigma \int_{\Omega}\text{Re}\left(\frac{1}{1+\tau D_p}\right)\\ 
			&=\,L_pL^*_{p}\langle n n\rangle_p+\frac{i T \sigma}{\tau}L^*_{p}(L^*_{p}-L_p)\langle\phi_a n\rangle_p \\
			& \quad +\frac{i T \sigma}{\tau}L_p(L_p - L^*_{p})\langle n \phi_a \rangle_p+\,\frac{T \sigma}{\tau} (L_p + L^*_{p})\;.
		\end{split}
	\end{equation}
	Using Eqs.~\eqref{n_phi_prop}$-$\eqref{L_p}, this then simplifies to
	\begin{equation}\label{G_nn_S}
		\begin{split}
			G^{S}_{J^0 J^0}(\omega, \textbf{k})
			&=\, \chi T \tau\bigg(\frac{L_p}{1-L_p}+\frac{L^*_{p}}{1-L^*_{p}}\bigg)\\
			&=-2 \chi T \tau+\frac{\chi\,T}{2 k}\,\frac{4 \tau k-i\ln\left(\frac{\frac{i}{\tau}-\omega+k}{\frac{i}{\tau}-\omega-k}\right)+i\ln\left(\frac{\frac{i}{\tau}+\omega-k}{\frac{i}{\tau}+\omega+k}\right)}{\left[1+\frac{1}{2 i \tau k}\ln\left(\frac{\frac{i}{\tau}-\omega+k}{\frac{i}{\tau}-\omega-k}\right)\right]\left[1-\frac{1}{2 i \tau k}\ln\left(\frac{\frac{i}{\tau}+\omega-k}{\frac{i}{\tau}+\omega+k}\right)\right]}\;.
		\end{split}
	\end{equation}
	This result is in agreement with the retarded density Green's function, $G^R_{J^0 J^0}(\omega, \textbf{k})$, found in Ref.~\cite{Romatschke:2015gic}.\footnote{In terms of $r-a$ correlators introduced by Eq.~\eqref{G_general}, $G^R_{J^0J^0}(\omega, \textbf{k})_{\mbox{\cite{Romatschke:2015gic}}}\equiv G_{ra}(\omega, \textbf{k})$.}
	In Ref.~\cite{Romatschke:2015gic}, Romatschke calculated the retarded Green's function of the charge density for kinetic theory in RTA in the framework of linear response.
	While Romatschke's result expresses the dissipative nature of the system, our EFT is able to feature the fluctuation effect as well, i.e., Eq.~\eqref{G_nn_S}. 
	However, to check the consistency of our results with Romatschke's work, we now apply the fluctuation-dissipation theorem,\footnote{The convention for defining $G^{R}$ in our Schwinger-Keldysh EFT differs from that in Ref.~\cite{Romatschke:2015gic} by a minus sign. In our Schwinger-Keldysh formalism, the fluctuation-dissipation theorem is expressed as $G_{rr}=\frac{2T}{\omega}\text{Im}G_{ra}$ \cite{Liu:2018kfw}. Thus Eq.~\eqref{FD} has a minus sign for consistency. }
	\begin{equation}\label{FD}
		G^S_{J^0J^0}(\omega, \textbf{k})=\,-\frac{2T}{\omega} \,\text{Im} \,G^{R}_{J^0J^0}(\omega, \textbf{k})\;,
	\end{equation}
	to Eq.~(14) in Ref.~\cite{Romatschke:2015gic}. 
	The latter can be written in the following form:
	\begin{equation}\label{G_R_Romatschke}
		G^R_{J^0J^0}(\omega, \textbf{k})_{[10]}=\,-\chi\Biggl[1- i \tau \omega+\frac{i \tau \omega}{1-\frac{1}{2 i \tau k}\ln\biggl(\frac{\frac{i}{\tau}+\omega-k}{\frac{i}{\tau}+\omega+k}\biggl)}\Biggl]\;.
	\end{equation}
	Substituting Eq.~\eqref{G_R_Romatschke} into Eq.~\eqref{FD} gives precisely $G^S_{J^0J^0}$ in Eq.~\eqref{G_nn_S}. \textcolor{black}{As expected, we obtain the same result for the retarded two-point function as Ref.~\cite{Romatschke:2015gic}, however, using the Schwinger-Keldysh EFT approach.}
	\begin{figure}
		\centering
		\includegraphics[width=0.48\textwidth]{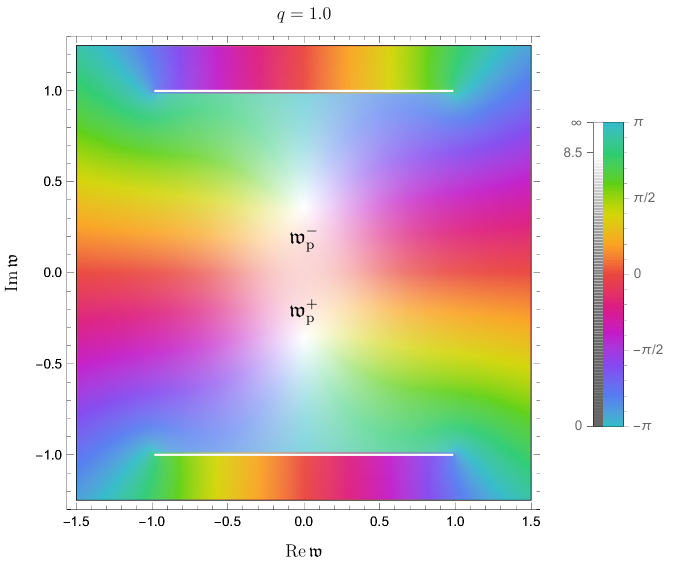}	\,\,
		\includegraphics[width=0.475\textwidth]{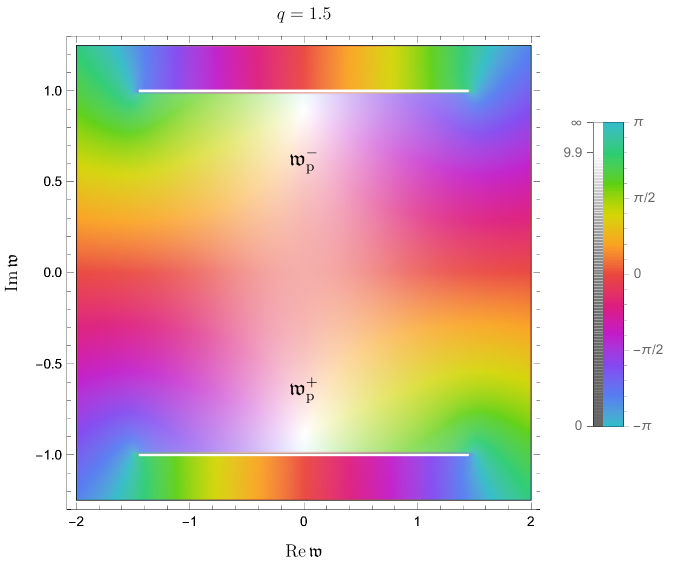}	\,\,
		\caption{The complex-valued function $G_{rr}(\wn)=|G_{rr}(\wn)|e^{i \alpha}$ in the complex-frequency plane, with $\wn\equiv\tau \omega$ and $\qn\equiv\tau \textbf{k}$. White dots illustrate the poles: $\wn=\mp i\pm i q \cot q$ with $q=|\qn|$. The thick white lines at $\wn=\pm i$ show the branch cuts.}
		\label{G_nn}
	\end{figure}

	In Fig.~\ref{G_nn}, we have illustrated the analytic structure of $G^S_{J^0J^0}$ in the complex-frequency plane for two different values of $k$. 
	As long as $k<k^*=\frac{\pi}{2\tau}$, there are two simple poles, $\omega_{\text{p}}^{+}$ (given by Eq.~\eqref{mode_non_pertrub}) and $\omega_{\text{p}}^{-}=(\omega_{\text{p}}^{+})^*$, associated with retarded and advanced Green's functions, respectively. 
	There are also two pairs of logarithmic branch points, each connected to each other by a branch cut. 
	The corresponding branch cuts can be chosen as
	\begin{equation}\label{branch_cuts}
		\mp\frac{i}{\tau}-k< \omega<\mp \frac{i}{\tau}+k\;.
	\end{equation}
	
	The higher the value of $k$, the farther the branch points are from each other and the closer the two poles to the branch cuts. 
	Eventually, when $k$ exceeds $k^*$, the two poles disappear. 
	These findings are in agreement with Romatschke's result about the analytic structure of the response function in the lower half of the complex $\omega$-plane.
	
	\section{Interactions: beyond the quadratic effective action}
	\label{SK_int}
	In our diffusive system, the interactions emerge through the self-interaction of the density. 
	This can be included in the Lagrangian by promoting the EFT coefficients in the free Lagrangian to be functions of the dynamical field   $\dot{\phi}\equiv \mu$. 
	There are three of these Wilsonian coefficients: $\tau$, $\chi$, and $\sigma$. 
	In terms of the fluctuations around a homogeneous state, that is, $\mu(t,\textbf{x})=\mu_0+\delta \mu(t,\textbf{x})$, we may then write:
	\begin{equation}\label{tau_chi}
		\begin{split}
			\delta n(\mu)=&\,n(\mu(t,\textbf{x}))-n(\mu_0)=\,\chi\,\delta \mu+\chi' \frac{\delta\mu^2}{2}  +\ldots\,\,\,\,\,\,\Longrightarrow\,\,\,\,\,\chi(\mu)=\frac{\partial n}{\partial \mu}=\,\chi+\chi' \delta\mu+\ldots\;,\\
			\tau(\mu)=&\,\tau+\tau'\, \delta\mu+\ldots\;,\\
			\sigma(\mu)=&\,\sigma+\sigma'\,\delta \mu+\ldots\;,
		\end{split}
	\end{equation}
	\textcolor{black}{Let us emphasize that Eq.~\eqref{tau_chi} does not provide the most general way of incorporating nonlinearities into our system. The coefficients $\tau'$ and $\sigma'$ can in principle be expanded in terms of derivative operators, leading to additional terms. 
    However, this process is not systematic, as it requires the introduction of infinitely many new coefficients to fully capture these effects.}
		
				\textcolor{black}{This contrasts with the case of linear terms in our system, where all coefficients are explicitly determined in terms of $\tau$ and $\sigma$, and nonlinear terms involve only two additional coefficients, $\tau'$ and $\sigma'$. Although the assumption underlying Eq.~\eqref{tau_chi} may seem restrictive, as we will demonstrate in Sec.~\ref{transport}, our cubic action successfully captures all possible linear and (quadratically) nonlinear transport effects that can be incorporated into the constitutive relation for the current, up to $\mathcal{O}(\epsilon^4)$. }
				
			\textcolor{black}{Thus, although this is a minimal model, it effectively reveals the physics associated with three-point functions. In the following, we will take advantage of the well-defined structure of our model to systematically study three-point functions at higher orders in derivatives. }
	
	For the sake of brevity,  we again drop the $\delta$'s and simply write $n$ and $\mu$ (in Eq.~\eqref{tau_chi}) in the following.
	Note that we are interested in the \textit{leading} interactions that arise as a result of the self-interactions described above. 
	For this, we truncate Eq.~\eqref{tau_chi} to first order and construct the cubic interactions in the Lagrangian. 

\textcolor{black}{To illustrate how this framework operates in practice, let us apply Eq.~\eqref{tau_chi} to the first term of the quadratic effective action \eqref{L_2}:	
	\begin{equation}\label{C_3}
		\begin{split}
		\textbf{D}\phi_a\frac{1}{1+ \tau(\mu)\textbf{D}} n\, =\,&	\textbf{D}\phi_a\frac{1}{1+ \tau\textbf{D}} n -	\textbf{D}\phi_a\frac{1}{1+ \tau\textbf{D}}\Big(\tau' \mu\,\textbf{D}\,\frac{1}{1+ \tau\textbf{D}}n\Big)+\ldots\\
			=\,&	\textbf{D}\phi_a\frac{1}{1+ \tau\textbf{D}} n -\,\frac{\tau' }{\chi}\,n\,\,\left(\frac{\textbf{D}}{1+ \tau\textbf{D}}n\right) \frac{1}{1- \tau\textbf{D}}	\textbf{D}\phi_a+\ldots\;,
		\end{split}
	\end{equation}
	where we have used the fact that $(A+a)^{-1}=A^{-1}-A^{-1}aA^{-1}+\mathcal{O}(a^2)$. 
	From the first to the second line we have replaced $\mu=\frac{n}{\chi}+\mathcal{O}(n^2)$ and also have performed an integration by parts.
	Let us note that the dots contain terms beyond the cubic order in the fields.
}

\textcolor{black}{
	Similarly, we can apply Eq.~\eqref{tau_chi} to the second term of the quadratic effective action \eqref{L_2}. This leads to the following resummed cubic effective Lagrangian: }
	\begin{equation}\label{L_3}\boxed{
			\mathcal{L}^{(3)}	=\int_{\Omega}\bigg[-\frac{\tau' }{\chi}n\left(\frac{\textbf{D}}{1+ \tau\textbf{D}}n\right)\frac{\textbf{D}}{1- \tau\textbf{D}}\phi_a+ i T \textcolor{black}{\frac{\sigma'}{\chi}}\,n\left(\frac{\textbf{D}}{1+ \tau\textbf{D}}\phi_a  \right) \frac{\textbf{D}}{1- \tau\textbf{D}}\phi_a\bigg]} \;.
	\end{equation}
\textcolor{black}{The final step is to impose the KMS condition on the above action. As a result, we obtain the relation $\sigma'= \chi \tau'$}.	See Appendix \ref{App_self_int} for details.
	
	\textcolor{black}{In order to calculate the three-point correlation functions, it is required to properly couple the above effective action to the external sources (see Eq.~\eqref{G_general}).}
		\textcolor{black}{Among all possible three-point functions, we focus on calculating $G^{000}_{rrr}$, which is the symmetrized correlation function of the charge density, i.e., $\langle J^0 J^0 J^0\rangle_{\text{sym.}}$. There are two primary reasons for this choice. \textit{First}, this correlator can be directly measured or simulated in real-time experiments \cite{Delacretaz:2023ypv}.   \textit{Second}, from a technical perspective, it is the simplest three-point correlator to compute. According to Eq.~\eqref{G_rrr}, we only need to turn on the scalar $a$-source $A_{a0}$. }
	\textcolor{black}{By implementing the replacement $\phi_a \rightarrow \phi_a +\partial_0 A_{a0}$, as suggested in Ref.~\cite{Liu:2018kfw}, we obtain the source part of the cubic action as follows}
\begin{equation}\label{L_3_source}
\begin{split}
	\mathcal{L}^{(3)}=&\,\int_{\Omega}\left[  -\frac{\tau' }{\chi}n\left(\frac{\textbf{D}}{1+ \tau\textbf{D}}n\right)\frac{1}{1- \tau\textbf{D}} \textcolor{black}{A_{a0}} + i T \tau'\,n\left(\frac{1}{1+ \tau\textbf{D}} \textcolor{black}{A_{a0}}\right)  \frac{\textbf{D}}{1- \tau\textbf{D}}\phi_a \right.
\\
& \qquad +\left. i T \tau'\,n\left(\frac{\textbf{D}}{1+ \tau\textbf{D}}\phi_a \right) \frac{1}{1- \tau\textbf{D}} \textcolor{black}{A_{a0}}+ i T \tau'\,n\left(\frac{1}{1+ \tau\textbf{D}} \textcolor{black}{A_{a0}}\right)  \frac{1}{1- \tau\textbf{D}} \textcolor{black}{A_{a0}}\right]\;.
	\end{split}
\end{equation}
\textcolor{black}{
	As can be seen, even in this seemingly simple case, the coupling to external sources introduces additional complexity. In addition to the two vertices present in Eq.~\eqref{L_3}, we obtain four new vertices arising from the source coupling. This significantly increases the complexity of computing the three-point function.
		Moreover, it should be noted that each of these six vertices is itself a complicated function in momentum space, further complicating the analysis. In the next subsection, we first elaborate on the structure of the two vertices in Eq.~\eqref{L_3} before addressing the full computation.}

\textcolor{black}{In summary, the technical advantage of calculating $G^{000}_{rrr}$ lies in the fact that, at the very least, the coupling of the action to the sources is straightforward, as demonstrated in the transition from Eq.~\eqref{L_3} to Eq.~\eqref{L_3_source}, resulting in only six vertices to consider.
	In contrast, computing a correlator such as $G^{000}_{raa}$ requires first rewriting Eq.~$\eqref{L_3}$ in terms of $\phi$ and $\phi_a$ and then introducing  both $A_a$ and $A_r$ sources. This process generates a significantly larger number of vertices, including terms with field content such as $nA_{a0}A_{ri}$, $A_{ri}A_{ri}A_{a0}$, and similar additional interactions.	
	Due to this increased complexity, we focus exclusively on the calculation of $G^{000}_{rrr}$  in this work.} \textcolor{black}{We plan to calculate $G^{000}_{raa}$ in future work \cite{Abbasi:2025}.}

	Before proceeding further, 
	we need to elaborate on a technical point. 
	In Eq.~\eqref{L_2}, we can explicitly perform the integration over $\Omega$.
	However, the result turns out to be highly non-local and complicated.
	Then applying Eq.~\eqref{tau_chi} to this non-local form of the free Lagrangian will be technically difficult. 
	For this reason, we choose not to do so; instead, we first applied  Eq.~\eqref{tau_chi} to Eq.~\eqref{L_2} and found   Eq.~\eqref{L_3}. 
	Now both Eqs.~\eqref{L_2} and \eqref{L_3} are ready to be integrated over $\Omega$. 
	Needless to say that to read off the correlation functions, we will have to eventually evaluate the $\Omega$-integrals.  
	
	
	\subsection{Analytic structure of vertices}
	\label{sec4.1}
	One final goal of this work is to calculate the three-point function of the density.
	To this end,  we first need to perform the $\Omega$-integral in Eq.~\eqref{L_3}. 
	As it is seen in Fig.~\ref{vertices}, we have absorbed the whole  $\Omega$-integral into the vertices; there is a dimensionless integral expression in the vertices. 
	In Fourier space, in the absence of sources, the cubic action $S_{EFT}^{(3)}=\int_{x}\mathcal{L}^{(3)}$ takes the following simple form
	\begin{equation}\label{S_EFT_3}
		S_{EFT}^{(3)}= \int_{1,2,3}\delta(1+2+3)\Big[\lambda_1(1,2,3)\,n_{1}\,n_{2}\,\phi_{a,3}+ \lambda_2(1,2,3)\,n_{1}\, \phi_{a,2}\,\phi_{a,3} \Big]\;,
	\end{equation}
	where $n_{\ell}\equiv n_{\omega_{\ell}\textbf{k}_{\ell}}$, and we use the same convention for $\phi_{a, \ell}$. 
	In addition, $\int_{\ell}\equiv\int_{\omega_{\ell}\textbf{k}_{\ell}}$ and $\delta(1+2+3)\equiv\delta(\omega_1+\omega_2+\omega_3)\delta^{(3)}(\textbf{k}_1+\textbf{k}_2+\textbf{k}_3)$. 
	We have also defined
	\begin{equation}\label{lam12}
		\lambda_1(1,2,3)\color{black} \equiv\frac{\tau'}{\chi\tau^2}\,\int_{\Omega}\frac{\tau D_2}{1+\tau D_{2}}\frac{-\tau D_3}{1-\tau D_{3}}\;,\qquad
		\lambda_2(1,2,3)\color{black} \equiv -iT\frac{\tau'}{\tau^2} \int_{\Omega}\frac{\tau D_2}{1+\tau D_{2}}\frac{-\tau D_3}{1-\tau D_{3}}\;.
	\end{equation}
	The only thing that we now need to do is the evaluation of the integral over $\Omega$ in the  vertex structure. 
	For this, we parameterize the momenta as follows: 
	\begin{equation}\label{Second_int_Fourier}
		\begin{split}
			\int_{\Omega}\frac{\tau D_2}{1+\tau D_{2}}\frac{-\tau D_3}{1-\tau D_{3}}=	\int_{\Omega}\frac{\tau(- i \omega_2 + i k_2\,\sin \theta \cos \phi) }{1+\tau(- i \omega_2 + i k_2\,\sin \theta \cos \phi )}\frac{-\tau[- i \omega_3 + i k_3\,\sin \theta \cos (\phi-\alpha_{23})] }{1-\tau[- i \omega_3 + i k_3\,\sin \theta \cos (\phi-\alpha_{23})  ]}\;.
		\end{split}
	\end{equation}
	As before $\int_{\Omega}\equiv\int\frac{d\Omega}{4\pi}=\,\frac{1}{4\pi}\int_{-1}^1d(\cos\theta) \int_{0}^{2\pi}d\phi$.  In the parameterization above
	\begin{itemize}
		\item The angle between $\textbf{k}_2$ and $\textbf{k}_3$ is taken to be $\alpha_{23}$.
		\item We have taken $\textbf{k}_2$ along the $x$-axis and $\textbf{k}_3$ in the $xy$-plane.
		\item The unit vector $\boldsymbol{v}$ in $\textbf{D}=\partial_t + \boldsymbol{v}\cdot \boldsymbol{\nabla}$ is then $\boldsymbol{v}=(\sin \theta \cos \phi, \sin \theta \sin \phi, \cos \theta)$.
	\end{itemize}
	The advantage of this parameterization is that we can perform the $\theta$-integral analytically. 
	However, the resulting expression turns out to be a complicated function of $\phi$ that we call  $\mathcal{I}(\phi;  1,2,\bar{1}+\bar{2})$, with  $j\equiv p_j$ and $\bar{j}\equiv-p_j$ (see Appendix \ref{I}).
	\begin{figure}\label{vertex_lambda_1_2}
		\,\,\,\,\,\,\,\,\,\,\,\,\,\,\,\,\,\,\,\,\,\,\,\,\,\,\,\,\begin{tikzpicture}
			\begin{feynman}
				\vertex (a) at (-1.5, -1.5);
				\vertex (b) at (1.5, -1.5);
				\vertex (c) at (0, 1.5);
				\vertex [dot] (d) at (0, 0);
				\diagram* {
					(a) -- [edge label={$n_{\omega_2,\textbf{k}_2}$}] (d),
					(b) -- [edge label={$n_{\omega_1,\textbf{k}_1}$}] (d),
					(c) -- [boson, edge label={$\phi_{a\,\omega_3,\textbf{k}_3}$}] (d),
				};
				\node[below=2cm of d] {\,\,\,\,\,\,\,\,\,\,\,\,\,\(\lambda_1(1,2,3) \)	\,\,\,\,\,\,\,	};
			\end{feynman}
		\end{tikzpicture}
		\,\,\,\,\,\,\,\,\,\,\,\,\,\,\,\,\,\,\,\,\,\,\,\,\,\,\,\,\,\,\,\,\,\,\,\,\,\,\,\,
		\begin{tikzpicture}
			\centering 
			\begin{feynman}
				\vertex (a) at (-1.5, -1.5);
				\vertex (b) at (1.5, -1.5);
				\vertex (c) at (0, 1.5);
				\vertex [dot] (d) at (0, 0);
				\diagram* {
					(a) -- [boson, edge label={$\phi_{a\,\omega_2,\textbf{k}_2}$}] (d),
					(b) -- [edge label={$n_{\omega_1,\textbf{k}_1}$}] (d),
					(c) -- [boson, edge label={$\phi_{a\,\omega_3,\textbf{k}_3}$}] (d),
				};
				\node[below=2cm of d] {\( \lambda_2(1,2,3) \)};
			\end{feynman}
		\end{tikzpicture}
		\caption{Cubic vertices  formed by $n$ and $\phi_a$.}
		\label{vertices}
	\end{figure}
	
	It is hard to analytically evaluate the $\phi$-integral of the $\mathcal{I}$ function. 
	However, we can determine the singularity of the integral, or equivalently the singularity of the coupling constants $\lambda_1$ and $\lambda_2$ introduced in Fig.~\ref{vertices}, without explicitly evaluating it. 
	Singularities can potentially occur either at pinched points of the integrand or at the endpoints of the integration contour (see Ref.~\cite{Abbasi:2022rum} for introducing this issue in hydrodynamics). 
	We discuss these two types of singularities separately in the following.
	\begin{itemize}
		\item[(i)] The \textit{pinch singularity} of $\lambda_{1,2}$ is determined by a set of $(p_1,p_2,-p_1-p_2)$  for which the integration contour gets pinched between two singular points of the integrand given by the function $\mathcal{I}$.
		
		Considering the function $\mathcal{I}$ (given by Eq.~\eqref{mathcal_I}), we search for the pinched points in four distinct parts.
		Except for the first line in Eq.~\eqref{mathcal_I}, each of the other four lines contains two or more potentially singular expressions. 
		A pinch singularity corresponds to a set $(p_1,p_2,-p_1-p_2)$ for which two of the latter expressions become simultaneously singular. Here are the results corresponding to the last four lines of Eq.~\eqref{mathcal_I}:
		\begin{itemize}
			\item $2^{nd}$ line: There are two sets of singular points,
			\begin{equation}\label{set_sing}
				\begin{split}
					\{\omega_2= &- \frac{i}{\tau} - k_2  \cos \phi\,,\omega_3= \frac{i}{\tau} - k_3 \cos(\alpha-\phi)\}\;,\\
					\{\omega_2= &- \frac{i}{\tau} + k_2  \cos \phi\,,\omega_3= \frac{i}{\tau} + k_3 \cos(\alpha-\phi)\} \;.
				\end{split}
			\end{equation}
			Note that $\phi$ can assume any value in the interval $0\le \phi<2 \pi$. 
			This means the above two sets contain an infinite number of singular points, forming a branch cut:
			\begin{equation}\label{vertex_2_bc}
				\begin{split}
					- \frac{i}{\tau} - k_2  <\omega_2&<- \frac{i}{\tau} + k_2 \;, \\
					\frac{i}{\tau} - k_3  <\omega_3&< \frac{i}{\tau} + k_3 \;.
				\end{split}
			\end{equation}
			
			\item $3^{rd}$ line: The same as the $2^{nd}$ line.
			\item $4^{th}$ and $5^{th}$ line: The same as the $2^{nd}$ line together with the following two points $	\{\omega_2 \tau= - i ,\omega_3 \tau= i \}$.
		\end{itemize}
		\item[(ii)] The \textit{endpoint singularity} of $\lambda_{1,2}$ is determined by a set of $(p_1,p_2,-p_1-p_2)$  for which the integrand becomes singular at $\phi=0$ or $\phi=2\pi$. 
		It can be easily checked that such points are nothing but the branch-point singularities located at the endpoints of the branch cuts given by Eq.~\eqref{vertex_2_bc}.
	\end{itemize}

	\textit{Note:} We will use these vertices to calculate the three-point correlators of $n$ and $\phi_a$. 
	However, to find the physical three-point functions, e.g., $G_{rrr}$, we will need more information. 
	In fact  $n$ and $\phi_a$ must appropriately couple to external sources. 
	This is explained in detail in Appendix \ref{App_self_int}.
	\subsection{Tree-level three-point functions}
	\label{sec4.2}
	Having elaborated on the interactions and the analytic structure of vertices, we are now ready to calculate the density three-point function. 
	We choose to calculate $G_{rrr}$:
	\begin{equation}\label{G_rrr}
		\begin{split}
			G_{rrr}(p_1,p_2)&=\,\frac{1}{i^3}\frac{\delta^3 \ln Z}{\delta A_{a0}(-p_1)\delta A_{a0}(-p_2)\delta A_{a0}(-p_3)}\\
			&=\,\frac{1}{Z}\frac{1}{i^3}\frac{\delta^3  }{\delta A_{a0}(-p_1)\delta A_{a0}(-p_2)\delta A_{a0}(-p_3)}\int \mathcal{D}n \mathcal{D}\phi_a\,e^{i S_{\text{eff}}}\frac{1}{6}\big(i S^{\text{f-s}}+ i S^{\text{int-s}}\big)^3\\
			&\equiv\,G^{\text{f-s}}_{rrr}(p_1,p_2,p_3)+G^{\text{int-s}}_{rrr}(p_1,p_2,p_3) \;,
		\end{split}
	\end{equation}
	where $p_1+p_2+p_3=0$.
	According to the coupling of the $n$- and $\phi_a$-fields to background sources (see Appendix \ref{App_self_int}), we have divided the above expression into two parts. 
	First, $G^{\text{f-s}}_{rrr}(p_1,p_2,p_3)$ corresponds to the case where the three $A_{a0}(p_i),\, i=1,2,3$ come from $\mathcal{L}^{\text{free}}_{\text{source}}$. 
	Second, $G^{\text{int-s}}_{rrr}(p_1,p_2,p_3)$ indicates a situation where one of the three $ A_{a0}(p_i)$ comes from $\mathcal{L}^{(3)}_{\text{source}}$ with the other two ones coming from $\mathcal{L}^{\text{free}}_{\text{source}}$. 
	
	Using standard perturbative methods, for $G^{\text{f-s}}_{rrr}(p_1,p_2,p_3)$  we find
	\begin{equation}\label{G_rrr_f_s}
		\begin{split}
			&G^{\text{f-s}}_{rrr}(p_1,p_2)=\,\,\,
			\,L_1L_2L_3\,\langle nnn\rangle\\
			&+i\chi T \,\,\,
			\bigg[
			L_1L_2\big(L_3^*-L_3\big)\,\langle n n \phi_a \rangle
			+L_1\big(L_2^*-L_2\big)L_3\,\langle n  \phi_a n\rangle
			+\big(L_1^*-L_1\big)L_2L_3\,\langle  \phi_a n n \rangle
			\bigg]\\
			&+  (i \chi T)^2\,\,
			\bigg[
			L_1\big(L_2^*-L_2\big)\big(L_3^*-L_3\big)\,\langle n\phi_a\phi_a\rangle+ \big(L_1^*-L_1\big)L_2\big(L_3^*-L_3\big)\,\langle \phi_an\phi_a\rangle\\
			&\hspace*{7.6cm} +\big(L_1^*-L_1\big)\big(L_2^*-L_2\big)L_3\,\langle \phi_a\phi_an\rangle
			\bigg]\;,
		\end{split}
	\end{equation}
	where for the sake of brevity we have used the notation $L_{j}\equiv L_{p_j}$. 
	Note that in any of the three-point functions above, the three legs have momenta $p_1$, $p_2$, and $p_3$, respectively, e.g., $\langle n\phi_a\phi_a\rangle\equiv \langle n^{p_1}\phi_a^{p_2} \phi_a^{p_3}\rangle\equiv \langle n\phi_a\phi_a\rangle(p_1,p_2)$; note that $p_3=-p_1-p_2$.

	Considering $\varphi_{1},\varphi_2 \in \{n, \phi\}$, we have also used
	\begin{equation}\nonumber
		\langle \varphi_{i}\varphi_{j}\varphi_{k}\rangle(p_1,p_2)=\delta(1+2+3)\int_{x_2,x_1,x_0} \!\!\!\!\!e^{i( \omega_1 t_2+\omega_2 t_1+\omega_3 t_0)-i( \textbf{k}_1\cdot \textbf{x}_2+ \textbf{k}_2\cdot \textbf{x}_1+ \textbf{k}_3\cdot \textbf{x}_0)}\langle\varphi_{i}(x_2)\varphi_{j}(x_1)\varphi_{k}(x_0)\rangle\,.
	\end{equation}
	Let us now present the expressions associated with the correlators contributing to Eq.~\eqref{G_rrr_f_s}. First, $\langle n n n \rangle$ is calculated as follows 
	\begin{equation}\label{nnn}
		\begin{split}
			\langle n^{p_1} n^{p_2} n^{p_3} \rangle=&\quad i \Big(\lambda_1(\bar{1},\bar{2},\bar{3})+\lambda_1(\bar{2},\bar{1},\bar{3})\Big)	\langle n n \rangle_{p_1}\langle n n \rangle_{p_2}\langle n \phi_a \rangle_{p_3}\\
			&+i \Big(\lambda_2(\bar{1},\bar{2},\bar{3})+\lambda_2(\bar{1},\bar{3},\bar{2})\Big) \langle n n \rangle_{p_1}	\langle  n \phi_a \rangle_{p_2}\langle n \phi_a \rangle_{p_3}\\
			&+(231)+(312)\;,
		\end{split}
	\end{equation}
	which corresponds to the following diagrams
	\begin{equation}\label{three_point}
		\Bigg[	\Big(\vcenter{
			\hbox{\begin{tikzpicture}
					\begin{feynman}
						\vertex (a) at (-1.5, -1.5);
						\vertex (b) at (1.5, -1.5);
						\vertex (c) at (0, .75);
						\vertex (e) at (0, 1.5);
						\vertex [dot] (d) at (0, 0);
						\diagram* {
							(a) -- [fermion, edge label={$p_2$}] (d),
							(b) -- [fermion, edge label={$p_1$}] (d),
							(c) -- [boson, edge label={}] (d),
							(e) -- [fermion, edge label={$p_3$}] (c),};
					\end{feynman}
		\end{tikzpicture}}}\Big)_{\lambda_1(\bar{1},\bar{2},\bar{3})+\lambda_1(\bar{2},\bar{1},\bar{3})}
		+
		\Big(\vcenter{
			\hbox{		\begin{tikzpicture}
					\centering 
					\begin{feynman}
						\vertex (a) at (-.75, -.75);
						\vertex (f) at (-1.5, -1.5);
						\vertex (b) at (1.5, -1.5);
						\vertex (c) at (0, .75);
						\vertex (e) at (0, 1.5);
						\vertex [dot] (d) at (0, 0);
						\diagram* {
							(a) -- [boson, edge label={}] (d),
							(f) -- [fermion, edge label={$p_2$}] (a),
							(b) -- [fermion, edge label={$p_1$}] (d),
							(c) -- [boson, edge label={}] (d),
							(e) -- [fermion, edge label={$p_3$}] (c),
						};
					\end{feynman}
		\end{tikzpicture}}}
		\Big)_{\lambda_2(\bar{1},\bar{2},\bar{3})+\lambda_2(\bar{1},\bar{3},\bar{2})}\Bigg]+ (231) + (312)\;.
	\end{equation}
	Here and in what follows: $\lambda_{1,2}(\bar{i},\bar{j},\bar{k})\equiv\lambda_{1,2}(-p_i,-p_j,-p_k)$. Similarly we have 
	\begin{equation}\label{n_n_phi}
		\begin{split}
			\langle n^{p_1} n^{p_2} \phi_a^{p_3} \rangle=&\quad i \Big(\lambda_1(\bar{2},\bar{3},\bar{1})+\lambda_1(\bar{3},\bar{2},\bar{1})\Big)	\langle n \phi_a \rangle_{p_1}\langle n n \rangle_{p_2}\langle  \phi_a n \rangle_{p_3}\\
			&+i \Big(\lambda_1(\bar{1},\bar{3},\bar{2})+\lambda_1(\bar{3},\bar{1},\bar{2})\Big)	\langle n n \rangle_{p_1}\langle n \phi_a \rangle_{p_2}\langle  \phi_a n \rangle_{p_3}\\
			&+i \Big(\lambda_2(\bar{3},\bar{1},\bar{2})+\lambda_2(\bar{3},\bar{2},\bar{1}))\Big)	\langle  n \phi_a \rangle_{p_1}\langle n \phi_a \rangle_{p_2}\langle \phi_a n \rangle_{p_3}\;,
		\end{split}
	\end{equation}
	with 
	\begin{equation}\label{vertices_eq}
		\begin{split}
			\Bigg[	\Big(\vcenter{
				\hbox{\begin{tikzpicture}
						\begin{feynman}
							\vertex (a) at (-1.5, -1.5);
							\vertex (f) at (.75, -.75);
							\vertex (b) at (1.5, -1.5);
							\vertex (c) at (0, .75);
							\vertex (e) at (0, 1.5);
							\vertex [dot] (d) at (0, 0);
							\diagram* {
								(a) -- [fermion, edge label={$p_2$}] (d),
								(b) -- [boson, edge label={$p_3$}] (f),
								(f) -- [fermion] (d),
								(c) -- [boson, edge label={}] (d),
								(e) -- [fermion, edge label={$p_1$}] (c),};
						\end{feynman}
			\end{tikzpicture}}}\Big)_{\lambda_1(\bar{2},\bar{3},\bar{1})+\lambda_1(\bar{3},\bar{2},\bar{1})}
			+\,\,	\centering 
			\Big(\vcenter{
				\hbox{			\begin{tikzpicture}\begin{feynman}
							\vertex (a) at (-.75, -.75);
							\vertex (h) at (.75, -.75);
							\vertex (f) at (-1.5, -1.5);
							\vertex (b) at (1.5, -1.5);
							\vertex (c) at (0, .75);
							\vertex (e) at (0, 1.5);
							\vertex [dot] (d) at (0, 0);
							\diagram* {
								(a) -- [boson, edge label={}] (d),
								(f) -- [fermion, edge label={$p_1$}] (a),
								(h) -- [fermion] (d),
								(b) -- [boson, edge label={$p_3$}] (h),
								(c) -- [boson, edge label={}] (d),
								(e) -- [fermion, edge label={$p_2$}] (c),
							};
						\end{feynman}
			\end{tikzpicture}}}
			\Big)_{\lambda_2(\bar{3},\bar{1},\bar{2})
			}	\Bigg]+ (1 \leftrightarrow 2)\;.
		\end{split}
	\end{equation}
	And finally
	\begin{equation}\label{n_phi_phi} 
		\langle n^{p_1}\phi_a^{p_2} \phi_a ^{p_3}\rangle=\,i \Big(\lambda_1(\bar{2},\bar{3},\bar{1})+\lambda_1(\bar{3},\bar{2},\bar{1})\Big)	\langle n \phi_a \rangle_{p_1}\langle  \phi_a n \rangle_{p_2}\langle  \phi_a n \rangle_{p_3}\;,
	\end{equation}
	which is diagrammatically given by 
	\begin{equation}\label{vertex}
		\begin{split}
			\Big(\vcenter{
				\hbox{\begin{tikzpicture}
						\begin{feynman}
							\vertex (a) at (-1.5, -1.5);
							\vertex (h) at (-.75, -.75);
							\vertex (f) at (.75, -.75);
							\vertex (b) at (1.5, -1.5);
							\vertex (c) at (0, .75);
							\vertex (e) at (0, 1.5);
							\vertex [dot] (d) at (0, 0);
							\diagram* {
								(h) -- [fermion] (d),
								(a) -- [boson, edge label={$p_3$}] (h),
								(b) -- [boson, edge label={$p_2$}] (f),
								(f) -- [fermion] (d),
								(c) -- [boson, edge label={}] (d),
								(e) -- [fermion, edge label={$p_1$}] (c),};
						\end{feynman}
			\end{tikzpicture}}}\Big)_{\lambda_1(\bar{2},\bar{3},\bar{1})}
			\centering + (321)\,.
		\end{split}
	\end{equation}
	Now we have all ingredients needed to calculate  $G^{\text{f-s}}_{rrr}(p_1,p_2,p_3)$.  So far, we have discussed the contribution arising from purely cubic interaction terms, as shown in Fig.~\ref{vertices}. 
	In order to explicitly calculate $G^{\text{f-s}}_{rrr}(p_1,p_2,p_3)$ we only need to substitute Eqs.~\eqref{nnn}, \eqref{n_n_phi}, \eqref{n_phi_phi} and the corresponding permutations into Eq.~\eqref{G_rrr_f_s}.
	
	However, there are also some cubic interactions that include source terms, as shown in Fig.~\ref{vertex_A} (see Appendix \ref{App_self_int} for details). 
	We introduce two new coupling constants $\lambda_1^{s}$, $\lambda_2^{s}$, corresponding to the vertices shown in this figure. 
	These interactions contribute to Eq.~\eqref{G_rrr} through the $\big(S^{\text{f-s}}\big)^2 S^{\text{int-s}}$ combination in the path integral. 
	Note that $S^{\text{f-s}} \big(S^{\text{int-s}}\big)^2$ or $\big(S^{\text{int-s}}\big)^3$ do not contribute to first order in perturbation theory.
	Therefore, the only contribution of the vertices with a single background field to $G_{rrr}$ is found to be
	\begin{equation}\label{G_rrr_int_s}
		\begin{split}
			&G^{\text{int-s}}_{rrr}(p_1,p_2)\\
			&=\frac{1}{Z}\frac{i^3}{i^3}\frac{3\times 2}{6}\int \mathcal{D}n \mathcal{D}\phi_a\,e^{i S_{\text{int}}}\int_{p'_1,p'_2}\Big[\lambda^{s}_1(1',2',\bar{3})n_{p'_1}n_{p'_2}+\big(\lambda^{s}_2(1',2',\bar{3})-\lambda_2^s(\bar{1}',\bar{2}',3)\big)n_{p'_1}\phi_{a,p'_2}\Big]\\
			&\hspace*{1cm} \times \Big[L_1 n_{p_1}+ i T \chi (L_{1}^*-L_1)\phi_{a,p_1}\Big] \Big[L_2 n_{p_2}+ i T \chi (L_{2}^*-L_2)\phi_{a,p_2}\Big]\delta(1+2+3)+ \,2\,\text{perm}\\
			&=\,\big(\lambda^{s}_1(\bar{1},\bar{2},\bar{3})+\lambda^{s}_1(\bar{2},\bar{1},\bar{3})\big) \\
			&\quad  \times \Big[L_1\langle n n \rangle_{p_1}+(i T \chi)(L_1^*-L_1)\langle \phi_a n \rangle_{p_1}\Big]\Big[L_2\langle n n \rangle_{p_2}+(i T \chi)(L_2^*-L_2)\langle \phi_a n \rangle_{p_2}\Big]\\
			&+\,\big(\lambda^{s}_2(\bar{1},\bar{2},\bar{3})-\lambda^{s}_2(1,2,3)+\lambda^{s}_2(\bar{2},\bar{1},\bar{3})-\lambda^{s}_2(2,1,3)\big) \\
			& \quad \times \Big[L_1\langle n n \rangle_{p_1}+(i T \chi)(L_1^*-L_1)\langle \phi_a n \rangle_{p_1}\Big] L_2\langle n \phi_a \rangle_{p_2}\\
			&\,\,\,\,+ 2 \,\text{perm}\,.
		\end{split}
	\end{equation}
	Then we arrive at the main result in this section:
	\begin{equation}\label{G_rrr_Final}\boxed{
			G_{rrr}(p_1,p_2)= \mbox{Eq.}\, \eqref{G_rrr_f_s}+ \mbox{Eq.}\, \eqref{G_rrr_int_s}} \;.
	\end{equation}
	Next, we first discuss how to trust Eq.~\eqref{G_rrr_Final}, 
	since it is a very complex and lengthy expression. 
	We then discuss the analytical structure of $G_{rrr}$ by explicitly evaluating it for some given values of the external momenta.

	\begin{figure}
		\begin{tikzpicture}
			\begin{feynman}
				\vertex (a) at (-1.5, -1.5);
				\vertex (b) at (1.5, -1.5);
				\vertex (c) at (0, 1.5);
				\vertex [dot] (d) at (0, 0);
				\diagram* {
					(a) -- [edge label={$n_{\omega_2,\textbf{k}_2}$}] (d),
					(b) -- [edge label={$n_{\omega_1,\textbf{k}_1}$}] (d),
					(c) -- [dashed, edge label={$A_{a0\,\omega_3,\textbf{k}_3}$}] (d),
				};
				\node[below=2cm of d] {\,\,\,\,\,\,\,\,\,\,\,\,\,\( \lambda^{s}_1(1,2,3)\color{black}\equiv -\frac{\tau'}{\tau\chi}\int_{\Omega}\frac{\tau D_2}{1+\tau D_{2}}\frac{1}{1-\tau D_{3}} \)		};
			\end{feynman}
		\end{tikzpicture}
		\,\,\,\,\,\,\,\,\,\,\,\,
		\begin{tikzpicture}
			\centering 
			\begin{feynman}
				\vertex (a) at (-1.5, -1.5);
				\vertex (b) at (1.5, -1.5);
				\vertex (c) at (0, 1.5);
				\vertex [dot] (d) at (0, 0);
				\diagram* {
					(a) -- [boson, edge label={$\phi_{a\,\omega_2,\textbf{k}_2}$}] (d),
					(b) -- [edge label={$n_{\omega_1,\textbf{k}_1}$}] (d),
					(c) -- [dashed, edge label={$A_{a0\,\omega_3,\textbf{k}_3}$}] (d),
				};
				\node[below=2cm of d] {\(\lambda^s_2(1,2,3)\color{black}\equiv \frac{i T \tau'}{\tau}\int_{\Omega}\frac{\tau D_2}{1+\tau D_{2}}\frac{1}{1-\tau D_{3}} \)};
			\end{feynman}
		\end{tikzpicture}
		\caption{Cubic vertices containing a single background field.}
		\label{vertex_A}
	\end{figure}
	
	\subsection{KMS consistency of the result}
	\label{sec4.3}
	Physical correlators in thermal systems reflect some specific features of the thermal state.
	In this work, we followed Ref.~\cite{Crossley:2015evo} and first applied KMS conditions to the effective Lagrangian as a feature of the thermal state, and then computed $G_{rr}$ and $G_{rrr}$. 
	However, the KMS conditions are more general. 
	The thermal physical correlators must satisfy the KMS conditions regardless of which EFT we study. 
	This is indeed what Heinz and Wang originally studied in Ref.~\cite{Wang:1998wg}.
	The complete set of KMS conditions between all physical two-point, three-point, and four-point correlators was derived in Ref.~\cite{Wang:1998wg}. Specifically, it is shown that $G_{rrr}$ must satisfy the following relation
	\begin{equation}\label{KMS_Heinz_Wang}
		\begin{split}
			\text{Im}\left[G_{rrr} + G_{raa} - G_{ara} - G_{aar}\right] =& \tanh\left(\frac{\beta k_1^0}{2}\right) \text{Im}\left[G_{rra} + G_{rar} - G_{arr}\right]\;,\\
			\text{Im}\left[G_{rrr} - G_{raa} + G_{ara} - G_{aar}\right] =& \tanh\left(\frac{\beta k_2^0}{2}\right) \text{Im}\left[G_{rra} - G_{rar} + G_{arr}\right]\;,
			\\
			\text{Im}\left[G_{rrr} - G_{raa} - G_{ara} + G_{aar}\right] =& \tanh\left(\frac{\beta k_3^0}{2}\right) \text{Im}\left[-G_{rra} + G_{rar} + G_{arr}\right]\;.
		\end{split}
	\end{equation}
	These relations are valid in the full quantum limit, including the statistical fluctuations as well. 
	However, the fluctuations in our work are statistical. 
	Following Ref.~\cite{Crossley:2015evo} to remove the quantum effects, we apply the transformation 
	\begin{equation}\label{quantum_to_classical}
		\phi_r\rightarrow \phi_r\,\,\,\,\,\,\,\phi_a\rightarrow \hbar \phi_a\,.
	\end{equation}
	Applying this to Eq.~\eqref{KMS_Heinz_Wang}, and taking the $\hbar\rightarrow 0$ limit, we arrive at the following simple relation
	\begin{equation}\label{KSM_constarint_Im_G_rrr}
		\text{Im}\,G_{rrr} =0\;.
	\end{equation}
	It should be noted that this relation is exact to all orders in the derivative expansion \cite{Crossley:2015evo}. 
	On the other hand, our $G_{rrr}$ is a resummed expression that can be expanded to higher order in derivatives. 
	Hence, we expect Eq.~\eqref{G_rrr_Final} to satisfy Eq.~\eqref{KSM_constarint_Im_G_rrr}. 
	Interestingly, we have checked it and found that our $G_{rrr}$ is purely real-valued. 
	We believe that this is a very non-trivial check for the correctness of our results.
	
	Let us note that there are other KMS constraints in Ref.~\cite{Wang:1998wg} that also contain $\text{Re} \,G_{rrr}$. 
	However, they all relate $\text{Re} \,G_{rrr}$ to other correlators such as $G_{rra}$, etc.. Since these correlators are not computed in this paper, we do not check the corresponding KMS constraints. 
	
	As another check, we have also explicitly found that our $G_{rrr}$ exactly reduces to the one calculated in Ref.~\cite{Delacretaz:2023ypv}, in leading order in $\epsilon$, where $\omega\sim k^2\sim \epsilon^2$.

	\subsection{Analytic structure of $G_{rrr}$  }
	\label{sec4.4}
	What we want to do in the following is to discuss the analytic structure of $G_{rrr}$, including its simple poles, singularities, and discontinuities.\footnote{See Ref.~\cite{Huber:2023uzd} for a field-theory framework that takes into account loop effects and is used to study the branch points of three-point functions.}
	
	Let us start by exploring the simple poles of $G_{rrr}$.
	Taking a closer look at Eq.~\eqref{G_rrr_Final}, we observe that $G_{rrr}$ can have at most six single simple poles. 
	In other words each external leg can produce two poles (see Fig.~\ref{G_nn}),
	\begin{equation}\label{pole_G_nnn}
		\begin{split}
			\omega_{1}=&\mp\frac{i}{\tau}\pm i k_1 \cot (\tau k_1)\;,\\
			\omega_{2}=&\mp\frac{i}{\tau}\pm i k_2 \cot (\tau k_2)\;,\\
			-\omega_{1}-\omega_{2}=&\mp\frac{i}{\tau}\pm i k_3 \cot (\tau k_3)\;,
		\end{split}
	\end{equation}
	with $k_3=\sqrt{(\textbf{k}_1+\textbf{k}_2)^2}$. 
	Each of these poles is the frequency at which one of the external legs goes on shell. 
	Thus, when any of $k_1$, $k_2$, or $k_3$ exceeds $k^*$, two of the six poles mentioned above will disappear. 
	This is why we said that there are ``at most" six of these poles.

	Below, we will illustrate the analytic structure of $G_{rrr}$ for fixed $k_1$, $k_2$, and $\omega_2$ in the complex $\omega_1$-plane. 
	Thus we expect Eq.~\eqref{pole_G_nnn} to reduce to at most four distinct single poles; we call them $	\omega_{\text{p}1}^{\pm}$ and $	\omega_{\text{p}2}^{\pm}$:
	\begin{equation}\label{p_G_nnn}
		\begin{split}
			\omega_{\text{p}1}^{\pm}=&\mp\frac{i}{\tau}\pm i k_1 \cot (\tau k_1)\;,\\
			\omega_{\text{p}2}^{\pm}=&\mp\frac{i}{\tau}\pm i k_3 \cot (\tau k_3)-\omega_2\;.
		\end{split}
	\end{equation}
	The other analytic feature of $G_{rrr}$ is the existence of singularities. 
	Each external leg has two branch-point singularities connected by a branch cut. 
	At fixed $k_1$, $k_2$, and $\omega_2$, however, there are two of them: 
	\begin{equation}\label{branch_cut_G_nnn}
		\begin{split}
			\pm \frac{i}{\tau}-k_1<\omega_1&< \pm \frac{i}{\tau}+k_1\;,\\
			\pm \frac{i}{\tau}-k_3-\omega_2<\omega_1&< \pm \frac{i}{\tau}+k_3-\omega_2\;.
		\end{split}
	\end{equation}
	Let us note that Eqs.~\eqref{p_G_nnn} and \eqref{branch_cut_G_nnn} are associated with the external legs of $\langle nnn\rangle$. 
	
\begin{figure}
		\centering
		\includegraphics[width=0.47\textwidth]{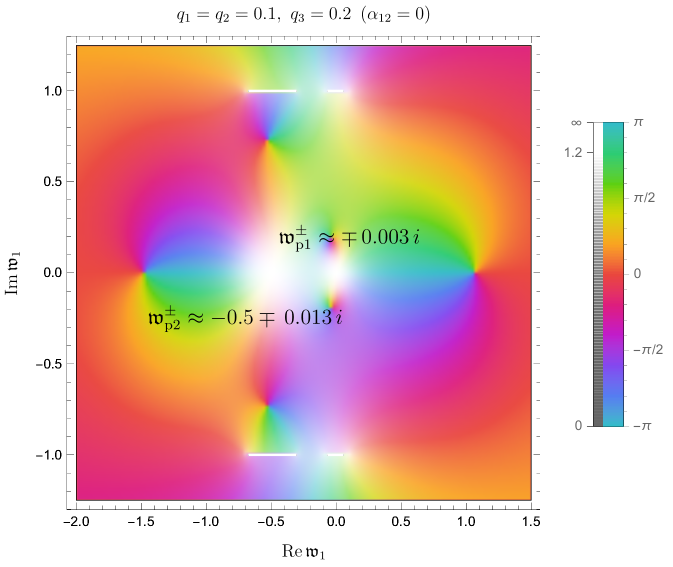}\,\,
		\includegraphics[width=0.47\textwidth]{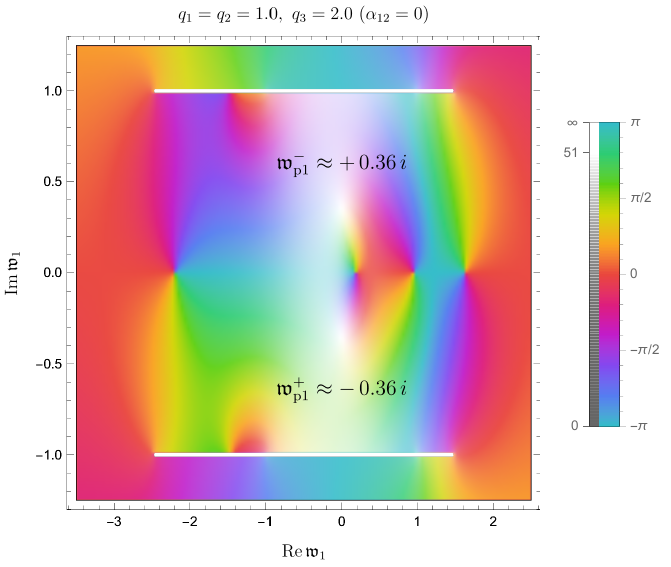}	\,\,
		\includegraphics[width=0.47\textwidth]{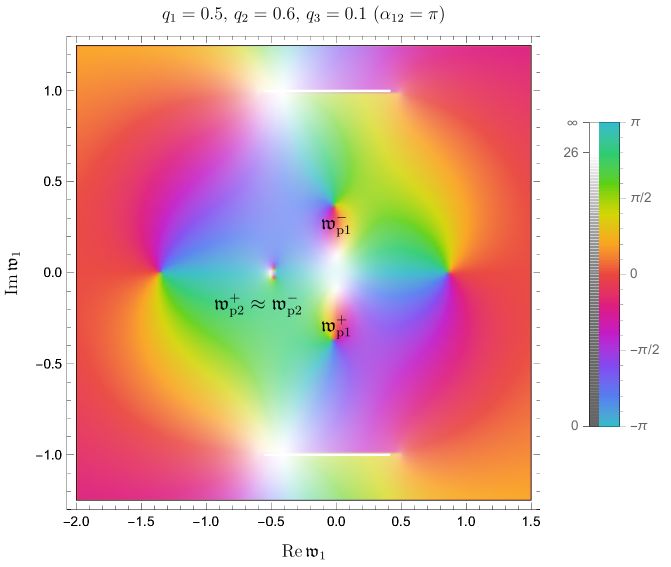}	\,\,
		\includegraphics[width=0.47\textwidth]{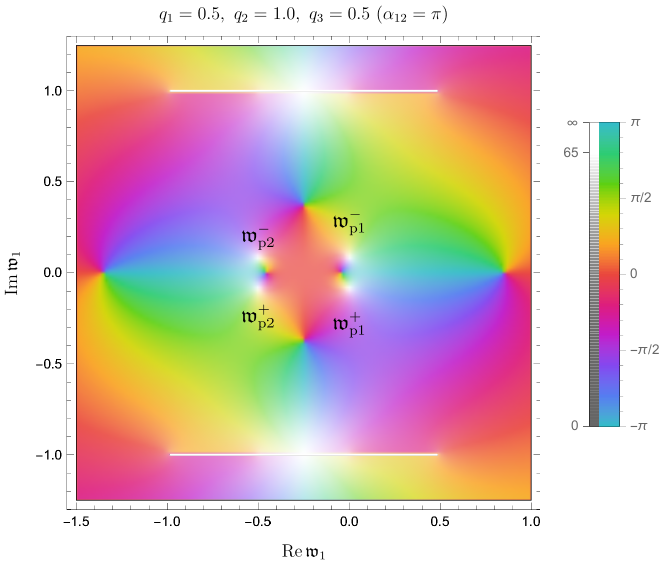}	\,\,
		\caption{The complex-valued function $G_{rrr}(\wn_1)=|G_{rrr}(\wn_1)|e^{i \alpha}$ in the complex-frequency place. \textcolor{black}{The modulus $|G_{rrr}(\wn_1)|$ is shown by a grey scale (with singularities appearing white, while roots appear black) and the phase $\alpha$ is shown in a color scale.}
			In all panels $\wn_2=0.5$. 
			The poles $\wn_{\text{p} 1,2}^{\pm}$ correspond to the isolated white dots.
			Top panel: the momenta $\qn_1=\tau \textbf{k}_1$ and $\qn_2=\tau \textbf{k}_2$ are taken to be equal and parallel. Left: for small momenta, right: for relatively large momenta such that  $q_3>q^*=\frac{\pi}{2}$. 
			Bottom panel:  the momenta $\qn_1=\tau \textbf{k}_1$ and $\qn_2=\tau \textbf{k}_2$ are taken to be anti-parallel. Left: $q_1, q_2<q^*$ with a small $q_3$, right: $q_2=2q_1<q^*$. 
			We have also defined $\alpha_{12}=\cos^{-1}\Big(\frac{q_3^2-q_1^2-q_2^2}{2 q_1 q_2}\Big)$.}
		\label{analytic_structure}
	\end{figure}

	In Fig.~\ref{analytic_structure} we show the analytic structure of $G_{rrr}$ by explicitly calculating the complicated function \eqref{nnn}. 
	Here are some comments about these plots.
	\begin{itemize}
		\item In the \textit{top left panel}, we have chosen $q_1$, $q_2$, and $\wn_2$ such that the two intervals given by Eq.~\eqref{branch_cut_G_nnn} do not overlap. 
		The locations of the white branch cuts are exactly as we have predicted; there are two separate horizontal cuts in each half-plane. 
		There are also four poles corresponding to Eq.~\eqref{p_G_nnn}.
		\item In the \textit{top right panel}, we have chosen $q_1$, $q_2$, and $\wn_2$ such that the two intervals given by Eq.~\eqref{branch_cut_G_nnn} overlap. 
		Using the values on the figure we predict two discontinuities, $\pm i-1.0<\wn_1< \pm i + 1.0$ and $\pm i-2.5<\wn_1< \pm i + 1.5$, with the latter interval encompassing the former. 
		However, since $q_3>q^*$ in this case, the leg corresponding to  $(\wn_3,\qn_3)$ does not go on-shell; as a result, we see only two simple poles corresponding to the on-shell condition associated with $(\wn_1,\qn_1)$, i.e., the first equation in Eq.~\eqref{p_G_nnn}.
		\item In the \textit{bottom left panel}, we have chosen $q_1$, $q_2$, and $\wn_2$ such that the two intervals given by Eq.~\eqref{branch_cut_G_nnn} overlap. 
		Using the values on the figure we predict two discontinuities, $\pm i-0.5<\wn_1< \pm i + 0.5$ and $\pm i-0.6<\wn_1< \pm i - 0.4$, with the latter interval encompassing the former.  
		Since $q_3$ is taken to be small in this case, the two poles corresponding to the second equation in Eq.~\eqref{p_G_nnn} are close to each other. 
		It is easy to see that in the limit $q_3\rightarrow 0$, $\wn_{\text{p}2}^{\pm}\rightarrow -\wn_2 $, which is identified by the value $-0.5$ in the figure. 
		\item In the \textit{bottom right panel}, we have chosen $q_1$, $q_2$, and $\wn_2$ such that the two intervals given by Eq.~\eqref{branch_cut_G_nnn} overlap. 
		The four poles are clearly separate. 
		The accidental symmetry observed between the location of poles is simply due to the fact that we have chosen $q_1=q_3$. 
		Considering Eq.~\eqref{branch_cut_G_nnn} for these values of momenta, $\omega_{\text{p}1}^{\pm}$ should map into $\omega_{\text{p}2}^{\pm}$ by a shift $-\wn_2$, which is $-0.5$ in the figure.
	\end{itemize}

\begin{figure}[H]
		\centering
		\includegraphics[width=0.45\textwidth]{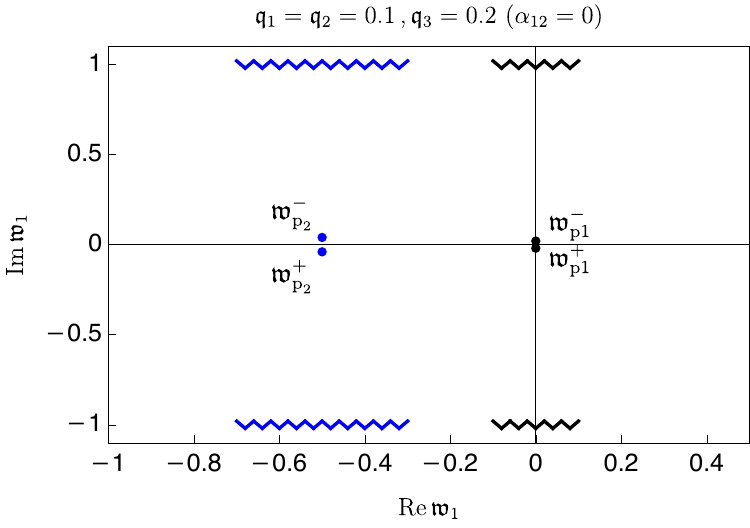}	\,\,
		\includegraphics[width=0.45\textwidth]{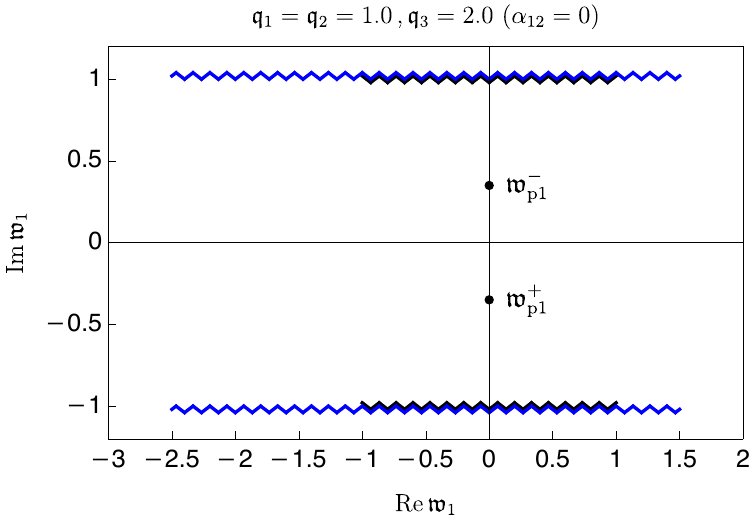}	\,\,
		\includegraphics[width=0.45\textwidth]{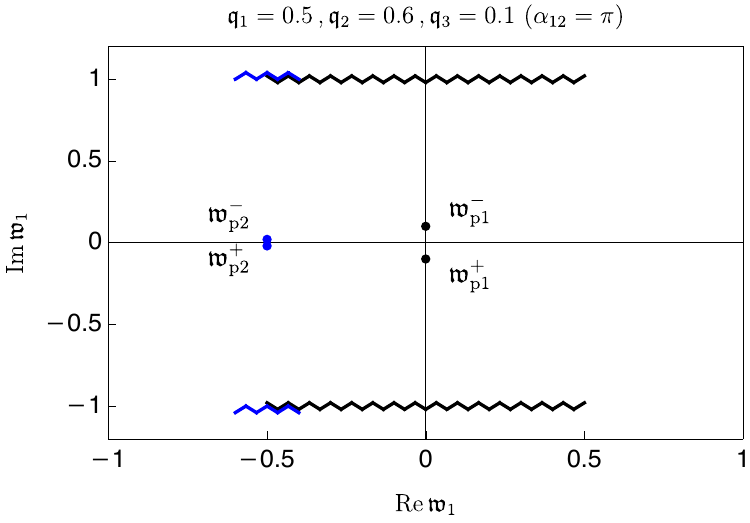}	\,\,
		\includegraphics[width=0.45\textwidth]{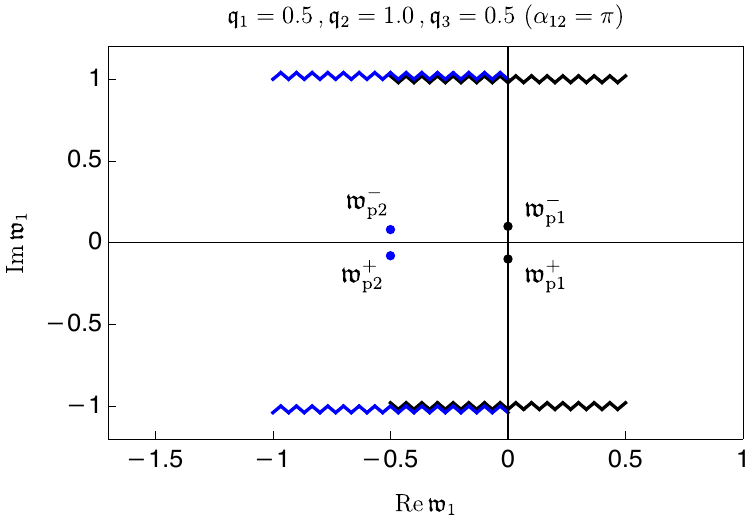}	\,\,
		\caption{\textcolor{black}{A simple illustration of branch cuts and poles of Fig.~\ref{analytic_structure}. Black and blue correspond to the two sets of solutions of Eqs.~\eqref{pole_G_nnn} and \eqref{p_G_nnn}.}}
		\label{simple}
	\end{figure}

	\textcolor{black}{For the sake of clarity, in Fig.~\ref{simple} we only show the cuts and poles of $G_{rrr}(\wn_1)$.}
	 In summary, the figures are consistent with the discussion around Eqs.~\eqref{pole_G_nnn} and \eqref{p_G_nnn}.
	This means that the analytic structure of $G_{rrr}$ is completely determined by the analytic structure of the external legs. 

	As long as any momentum $\textbf{k}_i$ satisfies $k_i<k^*$, the corresponding leg will produce two poles in the $G_{rrr}$ structure (one in the upper half-plane and one in the lower half-plane). 
	The branch-cut discontinuity of $G_{rrr}$ is also the union of branch cuts associated with the external legs.
	
	\subsection{Comments on transport}
	\label{transport}
In this work, we have so far focused on the structure of two-point and three-point correlation functions. 
These correlation functions inherently capture both dissipation and fluctuation effects and, therefore, may have significant implications for transport phenomena. 
In particular, in the small-frequency and small-momentum limit, they encode important information about hydrodynamic transport.

	Let us consider the dissipative part of the current as a double expansion near equilibrium in terms of the dynamical variable, the chemical potential $\mu$.
	The expansion occurs in two directions: In powers of  $\mu$ and in powers of derivatives acting on it.
	According to Ref.~\cite{Moore:2010bu}  (see also Refs.~\cite{Erdmenger:2008rm,Haack:2008xx,Grozdanov:2014kva}), the physical expectation is that two-point functions are associated with transport coefficients appearing in front of linear terms in $\mu$, at least at first order in gradients, such as $\nabla_i\mu$ and its higher derivatives. We refer to these as \textbf{linear transport coefficients}.
	
On the other hand, three-point functions are associated with transport coefficients multiplying nonlinear terms in $\mu$ as well, including those quadratic in $\mu$ and at least second order in gradients, such as 
 $\nabla_i\mu\nabla_i\mu$ and its higher derivatives. We refer to these as \textbf{quadratic transport coefficients}.
 
Having obtained resummed two- and three-point functions, the simple discussion above suggests that, in principle, we could compute all linear and quadratic transport coefficients appearing in the constitutive relations (of the charge density and the current), to arbitrarily high orders in derivatives.\footnote{Here, we discuss transport in the Landau-Lifshitz frame, where gradient corrections do not enter the constitutive relation of the charge density, but of the charge-current density.}
  However, to the best of our knowledge, a systematic hydrodynamic framework introducing the constitutive relations (containing the associated transport coefficients) for the system discussed in this work has not yet been constructed.
  In fact, relativistic hydrodynamics without a diffusive charge has been systematically developed up to third order in the derivative expansion \cite{Grozdanov:2015kqa,Diles:2023tau}. Three-point functions and their connection to quadratic transport coefficients have been investigated in Ref.~\cite{Moore:2010bu}. For a charged fluid, it has yet to be explored, as a systematic framework extending hydrodynamic theory to second order in derivatives has not yet been developed for charged fluids \cite{Kovtun:2012rj}.

 The system under study in this work also remains unexplored in the same way.
  To fill this gap, we begin by constructing the most general form of the constitutive relation for the current in a diffusive system, up to second order in $\mu$ and third order in derivatives ($\sim \epsilon^3$).   
  In the presence of an external electric field $E^{i}$, in the Landau-Lifshitz frame:
	\begin{equation}\label{Landau}
		\begin{split}
			J^0=&\,n(\mu)\equiv \chi \mu+\frac{\chi'}{2}\mu^2+\,\mathcal{O}(\epsilon^4\mu^3)\;,\\
		J^{i}=&D_{11}V^i-D_{12}\,\mu V^i\\
		&\,+D_{31}\nabla^i\nabla^2 V^i+D_{32}^{(1)}\mu\nabla^2 V^i+D_{32}^{(2)}V^i(\nabla\cdot V) +D_{32}^{(3)}\nabla^i(V\cdot V) \,+\mathcal{O}(\epsilon^4\mu^3)\;,
		\end{split}
	\end{equation}
	where $V^i=E^i-\nabla^i \mu$, and as before, $\chi'$ is given by Eq.~\eqref{tau_chi}. Here, $D_{jk}$ is the coefficient of a term that is of 
$j^{th}$ order in derivatives and contains $\mu^k$.
    Note that the above expressions are written in the rest frame of the system.
    While they can also be formulated in a covariant form, for consistency with the rest of this work, we will continue using the above representation.\footnote{We further elaborate on this structure and related aspects in an upcoming work \cite{Abbasi:2025}.} 
	
Now let us compare Eq.~\eqref{Landau} with  Eq.~\eqref{Current_re_group} in our system.  
One can rewrite Eq.~\eqref{Current_re_group} in  exactly the same form as Eq.~\eqref{Landau}. 
As discussed below Eq.~\eqref{Current_re_group}, achieving this equivalence requires applying the conservation equation at $\mathcal{O}(\epsilon^2)$ to ensure that $J^0=n+\mathcal{O}(\epsilon^4)$. 
This result has been rigorously demonstrated for the linear case in Appendix~\ref{diff_familiar}.
However, to generalize this discussion to include nonlinearities, we must extend the analysis beyond the linear regime. 
Considering Eq.~\eqref{tau_chi}, the nonlinear equation up to $\mathcal{O}(\epsilon^2)$ takes the form:
	\begin{equation}\label{}
(\chi+\chi' \mu)\dot{\mu}-\frac{1}{3}\Big[\tau \chi+(\tau \chi'+\tau \chi')\mu\Big]\nabla^2 \mu-\frac{1}{3}(\tau \chi'+\tau \chi')\big(\nabla \mu\big)^2=\,0\;.
	\end{equation}
	Deriving $\dot{\mu}$ from this equation and substituting into Eq.~\eqref{Current_re_group}, we find exactly Eq.~\eqref{Landau} with the following coefficients:
	\begin{subequations}\label{D_i_j}
		\begin{align}\label{threshold_1}
D_{11}&=\frac{\tau \chi}{3}\;,\\
D_{12}&=\frac{1}{3}(\tau \chi'+\tau \chi')\;,\\
D_{31}&=-\frac{\tau^3\chi}{45}\;,\\
D_{32}^{(1)}&=-\frac{\tau^2}{15}\Big(\tau' \chi+\frac{1}{3}\tau \chi'\Big)\;,\\
D_{32}^{(2)}&=\frac{\tau^2}{45}\Big(\tau \chi'-2\tau' \chi\Big)\;,\\
D_{32}^{(3)}&=\frac{2\tau^2}{90}\Big(\tau \chi'+8\tau' \chi\Big)\;.
		\end{align}
	\end{subequations}

This highlights the advantage of working with the expanded formulas \eqref{higher_diffusion} and \eqref{Current_re_group}. From these equations, we can directly extract linear, quadratic, and higher-order transport coefficients, without the need to compute correlation functions or rely on Kubo formulas.
This approach remains effective as long as the hydrodynamic constitutive relations can be systematically extended as a derivative expansion in $\mu$ and its higher powers.

Now, the key question is: what is the significance of these transport coefficients? To answer this question, let us recall that one important experimentally measurable quantity is the current $J^i$. Depending on whether the system describes the diffusion of charge, mass, or another conserved quantity, this current can be directly measured in experiments.

Let us assume that such a measurement has been performed for small perturbations (in amplitude) within the system. By setting the $D_{j2}$
coefficients in Eq.~\eqref{Landau} to zero, we can fit the experimental data for $J^i$
  to extract $\tau$. The static charge susceptibility is assumed to be known from thermodynamics. Repeating this experiment for different values of $\mu$, or equivalently $n$, allows us to determine $\tau(\mu)$ or $\tau(n)$. From this, $\tau'$ can be obtained as 
$\tau'=d\tau/d\mu$.  Similarly, $\chi'$ can be determined from thermodynamics.

This demonstrates that an experiment with small perturbations enables the extraction of transport coefficients that appear in higher-order terms, such as $D_{j2}$. Once these coefficients are known, we can use Eq.~\eqref{Landau}, along with the transport coefficients given in Eq.~\eqref{D_i_j}, to predict the system's response when perturbations are not small in amplitude. This highlights the predictive power of our theory.

Correlation functions provide an indirect method for measuring transport coefficients. To extract $D_{jk}$, one must construct Kubo formulas that relate the correlation functions of the current (or density) to these transport coefficients. Although our system is solvable and all transport coefficients are already known, deriving Kubo formulas for quadratic transport coefficients and comparing them with the correlation functions computed in this paper would serve as an important validation of our results in the small-momentum limit (see Ref.~\cite{Abbasi:2025}).

\color{black}	
	\subsection{1-loop correction to the two-point function}
	\label{2_loop}
	The theory we are studying has the property that its correlation functions have branch-point singularities, developing branch-cut discontinuities,  even at the linear-response level. 
	On the other hand, a well-known feature of the correlation functions in quantum field theory is having branch-point singularities (and consequently branch-cut discontinuities) due to interaction effects. 
	Here we want to explore such singularities for a simple two-point function, $G_{n\phi_a}$. 
	We expect that $G^{R}_{J^0J^0}$ has the same analytic structure. 
	The leading interaction effect is the following one-loop correction  (see Eq.~\eqref{lam12})
	\begin{equation}\label{loop}
		\begin{split}
			&	\scalebox{0.6}{	\begin{tikzpicture}[baseline=(a.base)]
					\begin{feynman}
						\vertex (a) ;
						\vertex [right=of a] (a1) ;
						\vertex [right=of a1] (a2);
						\vertex [above right=of a2] (a3);
						\vertex [below right=of a2] (a4);
						\vertex [above right=of a4] (a5); 
						\vertex [ right=of a5] (a6);
						\vertex [ right=of a6] (a7);
						\diagram* {
							(a) --[very thick](a1)-- [boson,very thick] (a2) -- [quarter left, very thick] (a3)--[boson,quarter left,very thick](a5)--[quarter left, very thick](a4)--[quarter left,very  thick](a2) ,
							(a5) -- [very thick] (a6)--  [boson,very thick](a7),
						};
					\end{feynman}
					\node[above] at (1,.3) {$p=(\omega, \textbf{k})$};
					\node[above] at (4,-1.8) {$p-p'$};
					\node[above] at (4,1.2) {$p'$};
			\end{tikzpicture}}\,\\
			&\sim\,G_{n\phi_a}(p)^2\int_{p'}\lambda_1(p-p', p', p)\lambda_1(p-p', p, p')G_{n\phi_a}(p')G_{nn}(p-p')\;.
		\end{split}
	\end{equation}
	Performing the integral in Eq.~\eqref{loop} is difficult analytically. 
	However, we can find the singular points of the  integral  (as a function of $\omega$ and $\textbf{k}$) without explicitly evaluating it. 
	Following 	Ref.~\cite{Abbasi:2022rum}, it just occurs at \textit{pinch singularities} of the integrand. 
	
	Let us first check if the singularities of $\lambda_1(p-p', p', p)\lambda_1(p-p', p, p')$ can become pinched.
	Based on earlier results about the vertices,  $\lambda_1(p-p', p', p)\lambda_1(p-p', p, p')$ is analytic everywhere but at the four points $\omega'=\pm\frac{i}{\tau}\pm k'$. 
	These are precisely the singular points of the logarithm in $G_{n\phi_a}(\omega',\textbf{k}')$. 
	Thus they do not produce a new kind of singularity for the integral. 
	The only potential pinch singularities are then located where the poles of $G_{n\phi_a}$ and $G_{nn}$ coincide. 
	From Eqs.~\eqref{n_phi_prop} and \eqref{n_n_prop}, we have
	\begin{subequations}\label{}
		\begin{align}\label{loop_up}
		\omega'&=\, -\frac{i}{\tau}+ ik'\cot (\tau k')\;, \\\label{loop_bottom}
		\omega-\omega'&=\,\mp\frac{i}{\tau}\pm i|\textbf{k}-\textbf{k}'|\cot (\tau|\textbf{k}-\textbf{k}'|)\;.
		\end{align}
	\end{subequations}
    
	The second pole (with lower signs) in Eq.~\eqref{loop_bottom} cannot coincide with Eq.~\eqref{loop_up}. 
	But the other one collides Eq.~\eqref{loop_up} if
	\begin{equation}\label{condition}
		\omega=\, -\frac{2i}{\tau}+ ik'\cot (\tau k')+ i|\textbf{k}-\textbf{k}'|\cot (\tau|\textbf{k}-\textbf{k}'|)\;.
	\end{equation}
	This function has an extremum at $k'=\frac{k}{2}$, which gives the branch-point singularity of the integral:
	\begin{equation}\label{bp}\boxed{
			\omega_{\text{b.p.}}^{\text{int}}=\, -\frac{2i}{\tau}+ ik\cot \left( \frac{\tau k}{2} \right)} \;.
	\end{equation}
	Do not confuse this equation with Eq.~\eqref{mode_non_pertrub}.
	Assuming that $\omega\sim k^2\sim \epsilon^2$, to leading order in $\epsilon$, and upon identifying $D\equiv\frac{\tau}{3}$, we find
	\begin{equation}\label{correspondence}
		\begin{split}
			\mbox{Eq.}\,\eqref{mode_non_pertrub}\,\,\,\rightarrow\,\,\,\,\omega&=-i D  k^2\;,\\
			\mbox{Eq.}\,\eqref{bp}\,\,\,\rightarrow\,\,\,\,\omega&=- \frac{i}{2} D k^2\;.
		\end{split}
	\end{equation}
	As one expects, we see that in the leading order of the $\epsilon$-expansion, Eq.~\eqref{mode_non_pertrub} gives the diffusion pole. 
	Interestingly, in the same limit Eq.~\eqref{bp} reduces to the branch-point singularity associated with the nonlinear effects in the theory of diffusive fluctuations found in Ref.~\cite{Chen-Lin:2018kfl}. 

\begin{figure}[H]
	\centering
	\includegraphics[width=0.4\textwidth]{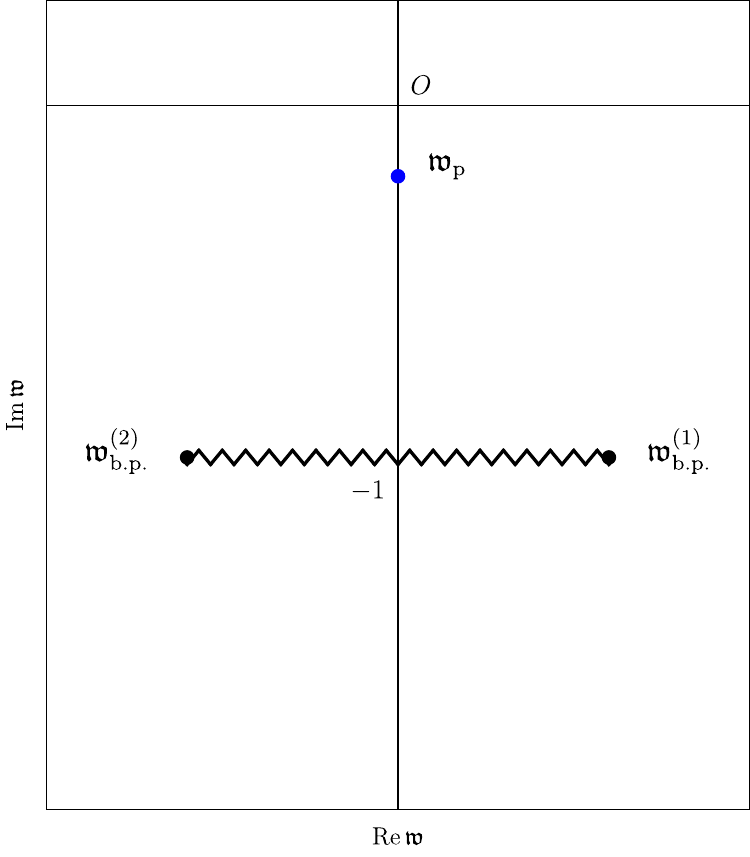}\,\,\,\,\,\,\,\,\,\,\includegraphics[width=0.4\textwidth]{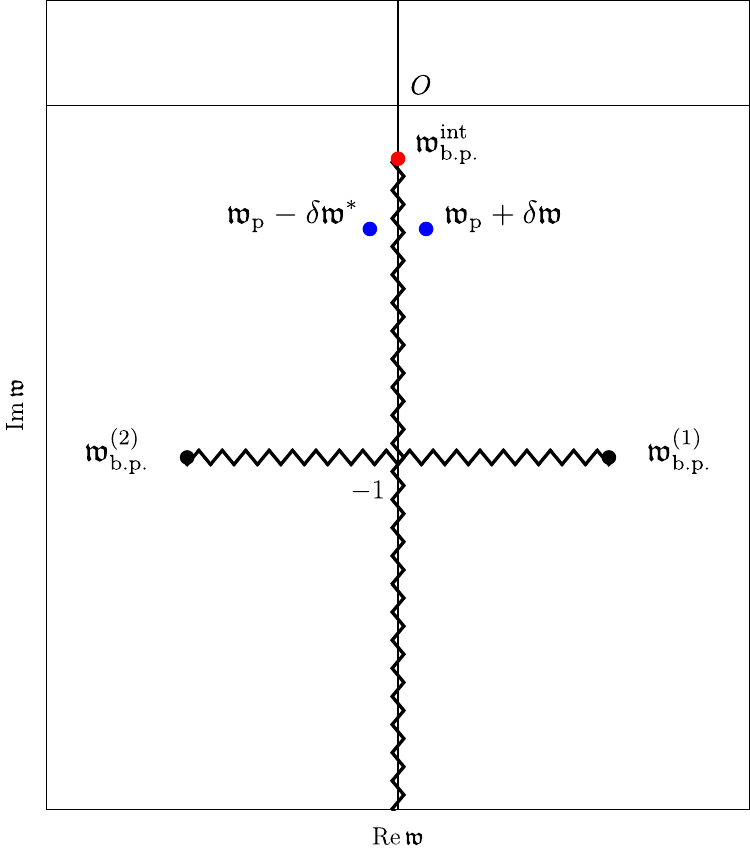}
	
	\caption{One-loop correction to the analytic structure of the response function $G^{R}_{nn}(\wn,\qn)$ for $q<\frac{\pi}{2}$. Left panel: free theory; there is a single pole $\wn_{\text{p}}=-i+i q\cot q$, together with two branch points $\wn_{\text{b.p.}}^{(1,2)}=-i \pm q$ connected with a branch cut \cite{Romatschke:2015gic}. Right panel: interacting theory. Due to the interactions, another branch-point singularity emerges at $\wn_{\text{b.p.}}^{\text{int}}=-2i+i q\cot \frac{q}{2}$, causing the pole $\wn_{p}$ to  split into two poles lying on the two sides of the  branch cut emanating from  $\wn_{\text{b.p.}}^{\text{int}}$ to infinity. This splitting was first discussed in Ref.~\cite{Chen-Lin:2018kfl}  for diffusion at leading order in derivatives. Then it was shown in Ref.~\cite{Abbasi:2021fcz} that this would also be the case at higher orders in derivatives (see also Fig.~4 in the Supplemental Material of Ref.~\cite{Abbasi:2022aao}).}
	\label{analytic_G_rr}
\end{figure}

This is illustrated in Fig.~\ref{analytic_G_rr} for the lower-half complex $\wn$-plane. There are two cuts: the first is the genuine cut of the theory (a horizontal cut connecting $\wn_{\text{b.p.}}^{(1)}$ and $\wn_{\text{b.p.}}^{(2)}$), while the second is a cut that emerges at the one-loop level, emanating from $\wn_{\text{b.p.}}^{\text{int}}$ and extending to infinity in the lower half of the complex $\wn$-plane.

An interesting question that arises here is how loop effects influence the linear (classical) dispersion relations. As mentioned earlier, hydrodynamic loops are difficult to calculate analytically in our model. However, one can qualitatively explore the relevant physics.

The effect of hydrodynamic loops on classical dispersion relations was first pointed out in Ref.~\cite{Chen-Lin:2018kfl} for a diffusive system at leading order in the derivative expansion. It was noted that the classical pole $\wn_{\text{p}}$, shown in the left panel of Fig.~\ref{analytic_G_rr}, would split into two modes symmetrically positioned on either side of the vertical branch cut, as depicted in the right panel of the figure.

Subsequently, one of the authors investigated this phenomenon within the Schwinger-Keldysh EFT framework in an SYK chain, extending the analysis to higher orders in the derivative expansion \cite{Abbasi:2021fcz}. The SYK chain serves as a model for energy diffusion in $(1+1)$-dimensional systems. This study revealed that the same mode splitting occurs for the entire range of momenta up to the EFT cutoff.

This line of research was further pursued in Ref.~\cite{Abbasi:2022aao}, again within the EFT formalism, but in a different system incorporating higher-derivative corrections. The model examined in 		Ref.~\cite{Abbasi:2022aao}
 was the telegrapher equation. Compared to energy diffusion in the SYK chain, which describes purely infrared (IR) dynamics, diffusion in the telegrapher model also involves a ultraviolet (UV) scale. In other words, in addition to the diffusive mode, the spectrum of the classical theory includes a non-hydrodynamic mode. This study extends the findings of Ref.~\cite{Abbasi:2021fcz} by examining the interplay between the following three effects:

\begin{enumerate} \item The effect of hydrodynamic loops on the dispersion relation of the diffusive mode
	\item Higher-derivative effects
	\item The effect of a non-hydrodynamic mode
\end{enumerate}
Reference 		\cite{Abbasi:2022aao}
 finds that, in addition to the diffusion mode, the non-hydrodynamic mode also undergoes splitting due to loop effects.

From the perspective of the above discussion, for $k < k^*$, the model studied in the present work is analogous to the SYK chain. In this regime, there is only a single diffusive pole in the spectrum of the classical theory \cite{Romatschke:2015gic}, and thus, we expect the same mode splitting to occur. However, a key distinction between our model and the SYK chain is the presence of a horizontal cut in Fig.~\ref{analytic_G_rr}.

Naïvely, one would expect that as the value of $k$ increases, the two blue poles shown in the right panel of the figure move farther from the vertical cut and gradually asymptote to $\wn = -i/\tau$. Whether this expectation holds requires a more careful analysis of Eq.~\eqref{loop}. We leave a detailed investigation of this issue for future work.
\color{black}
	\section{Effective action directly from the noisy Boltzmann equation }
	\label{kin_theory}
    \textcolor{black}{The Schwinger-Keldysh EFT constructed in Sections 4 and 5 systematically incorporates the effects of fluctuations (noise) in the system, which are then reflected in the correlation functions. A natural question arises: can the same correlation functions be obtained from the Boltzmann equation?}

\textcolor{black}{In the linear-response regime, this can be achieved indirectly by first computing the response functions \cite{Romatschke:2015gic, Bajec:2024jez} and then applying the fluctuation-dissipation theorem, as discussed around Eq.~\eqref{FD}. Note that the relativistic Boltzmann equation \eqref{Boltzmann_rel} is inherently deterministic and does not explicitly account for fluctuations.}

\textcolor{black}{To directly incorporate the effects of fluctuations, a traditional approach is to phenomenologically introduce a noise term on the right-hand side of the Boltzmann equation and fix the strength of the noise correlation function via the fluctuation-dissipation theorem. In this section, we follow this approach and then attempt to construct an MSR effective action corresponding to the integrated noisy Boltzmann equation. The latter equation is, in fact, the noisy diffusion equation. Therefore, the resulting effective action should match the effective action for diffusion constructed in the previous sections. }
    
	Following Landau-Lifshitz \cite{landau1981kinetic} (see also Ref.~\cite{Calzetta:1999xh}), the Boltzmann equation can be transformed into a stochastic (noisy) equation. 
	This is done by simply placing a noise distribution function on the right-hand side of the Boltzmann equation. 
	In RTA, we can write
	\begin{equation}\label{Boltzman_noisy}
		\partial_t f(x,\boldsymbol{p})+\boldsymbol{v \cdot \nabla} f(x,\boldsymbol{p})=-\frac{f(x,\boldsymbol{p})-f^{(0)}(x,\boldsymbol{p})}{\tau}+\xi_{\textcolor{black}{0}}(x,\boldsymbol{p})\;.
	\end{equation}
	Considering $\langle \xi(x,\boldsymbol{p}) \rangle=0$, then the noise-averaged equation is the one we started with in Sec.\ \ref{sec_diff}. 
	In fact $f$ in that section is $\langle f\rangle$,
	\begin{equation}\label{Boltzmann_av}
		\partial_t \langle f(x,\boldsymbol{p})\rangle+\boldsymbol{v \cdot \nabla} \langle f(x,\boldsymbol{p})\rangle =-\frac{\langle f(x,\boldsymbol{p})-f^{(0)}(x,\boldsymbol{p})\rangle}{\tau}\;.
	\end{equation}
	However, for the sake of brevity, we have removed the brackets. 
   We want to continue with Eq.~\eqref{Boltzman_noisy} in a derivative expansion.
		For this, we put the noise term on the right-hand side of the noise-averaged Boltzmann equation \eqref{Boltzamnn_2}:
	\begin{equation}\label{noisy_Boltzman}
		\frac{\boldsymbol{\textbf{D}}}{1\textcolor{black}{+}\tau\textbf{D}} f_{\xi}^{(0)}(x,\boldsymbol{p})=-\frac{f_{\xi}(x,\boldsymbol{p})-f_{\xi}^{(0)}(x,\boldsymbol{p})}{\tau}+\xi(x,\boldsymbol{p})\;.
	\end{equation}
	To make the problem analytically tractable, we perform an integral over momentum space:
	\begin{equation}\label{noisy_Boltzman_inegrated}
		\int_{\boldsymbol{p}}	\frac{\boldsymbol{\textbf{D}}}{1\textcolor{black}{+}\tau\textbf{D}} f_{\xi}^{(0)}(x,\boldsymbol{p})=\,\int_{\boldsymbol{p}}\xi(x,\boldsymbol{p})\;,
	\end{equation}
	\textcolor{black}{where we have used Eq.~\eqref{matching} and defined $\xi=(1+\tau \textbf{D})^{-1}\xi_{0}$.} 
	Here the subscript $\xi$ for $f^{(0)}$ denotes that $f^{(0)}_{\xi}$ is a noisy solution.
	Now we want to find an effective action whose associated equation of motion is the equation above. 
	For this, we will use the MSR formalism \cite{martin1973statistical}. 
	First, from the equation above we can read off the correlation function of the noise:
	\begin{equation}\label{noise_xi}
		\begin{split}
			\int_{\boldsymbol{p},\boldsymbol{p}'}\big\langle \xi_{\omega, \textbf{k}}(\boldsymbol{p})  \xi_{-\omega, -\textbf{k}}(\boldsymbol{p}')\big\rangle=&\int_{\boldsymbol{p},\boldsymbol{p}'} \,\frac{D}{1- \tau D}\frac{-D'}{1+\tau D'}\big\langle f^{(0)}_{\textcolor{black}{\xi;\,}\omega, \textbf{k}}(\boldsymbol{p})  f^{(0)}_{\textcolor{black}{\xi;\,}-\omega, -\textbf{k}}(\boldsymbol{p}')\big\rangle\\
			=&\,T\,\chi\,\int_{\Omega}\,\Big(\frac{D}{1+\tau D}-\frac{D}{1-\tau D}\Big)\;,
		\end{split}
	\end{equation}
	where $D= - i \omega + \hat{\boldsymbol{p}}\cdot (i \textbf{k})$ and $D'= - i \omega + \hat{\boldsymbol{p}}'\cdot (i \textbf{k})$.
	
	The next step is to rewrite the non-equal time correlation function of $f^{(0)}$ as the average over the noise distribution:
	
	\begin{equation}\label{Jacobian}
		\begin{split}
			\int_{\boldsymbol{p}_1,\boldsymbol{p}_2}\!\!\! \big\langle   f^{(0)}(x_1, \boldsymbol{p}_1)    f^{(0)}(x_2, \boldsymbol{p}_2)\big\rangle=&	\int_{\boldsymbol{p}_1,\boldsymbol{p}_2}\!\int \mathcal{D}\xi\, e^{-W[\xi]}\,   f_{\xi}^{(0)}(x_1, \boldsymbol{p}_1)    f_{\xi}^{(0)}(x_2, \boldsymbol{p}_2) \\
			=&	\int_{\boldsymbol{p}_1,\boldsymbol{p}_2}\!\int\mathcal{D}f^{(0)}\!\!\!\int \mathcal{D}\xi\, e^{-W[\xi]}\,\delta(\text{e.o.m.})\,J\, f^{(0)}(x_1, \boldsymbol{p}_1)    f^{(0)}(x_2, \boldsymbol{p}_2) \, ,
		\end{split}
	\end{equation}
	where  $J=\delta(\text{e.o.m.})/\delta f^{(0)}$ is the Jacobian and ``e.o.m.'' is the noisy equation of motion~\eqref{noisy_Boltzman}. 
	In addition
	\begin{equation}\label{noiseweight}
		W[\xi]=\frac{1}{2}\int_{\boldsymbol{p},\boldsymbol{p'}}\,\int_{x,x'}\,\xi(x';\boldsymbol{p}')A(x,x';\boldsymbol{p})\xi(x;\boldsymbol{p})\;,
	\end{equation}
	where 
	\begin{equation}\label{noisevariance}
		T\,\chi\,\Big(\frac{\textbf{D}}{1+\tau \textbf{D}}-\frac{\textbf{D}}{1-\tau \textbf{D}}\Big)A(x,x';\boldsymbol{p})=\,\delta(t-t')\delta^{3}(\textbf{x}-\textbf{x}')\;.
	\end{equation}
	The idea of finding the effective action is to exponentiate the delta function in Eq.~\eqref{Jacobian} and then integrate over the noise field $\xi$.
	At the linearized level, i.e., when the e.o.m. is linear, the Jacobian is field-independent.
	By introducing an auxiliary field $f_{a}(x, \boldsymbol{p})$, we exponentiate the delta function as
	\begin{align}\label{noisecorr}
		\int_{\boldsymbol{p}_1, \boldsymbol{p}_2}\big\langle   f^{(0)}(x_1, \boldsymbol{p}_1)    f^{(0)}(x_2, \boldsymbol{p}_2)\big\rangle & =\int_{\boldsymbol{p}_1 , \boldsymbol{p}_2}\int\mathcal{D}f^{(0)}\int \mathcal{D}\xi\, e^{-\frac{1}{2}\int \xi A\xi}\nonumber \\
		& \times \int \mathcal{D}f_a \, e^{i \int (\text{e.o.m.} )f_a}\,J\, f^{(0)}(x_1,\boldsymbol{p}_1)   f^{(0)}(x_2,\boldsymbol{p}_2) \; .
	\end{align}
	Considering Eq.~\eqref{noise_xi}, the integration over $\xi$ gives
	\begin{equation}\label{denscorr}
		\big\langle n(x_1)   n(x_2)\big\rangle=\int\mathcal{D}n\, \mathcal{D}n_a\,\, e^{i S_{\text{eff}}^{(2)}[n,\,n_a]}\,  n(x_1)   n(x_2)\; ,
	\end{equation}
	where 
	\begin{equation}\label{EFT_noisy}
		S^{(2)}_{\text{eff}}[n,\,n_a]=\,\int dt \, d^dx\bigg(i\,n_a \,A\,n_a-\, \int_{\Omega}n_a \frac{\boldsymbol{\textbf{D}}}{1-\tau\textbf{D}} n\bigg) \; ,
	\end{equation}
	with 
	\begin{equation}\label{noisevariance_2}
		A=\,T\,\chi\,\int_{\Omega}\,\Big(\frac{\textbf{D}}{1+\tau \textbf{D}}-\frac{\textbf{D}}{1-\tau \textbf{D}}\Big)\;.
	\end{equation}
	The effective action \eqref{EFT_noisy} is exactly the same as the Schwinger-Keldysh effective action \eqref{L_2}.
	
	Two comments:
	\begin{enumerate}
		\item We showed that the quadratic MSR effective action that can produce the integrated noisy Boltzmann equation \eqref{noisy_Boltzman_inegrated} is exactly the same as the quadratic Schwinger-Keldysh effective action for diffusion. 
		\item To find the interacting  Schwinger-Keldysh effective action, we promoted $\tau$ to become a function of $n$. 
		The resulting cubic action is given by Eq.~\eqref{L_3}. 
		It is obvious that this effective action also produces the nonlinear terms in the noisy Boltzmann equation as well.
	\end{enumerate}
	The equivalence of the Schwinger-Keldysh effective action and the MSR effective action for Gaussian noise is well known 	\cite{Jain:2023obu}. 
	Our calculations are in agreement with this point; thus we can use any of the two formalisms to study the correlation functions.
	
	\section{Review, conclusion, and outlook}
	\label{Review}
	In this work we have constructed a Schwinger-Keldysh EFT for the diffusion. 
	Considering the Boltzmann equation in RTA as the microscopic dynamics in the system, we have been able to construct the EFT to infinite order in derivatives, analytically. 
	We have five concrete results:
	\begin{enumerate}
		\item For the first time, we analytically found a diffusion equation that follows RTA Boltzmann dynamics to infinite order. 
		It is given by Eq.~\eqref{diffusion_all_order} or its resummed version Eq.~\eqref{Resummed}.
		\item The quadratic Schwinger-Keldysh effective action has been found analytically up to infinite order in derivatives. 
		In closed form it is given by Eq.~\eqref{L_2}. 
		From this effective action we computed explicitly $G^{S}_{\text{J}^0\text{J}^0}$. 
		For comparison with known results from the literature we also computed $G^{R}_{\text{J}^0\text{J}^0}$. 
		There is full agreement.
		\item The main part of this work was devoted to constructing the interaction part of the Schwinger-Keldysh EFT, to infinite order in derivatives. 
		It is given by the closed formula \eqref{L_3}. 
		From this, we explicitly computed the symmetrized three-point function in momentum space, namely $G_{rrr}$. 
		We then discussed the analytic structure of these correlators in detail. 
		We found that the singularities and poles structure of these correlators are the union of the singularities and pole structure of the external legs in momentum space.
		\item We analytically calculated branch-point singularities in the structure of the two-point function that arise from loop effects. 
		This reveals a rich analytic structure of the two-point function in our system. 
		In the absence of interactions, there are two branch points in the lower half of the complex-frequency plane, indicating a weak-coupling limit in the system.
		The hydrodynamic interaction changes this by adding another branch point, which we calculate. This rich structure is shown in  Fig.~\ref{analytic_G_rr}.
		\item Finally, we showed that our Schwinger-Keldysh effective action for diffusion can also be derived from the noisy Boltzmann equation using the MSR formalism. 
		At the level at which we presented results in this paper, the two formalisms are identical. 
		However, our Schwinger-Keldysh EFT can be systematically generalized to include the effects of non-Gaussian noise, which is not known in the case of the MSR formalism.
	\end{enumerate}
	One possible application of our EFT results could be a system at weak coupling with a large susceptibility. 
	This can be, for instance, described by a weakly coupled quantum field theory at relatively large $N$. 
	Following Ref.~\cite{Liu:2018kfw}, in this limit our Schwinger-Keldysh Lagrangian is formally given by
	\begin{displaymath}
		\begin{split}
			\mathcal{L}=&\,\mathcal{L}^{\text{free}}+\mathcal{L}^{(3)}+\mathcal{L}^{(4)}+\ldots\\
			\sim&\,\mathcal{O}\big(\epsilon+\epsilon^2+\ldots\big)+\frac{1}{N}\mathcal{O}\big(\epsilon^2+\epsilon^3+\ldots\big)+\frac{1}{N^2}\mathcal{O}\big(\epsilon^3+\epsilon^4+\ldots\big)+\ldots\;.
		\end{split}
	\end{displaymath}
	For large $N$, one can truncate the $\frac{1}{N}$ expansion at any arbitrary order, while keeping an infinite order of derivatives \cite{Abbasi:2021fcz}.  In this work we truncate the $\frac{1}{N}$ expansion at sub-leading order.
	
	Having a Schwinger-Keldysh EFT for diffusion in an analytical form and to infinite order in derivatives, it could be worthwhile to study \textit{real-time dynamics} of correlation functions.\footnote{\textcolor{black}{See Ref.~\cite{Vardhan:2024qdi} for a discussion on the real-time dynamics of one-point functions in the Schwinger-Keldysh EFT framework.}} 
	In this work, we focused on the structure of correlation functions in momentum space. 
	In principle one should be able to find the inverse Fourier transform of these correlators numerically. 
	Then this would provide us with an important tool to investigate the thermalization time scale in the systems discussed above. 
	In particular, by considering the late-time dynamics of three-point functions,  we can determine: \textit{first} how fast this correlator thermalizes, and \textit{second} at what time scale the leading-order results in the $\epsilon$-expansion are reliable.
	
	In another direction, it would be very interesting to construct the Schwinger-Dyson equations to compute the evolution of the correlators in the EFT. 
	This would allow us to study the real-time dynamics of the correlators in a time-dependent background. 
	The results could then be compared with those in large-$N$ holographic systems under strong coupling \cite{Cartwright:2019opv,Erdmenger:2012xu,Ammon:2016fru,Landsteiner:2017lwm,Wondrak:2020tzt}, hopefully revealing more aspects of the dynamical thermalization associated with the QGP \cite{Florkowski:2017olj}.
	
	From another direction, it would be very interesting to develop an EFT describing  causal and stable diffusion \cite{Jain:2023obu,Hoult:2021gnb,Pu:2009fj}, to all order in derivatives, for the system studied in this paper.
	
One of the main advantages of working within RTA is that it allows for substantial analytical progress. It would be interesting to explore extensions where the collision kernel is derived from microscopic quantum field theories, such as scalar $\lambda \phi^4$
	theory \cite{Moore:2018mma,Grozdanov:2018atb,Denicol:2022bsq,Ochsenfeld:2023wxz,Rocha:2024cge}, leading to a momentum-dependent relaxation time. While the momentum dependence complicates the analysis, it is hoped that by constructing an appropriate Schwinger-Keldysh EFT, one could systematically study fluctuations in this more realistic setting.

	Before concluding this article, we would like to emphasize that this work is only the first step towards the main goal, which is to study non-Gaussian correlation functions near critical points. 
	In this work, we only focused on developing methods and ideas in special microscopic systems far away from any critical points. 
	The next step is to implement critical properties in the equation of state and study the dynamical fluctuations.

	\section*{Acknowledgments}
	
	We would like to thank Carsten Greiner and Hendrik van Hees for useful discussions on noisy Boltzmann equation. 
	N.A.\ would like to thank Ali Davody for valuable discussions and comments during the completion of this work in Frankfurt. 
	We are grateful to Luca Delacr\'etaz  and Sa\v{s}o Grozdanov for comments on the draft of the paper. 
	N.A.\ would like to thank the Institute for Theoretical Physics of Goethe University for the warm hospitality during the completion of this work.
	N.A.\ was supported by grant number 561119208 “Double First Class” start-up funding of Lanzhou University, China. The research of N.A.\ was also supported in part by the ExtreMe Matter Institute EMMI at the GSI Helmholtzzentrum für Schwerionenforschung, Darmstadt, Germany.
	D.H.R.\ acknowledges support by the Deutsche Forschungsgemeinschaft (DFG, German Research Foundation) through the CRC-TR 211 ``Strong-interaction matter under extreme conditions'' – project number 315477589 – TRR 211 and
	by the State of Hesse within the Research Cluster ELEMENTS (Project ID 500/10.006). 
	
	\appendix

	\section{Useful integrals}
	\label{integrals}
	Recalling that $D_p\equiv D_{\omega, \textbf{k}}=- i \omega + i k \cos\theta$, we evaluate the following integrals:
	\begin{eqnarray}
		\int_{\Omega}\frac{1}{1+\tau D_p}&=&L_p\;, \\
		\int_{\Omega}\frac{\cos \theta}{1+\tau D_p}&=&\frac{1}{i\tau k}\,\Big[1-(1-i \tau \omega)L_p\Big]\;,\\
		\int_{\Omega}\frac{\cos^2 \theta}{1+\tau D_p}&=&\frac{1- i \tau \omega}{\tau^2 k^2}\,\Big[1-(1-i \tau \omega)L_p\Big]\;,\\
		\int_{\Omega}\frac{D_p}{1+\tau D_p}&=&\frac{1}{\tau}\,\Big(1-L_p\Big)\;,\label{A.4} \\
		\int_{\Omega}\frac{D_p^2}{1+\tau D_p}&=&\frac{1}{\tau^2}\,\Big(-1-i \tau \omega+L_p\Big)\;,\\
		\int_{\Omega}\bigg(\frac{1}{1+\tau D_p}\bigg)_{\Theta}&=&\frac{1}{2}\,\Big(L_p+L^*_p\Big)\;,\\
		\int_{\Omega}\bigg(\frac{1}{1+\tau D_p}\bigg)_{\Theta}D_p&=&\frac{1}{2\tau}\,\Big(-L_p+L^*_p\Big)\;,
	\end{eqnarray}
	with 
	\begin{equation}\label{L_p_2}
		L_p=\frac{1}{2 i \tau k}\ln \left(\frac{\frac{i}{\tau}+\omega - k}{\frac{i}{\tau}+\omega +k}\right) \;.
	\end{equation}
	
	\section{Diffusion equation in a more familiar form}
	\label{diff_familiar}
	According to the discussion below Eq.~\eqref{Current_re_group} and also the content of Table.~\ref{Table}, we can drop all the derivative corrections from $\text{J}^0$ in the diffusion equation. 
	It then takes the following form
	\begin{equation}\label{equ_diff}
		\partial_t n + \nabla_i\bigg[\,-\frac{\tau}{3}\nabla_i n+\tau^2\left(\frac{2}{3} \nabla_i\dot{n}-\frac{\tau}{5}\nabla_i\nabla^2n\right)-\tau^3\left(\nabla_i\ddot{n}-\frac{4\tau}{5}\nabla_i\nabla^2\dot{n}+\frac{\tau^2}{7}\nabla_i \nabla^4n\right)+\ldots\bigg]=0.
	\end{equation}
	The same procedure done to simplify $\text{J}^0$ in Tab.~\ref{Table} can be done for $\text{J}^i$, i.e., the expression in  the brackets in the above equation. Let us do it in more detail here. First we note that the perturbative solution to the above equation can be formally written as 
	\begin{equation}\label{}
		n=\,n_{(2)}+\epsilon^2\,n_{(4)}+\epsilon^4\,n_{(6)}+\cdots\equiv\, n_{(2)+(4)+(6)+\cdots} \;,
	\end{equation}
	where the subscripts point out the order of equation in the $\epsilon$-expansion. 
	Substituting this ansatz into Eq.~\eqref{equ_diff}, we can read off the corresponding equations:
	\begin{equation}\label{equations_in_epsilon}
		\begin{split}
			\mathcal{O}(\epsilon^2)&\,\,\,\,\,\,\,\,\,\,\,\,\,
			\dot{n}_{(2)}-\frac{\tau}{3}\nabla^2 n_{(2)}=0\;,\\
			\mathcal{O}(\epsilon^4)&\,\,\,\,\,\,\,\,\,\,\,\,\,
			\dot{n}_{(4)}-\frac{\tau}{3}\nabla^2 n_{(4)}+\tau^2\bigg(\frac{2}{3}\nabla^2\dot{n}_{(2)}-\frac{\tau}{5}\nabla^4n_{(2)}\bigg)=0\;,\\
			\mathcal{O}(\epsilon^6)&\,\,\,\,\,\,\,\,\,\,\,\,\,
			\dot{n}_{(6)}-\frac{\tau}{3}\nabla^2 n_{(6)}+\tau^2\bigg(\frac{2}{3}\nabla^2\dot{n}_{(4)}-\frac{\tau}{5}\nabla^4n_{(4)}\bigg)\\
			& \hspace*{0.5cm}-\tau^3\bigg(\nabla^2\ddot{n}_{(2)}-\frac{4\tau}{5}\nabla^4\dot{n}_{(2)}+\frac{\tau^2}{7}\nabla^6n_{(2)}\bigg)=0\;,
		\end{split}
	\end{equation}
	and similarly for higher orders in the expansion. 
	In order to impose the above equations to Eq.~\eqref{equ_diff}, we first substitute  $n=\,n_{(2)}+\epsilon^2n_{(4)}+\epsilon^6n_{(6)}$ into the equation; up to $\mathcal{O}(\epsilon^6)$, we have 
	\begin{equation}\label{full_equation_up_to_6}
		\begin{split}
			&\dot{n}_{(2)+(4)+(6)} \,-\frac{\tau}{3}\nabla^2 n_{(2)+(4)+(6)}\\
			&+\tau^2\left(\frac{2}
			{3} \nabla^2\dot{n}_{(2)+(4)}-\frac{\tau}{5}\nabla^4n_{(2)+(4)}\right)\\
			&-\tau^3\left(\nabla^2\ddot{n}_{(2)}-\frac{4\tau}{5}\nabla^4\dot{n}_{(2)}+\frac{\tau^2}{7} \nabla^6n_{(2)}\right)=0\;.
		\end{split}
	\end{equation}
	What we need to do is to eliminate the dot terms in the second and third lines. Before doing this we simplify the second equation  \eqref{equations_in_epsilon}, simply by imposing the first equation:
	\begin{equation}\label{}
		\begin{split}
			&\dot{n}_{(4)}-\frac{\tau}{3}\nabla^2 n_{(4)}+\tau^2\bigg(\frac{2}{3}\nabla^2\big(\frac{\tau}{3}\nabla^2 n_{(2)}\big)-\frac{\tau}{5}\nabla^4n_{(2)}\bigg)=0\\
			\rightarrow \,\,\,\,&\dot{n}_{(4)}=\,\frac{\tau}{3}\nabla^2 n_{(4)}-\frac{\tau^3}{45}\nabla^4n_{(2)}\;.
		\end{split}
	\end{equation}
	Applying this equation together with $\dot{n}_{(2)}=\frac{\tau}{3}\nabla^2 n_{(2)}$ to Eq.~\eqref{full_equation_up_to_6}, we find
	\begin{equation}\label{full_equation_up_to_6_2}
		\begin{split}
			&\dot{n}_{(2)+(4)+(6)} \,-\frac{\tau}{3}\nabla^2 n_{(2)+(4)+(6)}\\
			&+\tau^2\left(\frac{2\tau}
			{9} \nabla^4 n_{(2)+(4)}-\frac{2\tau^3}
			{135} \nabla^6 n_{(2)}-\frac{\tau}{5}\nabla^4n_{(2)+(4)}\right)\\
			&-\tau^3\left(\frac{\tau^2}{9}\nabla^6n_{(2)}-\frac{4\tau^2}{15}\nabla^6n_{(2)}+\frac{\tau^2}{7} \nabla^6n_{(2)}\right)=0\;,
		\end{split}
	\end{equation}
	which simplifies to 
	\begin{equation}\label{}
		\dot{n}_{(2)+(4)+(6)} \,-\frac{\tau}{3}\nabla^2 n_{(2)+(4)+(6)}
		+\frac{\tau^3}
		{45} \nabla^4 n_{(2)+(4)}-\frac{2\tau^5}{945} \nabla^6n_{(2)}=0\;.
	\end{equation}
	Note that up to $\mathcal{O}(\epsilon^6)$, the above equation can be also written as 
	\begin{equation}\label{}
		\dot{n}_{(2)+(4)+(6)} \,-\frac{\tau}{3}\nabla^2 n_{(2)+(4)+(6)}
		+\frac{\tau^3}
		{45} \nabla^4 n_{(2)+(4)\textcolor{red}{+(6)}}-\frac{2\tau^5}{945} \nabla^6n_{(2)\textcolor{red}{+(4)+(6)}}=0\;.
	\end{equation}
	The terms in red are beyond order $\mathcal{O}(\epsilon^6)$, however we have just added them to formally construct an equation for $n=\,n_{(2)}+\epsilon^2\,n_{(4)}+\epsilon^4\,n_{(6)}+\cdots\equiv\, n_{(2)+(4)+(6)+\cdots}$. The above equation is now exactly Eq.~\eqref{diffusion_changed} truncated at order $\mathcal{O}(\epsilon^6)$.
	\section{KMS constraints on the free effective action}
	\label{KMS_L_2}
	Having the classical equation of motion \eqref{Resummed} in hand, the dissipative part of the effective Lagrangian is written as 
	\begin{equation}\label{diss_L}
		\mathcal{L}	= -\int_{\Omega}\left(\frac{\mathbf{D}}{1+ \tau \textbf{D}}\, n\right) \phi_a+\,\ldots\;,
	\end{equation}
	where $\dot{\phi}\equiv\mu$. 
	Now we parameterize the action to quadratic order in the $\phi_a$-field as follows:
	\begin{equation}\label{L_2_bofore_KMS}
		\mathcal{L}^{\text{free}}	=  \int_{\Omega}\bigg[(\textbf{D}\phi_a)\frac{1}{1+ \tau\textbf{D}} n  +i\, (\textbf{D}\phi_a)\,M\,\textbf{D} \phi_a\bigg]\;.
	\end{equation}
	The operator $M$ is found from applying the KMS constraint (in the classical limit) to the action, i.e., requiring the action to be invariant under 
	\begin{equation}\label{KMS}
		\phi\textcolor{black}{(x)}\rightarrow \tilde{\phi}\textcolor{black}{(x)}=\Theta \phi\textcolor{black}{(x)}\;,\,\,\,\,\,\,\,\,\phi_a\textcolor{black}{(x)}\rightarrow \tilde{\phi}_a\textcolor{black}{(x)}=\Theta\left[ \phi_a\textcolor{black}{(x)}+ i \beta \dot{\phi}\textcolor{black}{(x)}\right]\;,
	\end{equation}
	with $\Theta=\mathcal{PT}$. 
	These relations are referred to as 
	$\mathbb{Z}_2$ \textit{dynamical KMS transformation} \cite{Liu:2018kfw} (note that $\Theta^2=1$).
	We now perform the $\mathbb{Z}_2$  dynamical KMS transformation of the action $\int_x \mathcal{L}^{\text{free}}$.
	It is advantageous to first substitute $x \rightarrow \Theta x \equiv - x$ in the space-time integral of the action before performing the transformation. 
	We then use
	\begin{equation}\label{KMS_3}
		\begin{split}
			\tilde{\phi}(\Theta x)=&\,\Theta \phi(\Theta x)\equiv\,-\phi(x)\,,\\
			\tilde{\phi}_a(\Theta x)=&\,\Theta\phi_a(\Theta x)- i \beta (\Theta\partial_t)\,\Theta \phi(\Theta x)=\,-\big[\phi_a(x)+ i \beta \dot{\phi}(x)\big]\;,
		\end{split}
	\end{equation}
	where we have employed Eq.~\eqref{KMS}.
	This specific transformation\footnote{See the explanations below Eq.~\eqref{gauging}.	
		For more details, see Table I in Ref.~\cite{Jain:2023obu}. }  of $\phi$ and $\phi_a$ under $\Theta$ will be important to satisfy the KMS invariance of $\int_x \mathcal{L}^{(3)}$.
	The result of the transformation of the action is then
	\begin{align}\label{L_T}
		\lefteqn{\int_x	\mathcal{L}^{\text{free}}\rightarrow 
			\,\int_x\mathcal{L}^{\text{free}}} \nonumber\\
		&+\int_x\int_{\Omega}\left[(\textbf{D}\phi_a)\bigg(\frac{1}{1- \tau \textbf{D}}-\frac{1}{1+ \tau \textbf{D}}\bigg)\, \chi\dot{\phi}\, -\beta \,(\textbf{D}\phi_a)\, (\Theta M) \textbf{D}\dot{\phi}-\beta\,(\textbf{D}\dot{\phi}) \,(\Theta M) \textbf{D}\phi_a\right]\nonumber\\
		&+\int_x\int_{\Omega}\,\left[i\beta\,(\textbf{D}\dot{\phi})\,\frac{1}{1- \tau \textbf{D}}\chi\dot{\phi} -\,i \beta^2\,(\textbf{D}\dot{\phi})\,(\Theta M)\,\textbf{D}\dot{\phi}\right]\nonumber\\
		&+\int_x\int_{\Omega}\,\big[i\, (\textbf{D}\phi_a) (\Theta M)\textbf{D}\phi_a - i\,(\textbf{D}\phi_a)\,M\,\textbf{D}\phi_a\big]\;.
	\end{align}
	Here, we have also made use of the fact that $n=\chi \dot{\phi}$, with $\chi = const.$.
	
	In order for the quadratic effective action $\int_x \mathcal{L}^{\text{free}}$ to satisfy the KMS condition, the sum of all  extra terms on the right-hand side of Eq.~\eqref{L_T} must vanish. 
	To show how this happens and what the result of this cancellation is, we have classified them in three groups: 
	\begin{enumerate}
		\item [$\underline{\phi_a\phi_a}$] To make  the terms in last line of Eq.~\eqref{L_T} cancel, $M$ must be $\Theta$-invariant
		\begin{equation}\label{M_property}
			\Theta M = M\;.
		\end{equation}
		\item [$\underline{\dot{\phi}\,\dot{\phi}}$] Requiring such terms to cancel each other (in the third line of Eq.~\eqref{L_T}), one can easily read off $M$. 
		However, the resulting $M$ must obey Eq.~\eqref{M_property}. 
		Thus we find it useful to express these terms in an explicit $\Theta$-invariant way. 
		To this end, let us see how the first term transforms by twice performing an integration by parts
		\begin{equation}\label{integration_by_part}
			\begin{split}
				\int_{\Omega}\,(\textbf{D}\dot{\phi})\,\frac{1}{1- \tau \textbf{D}}\dot{\phi}=&\,\int_{\Omega}\,(\textbf{D}\dot{\phi})\bigg[1+\tau\mathbf{D}+(\tau\mathbf{D})^2+(\tau\mathbf{D})^3+\ldots\bigg]\dot{\phi}\\
				\mbox{integration by parts}\,\rightarrow &\,\int_{\Omega}\,\left\{\bigg[1-\tau\mathbf{D}+(\tau\mathbf{D})^2-(\tau\mathbf{D})^3+\ldots\bigg]\textbf{D}\dot{\phi}\right\}	 \dot{\phi}  \\
				\mbox{integration by parts}\,\rightarrow &\,\int_{\Omega}\,\left(\frac{1}{1+ \tau \textbf{D}}\dot{\phi}\right)(-\mathbf{D}\dot{\phi}) \;.
			\end{split}
		\end{equation}
		This tells us that the third line of Eq.~\eqref{L_T} can be written in the following $\Theta$-invariant form:
		\begin{equation}\label{rewrite_B.5}
			\begin{split}
				&+i\beta\chi\int_{\Omega}\,\frac{1}{2}\,(\mathbf{D}\dot{\phi})\,\bigg(\frac{1}{1- \tau \textbf{D}}-\frac{1}{1+ \tau \textbf{D}}\bigg) \dot{\phi} -\,i \beta^2\,\int_{\Omega}\,(\textbf{D}\dot{\phi})\,(\Theta M)\,\textbf{D}\dot{\phi}\\
				=&+i\beta\chi\tau\int_{\Omega}\,(\mathbf{D}\dot{\phi})\,\frac{1}{1- (\tau \textbf{D})^2}\mathbf{D}\dot{\phi} -\,i \beta^2\int_{\Omega}\,(\textbf{D}\dot{\phi})\,(\Theta M)\,\textbf{D}\dot{\phi}\;.
			\end{split}
		\end{equation}
		Requiring the above expression to vanish, we find
		\begin{equation}\label{M_classical}
			\boxed{M=T\,\chi\,\tau\,\bigg(\frac{1}{1+ \tau \textbf{D}}\bigg)_{\Theta}}\;,
		\end{equation}
		with
		\begin{equation} \label{def_Thetasym}
			\bigg(\frac{1}{1+ \tau \textbf{D}}\bigg)_{\Theta}\equiv\frac{1}{2}\bigg(\frac{1}{1- \tau \textbf{D}}+\frac{1}{1+ \tau \textbf{D}}\bigg) \;.
		\end{equation}
		One readily observes that the expression \eqref{M_classical} for $M$ fulfills the requirement \eqref{M_property}.
		
		\item [$\underline{\phi_a\,\dot{\phi}}$] As the last requirement, one can straightforwardly show that Eq.~\eqref{M_classical} makes the terms in the second line of Eq.~\eqref{L_T} cancel each other. 
		To see this, note that the differential operator $M$ is invariant under integration by parts, i.e, it can act to the right onto $\textbf{D} \dot{\phi}$ as well as to the left onto $\textbf{D} \phi_a$, which makes the last two terms in the second line of Eq.~\eqref{L_T} equal.
		This completes the proof of KMS invariance of the action with the Lagrangian \eqref{L_2_bofore_KMS}, with $M$ given by Eq.~\eqref{M_classical}. 
	\end{enumerate}
	
	In order to read off the physical correlation functions, the effective action must be coupled to gauge-invariant sources.  
	Following Ref.~\cite{Crossley:2015evo}, we need to apply the following replacement to the Lagrangian,
	\begin{equation}\label{gauging}
		\partial_{\mu}\phi_{a,r}\rightarrow B_{a,r\mu}=\partial_{\mu}\phi_{a,r}+A_{a,r\mu}\;.
	\end{equation}
	This justifies the sign change of $\phi_r \equiv \phi$ and $\phi_a$ under the action of $\Theta$; since $\Theta A_{a,r\,\mu}=A_{a,r\,\mu}$ and $\Theta \partial_{\mu}=-\partial_{\mu}$, for $B_{a,r\,\mu}$ to transform consistently under $\Theta$, it is required that  $\Theta \phi_{a,r\,\mu}=-\phi_{a,r\,\mu}$.

	Since we only want to compute $r$-correlators in this work, the $A_{r\mu}$ sources can be turned off. 
	Doing so, Eq.~\eqref{L_2_bofore_KMS} takes the following form
	\begin{equation}\label{gauging_L_2}
		\begin{split}
			\mathcal{L}	=&  \int_{\Omega}\left\{ B_{a0}\frac{1}{1+ \tau\textbf{D}} n
			+ (\nabla_i\phi_a)\frac{v_i}{1+ \tau\textbf{D}} n+\,i \,T\sigma \bigg[ B_{a0}\frac{1}{1+  \tau\textbf{D}}B_{a0} \right.\\
			&\left.+ \,B_{a0}\frac{v_i}{1+  \tau\textbf{D}} (\nabla_i\phi_a)+ \, (\nabla_i\phi_a)\frac{v_i}{1+  \tau\textbf{D}}B_{a0}+\, (\nabla_i\phi_a)\frac{v_iv_j}{1+  \tau\textbf{D}} (\nabla_j\phi_a)\bigg]\right\},
		\end{split}
	\end{equation}
	where $\sigma\equiv \chi \tau$.
	Note that we have also dispensed with the $\Theta$-symmetrization of $M$, Eq.~\eqref{M_classical}, as one can show via an integration by parts that both terms in Eq.~\eqref{def_Thetasym} lead to the same result.
	\section{Self-interactions and background sources}
	\label{App_self_int}
	In order to find the cubic terms in the Lagrangian coupled to the background sources, we now proceed by incorporating self-interactions into the Lagrangian. 
	This is done by promoting the transport coefficients in the quadratic Lagrangian to become functions of the dynamical field $\mu\equiv\dot{\phi}$. 
	\begin{equation}\label{Full_S_eff}
		\begin{split}
			\mathcal{L}	=&  \int_{\Omega}\left\{ B_{a0}\frac{1}{1+ \tau(\mu)\textbf{D}} n
			+ (\nabla_i\phi_a) \frac{v_i}{1+ \tau(\mu)\textbf{D}} n +\,i \,T\sigma(\mu) 
			\bigg[ \,B_{a0}\frac{1}{1+  \tau(\mu)\textbf{D}}B_{a0} \right. \\
			&\left. + \,B_{a0}\frac{v_i}{1+  \tau(\mu)\textbf{D}} \nabla_i\phi_a+ \, (\nabla_i\phi_a)\frac{v_i}{1+  \tau(\mu)\textbf{D}}B_{a0}+\, (\nabla_i\phi_a)\frac{v_iv_j}{1+  \tau(\mu)\textbf{D}} \nabla_j\phi_a\bigg]\right\}\;,
		\end{split}
	\end{equation}
	where we have (see Eq.~\eqref{tau_chi} and the explanations below it)
	\begin{equation}\label{self_int_def}
		\begin{split}
			n(\mu)=&\,\chi\,\mu+\chi' \frac{\mu^2}{2}  +\ldots\,\,\,\,\,\,\rightarrow\,\,\,\,\,\chi(\mu)=\frac{\partial n}{\partial \mu}=\,\chi+\chi'\mu+\ldots\;,\\
			\tau(\mu)=&\,\tau+\tau'\, \mu+\ldots\;,\\
			\sigma(\mu)=&\,\sigma+\sigma'\,\mu+\ldots\;.
		\end{split}
	\end{equation}
	Since we are interested in working with the field $n$, not $\mu$, then $\chi'$ will not appear in our following expressions. 
	Note that $\chi'$ could appear if we turned on an $A_r$-source \cite{Jain:2020zhu}, however, as we said earlier, we need only the $A_{a0}$ source.
	
	We now determine how quadratic and cubic terms can be found by systematically truncating the above effective action. 
	Let us start with the dissipative term and apply Eq.~\eqref{self_int_def} to it 
	\begin{equation}\label{C_3}
		\begin{split}
			B_{a0}\frac{1}{1+ \tau(\mu)\textbf{D}} n\, =\,&B_{a0}\frac{1}{1+ \tau\textbf{D}} n -B_{a0}\frac{1}{1+ \tau\textbf{D}}\Big(\tau' \mu\,\textbf{D}\,\frac{1}{1+ \tau\textbf{D}}n\Big)+\ldots\\
			=\,&B_{a0}\frac{1}{1+ \tau\textbf{D}} n -\,\frac{\tau' }{\chi}\,n\,\,\left(\frac{\textbf{D}}{1+ \tau\textbf{D}}n\right) \frac{1}{1- \tau\textbf{D}}B_{a0}+\ldots\;,
		\end{split}
	\end{equation}
	where we have used the fact that $(A+a)^{-1}=A^{-1}-A^{-1}aA^{-1}+\mathcal{O}(a^2)$. 
	From the first to the second line we have replaced $\mu=\frac{n}{\chi}+\mathcal{O}(n^2)$ and also have performed an integration by parts.
	Let us note that the dots contain terms beyond the cubic order in the fields.
	
	Now we just need to apply the above procedure to the full nonlinear effective action given by Eq.~\eqref{Full_S_eff}, and find the quadratic and cubic orders, separately.
	For now, we temporarily turn off the sources and just present the effective action in the absence of them:
	\begin{equation}
		\begin{split}
			\mathcal{L}^{\text{free}}	=&  \int_{\Omega}\bigg[(\textbf{D}\phi_a)\frac{1}{1+ \tau\textbf{D}} n  +i T \sigma (\textbf{D}\phi_a)\bigg(\frac{1}{1+\tau \textbf{D}}\bigg)_{\Theta}\textbf{D} \phi_a \bigg]
			\\\label{L_3_exact}
			\mathcal{L}^{(3)}=&\int_{\Omega}\left\{-\frac{\tau' }{\chi}n\left(\frac{\textbf{D}}{1+ \tau\textbf{D}}n\right) \frac{\textbf{D}}{1- \tau\textbf{D}}\phi_a+ i T \sigma\frac{\tau' }{\chi}n\,\left[ \bigg(\frac{\textbf{D}}{1+ \tau\textbf{D}}\phi_a\bigg)  \frac{\textbf{D}}{1- \tau\textbf{D}}\right]_{\Theta}\textbf{D}\phi_a \right.\\
			&\qquad +\left.i T \frac{\sigma'}{\chi} n(\textbf{D}\phi_a)\bigg(\frac{1}{1+\tau \textbf{D}}\bigg)_{\Theta}\textbf{D} \phi_a
			\right\}\;.
		\end{split}
	\end{equation}

	\subsection*{KMS constraints on the cubic Lagrangian}
	In this section we investigate how the KMS condition constrains the cubic effective action 
	$\mathcal{L}^{(3)}$ in Eq.~\eqref{L_3_exact}.
	Performing the $\mathbb{Z}_2$ dynamical KMS transformation using similar steps as for the free action in Appendix \ref{KMS_L_2}, i.e., first substituting $x \rightarrow \Theta x$ in the space-time integral and then applying \textcolor{blue}{\eqref{KMS}}, we find 
	\begin{equation}\label{L_3_T}
		\begin{split}
			\int_x\mathcal{L}^{(3)}	\,\,\,\,\,\,\rightarrow \,\,\,\,\,&\int_x\int_{\Omega}\left( -\tau'\chi \dot{\phi}\left(\frac{-\mathbf{D}}{1- \tau \textbf{D}}\dot{\phi}\right)\frac{-\mathbf{D}}{1+ \tau \textbf{D}} (-\phi_a- i \beta \dot{\phi}) \right. \\
			&+ i T \sigma\tau' \,\dot{\phi}\,\left\{ \left[\frac{-\textbf{D}}{1- \tau\textbf{D}}(-\phi_a- i \beta \dot{\phi})\right]  \frac{-\textbf{D}}{1+ \tau\textbf{D}}\right\}_{\Theta}(-\textbf{D})(-\phi_a- i \beta \dot{\phi})\\
			&+i \left. T \sigma' \dot{\phi}\left[ (-\textbf{D})(-\phi_a- i \beta \dot{\phi})\right] \bigg(\frac{1}{1-\tau \textbf{D}}\bigg)_{\Theta}(-\textbf{D})(-\phi_a- i \beta \dot{\phi})\right)\;.
		\end{split}
	\end{equation}
	\begin{itemize}
		\item Let us first consider the $\dot{\phi}\dot{\phi}\dot{\phi}$ terms. 
		Since such terms did not appear in the original cubic Lagrangian \eqref{L_3_exact}, we have to require that all these terms cancel each other.
		Thus we obtain
		\begin{equation}\label{KSM_constarint}\boxed{
				\sigma=\,\chi\,\tau\,,\,\,\,\,\,\,\,\,\frac{\sigma'}{\sigma}=\frac{\tau'}{\tau}}\;.
		\end{equation}

		\item Using Eq.~\eqref{KSM_constarint}, one can show that the three $\dot{\phi}\dot{\phi}\phi_a$ terms sum up to the first term of $\mathcal{L}^{\text{(3)}}$ in Eq.~\eqref{L_3_exact}.
		\item Finally, it is seen that the two $\dot{\phi}\phi_a\phi_a$ terms (in the second and third lines of Eq.~\eqref{L_3_T}) are nothing but the last two terms of $\mathcal{L}^{(3)}$ in Eq.~\eqref{L_3_exact}.
	\end{itemize}
	Applying the above two constraints for $\mathcal{L}^{\text{(3)}}$ to Eq.~\eqref{L_3_exact}, it takes the following simpler form
	\begin{equation}\label{L_3_KMS_final}
		\begin{split}
			\mathcal{L}^{(3)}=\,\int_{\Omega}\bigg[	-\frac{\tau' }{\chi}n\left(\frac{\textbf{D}}{1+ \tau\textbf{D}}n\right)\frac{\textbf{D}}{1- \tau\textbf{D}}\phi_a+ i T \tau'\,n\left(\frac{\textbf{D}}{1+ \tau\textbf{D}}\phi_a  \right)\frac{\textbf{D}}{1- \tau\textbf{D}}\phi_a\bigg]\;.
		\end{split}
	\end{equation}
	\section{$\mathcal{I}$ functions}
	\label{I}
	Considering
	\begin{equation}\label{def_I_function}
		\begin{split}
			\int_{\Omega}\frac{\tau D_2}{1+\tau D_{2}}\frac{-\tau D_3}{1-\tau D_{3}}=\,\frac{1}{4\pi}	\int_{0}^{2\pi} d\phi \,\mathcal{I}(\phi;1,2,3)
		\end{split}
	\end{equation}
	we find
	\begin{equation}\label{mathcal_I}
		\begin{split}
			\mathcal{I}&(\phi;  1,2,3)=\,2-i\pi\frac{  q_2\cos \phi-q_3\cos(\alpha_{23}-\phi)}{q_2 q_3  \cos\phi \cos(\alpha_{23}-\phi)}\\
			&-\frac{2i \pi  \left(\wn_2+i\right)^2\sqrt{1-\frac{q_2^2  \cos ^2\phi }{\left(\wn_2+i\right){}^2}} \left[-q_2 \wn_3+q_3 \left(\wn_2+i\right) \frac{\cos(\alpha_{23}-\phi)}{\cos \phi } \right]}{q_2\Big[q_2^2-2(\wn_3+i)^2+q_2^2\cos(2\phi)\Big]\left[q_2 \left(\wn_3-i\right)-q_3 \left(\wn_2+i\right) \frac{\cos(\alpha_{23}-\phi)} {   \cos \phi }\right]}\\
			&+\frac{2i \pi  \left(\wn_3-i\right)^2  \sqrt{1-\frac{q_3^2 \cos ^2(\alpha_{23} -\phi )}{\left(\wn_3-i\right){}^3}} \left[-q_3 \wn_2+q_2 \left(\wn_3-i\right) \frac{\cos \phi }{\cos(\alpha_{23}-\phi)} \right]}{q_3\Big[q_3^2-2(\wn_3-i)^2+q_3^2\cos\big(2(\alpha_{23}-\phi)\big)\Big]\left[q_3 \left(\wn_2+i\right)-q_2 \left(\wn_3-i\right) \frac{   \cos \phi }{\cos(\alpha_{23}-\phi)} \right]}\\
			&-\,\frac{i \left(\wn_2+i\right)  \left[-q_2 \wn_3+q_3 \left(\wn_2+i\right) \frac{\cos(\alpha_{23}-\phi)}{\cos \phi}  \right]\ln \left(\frac{\sqrt{q_2^2 \cos ^2 \phi +\left(1+i \wn_2\right){}^2}-q_2 \cos \phi }{\sqrt{q_2^2 \cos ^2\phi  +\left(1+i \wn_2\right){}^2}+q_2 \cos \phi }\right)}{q_2\sqrt{q_2^2  \cos ^2\phi -\left(\wn_2+i\right){}^2} \left[q_3 \left(\wn_2+i\right) \cos(\alpha_{23}-\phi)-q_2 \left(\wn_3-i\right)\cos\phi\right]}\\
			&-\frac{i \left(\wn_3-i\right)  \left[-q_3 \wn_2+q_2 \left(\wn_3-i\right) \frac{\cos \phi}{\cos(\alpha_{23}-\phi)}  \right]\ln \left( \frac{\sqrt{q_3^2 \cos ^2(\alpha_{23} -\phi )+\left(1+i \wn_3\right){}^2}-q_3 \cos (\alpha_{23}-\phi) }{\sqrt{q_3^2 \cos ^2(\alpha_{23} -\phi )+\left(1+i \wn_3\right){}^2}+q_3 \cos (\alpha_{23}-\phi) }\right)}{q_3\sqrt{q_3^2  \cos ^2(\alpha_{23}-\phi) -\left(\wn_3-i\right){}^2} \left[q_3 \left(\wn_1+i\right) \cos(\alpha_{23}-\phi)-q_2 \left(\wn_3-i\right)\cos\phi\right]}\;,
		\end{split}
	\end{equation}
	with $\alpha_{23}=\cos^{-1}\Big(\frac{q_1^2-q_2^2-q_3^2}{2 q_2 q_3}\Big)$.
	\section{A quick check: reproducing the effective action of Chen-Lin, Delacr\'etaz, and Hartnoll \cite{Chen-Lin:2018kfl}}
	\label{quick_check}
	In Ref.~\cite{Chen-Lin:2018kfl} the EFT for charge diffusion for a general non-integrable QFT has been constructed. 
	Comparing that work and the present, each has one distinct advantage over the other:
	\begin{enumerate}
		\item While Ref.~\cite{Chen-Lin:2018kfl} constructs the EFT for diffusion in a general non-integrable QFT, our EFT is specific to a system of massless relativistic particles described by kinetic theory in RTA.
		\item In Ref.~\cite{Chen-Lin:2018kfl}, the EFT is constructed to leading order in the derivative expansion, prescribed by the derivative counting scheme $\omega\sim k^2\sim \epsilon^2$. 
		Our EFT, however, contains all orders in derivative.
	\end{enumerate}
	As a result, our effective action should  be consistent with that of Ref.~\cite{Chen-Lin:2018kfl} in  leading order of the  derivative expansion, i.e., to order $\epsilon^2$, upon appropriate identification of the transport coefficients between the two works. To check the consistency, let us point out that
	\begin{equation}\label{int_expansion}
		\begin{split}
			\int_{\Omega}\,\frac{\textbf{D}}{1+\tau \textbf{D}}=&\,\partial_t-\frac{\tau}{3}\nabla^2+\mathcal{O}(\epsilon^3)\;, \\
			\int_{\Omega}\,\bigg(\frac{\textbf{D}}{1+\tau \textbf{D}}\bigg)_{\Theta}=&-\frac{\tau}{3}\nabla^2+\mathcal{O}(\epsilon^3)\;.
		\end{split}
	\end{equation}
	Substituting this into Eq.~\eqref{L_2} we obtain
	\begin{equation}\label{S_eff_epsilon_2}
		S^{(2)}_{EFT}=\int_{x}\,\bigg[-\left(\partial_tn-\frac{\tau}{3}\nabla^2n\right)\,\phi_a-i\,T\,\chi\,\frac{\tau}{3}\phi_a\,\nabla^2\phi_a\bigg]\;,
	\end{equation}
	which is precisely the effective action of Ref.~\cite{Chen-Lin:2018kfl} with the diffusion constant of Ref.~\cite{Chen-Lin:2018kfl} identified by $D\equiv\frac{\tau}{3}$.
	
	To show the consistency between the two works at cubic order, we note that the function  $\mathcal{I}(\phi;1,2,3)$ takes a very simple form at order $\epsilon^2$:
	\begin{equation}\label{I_2_leading}
		\mathcal{I}(\phi;1,2,3)=\,k_2\,k_3\, \,\cos (\alpha-\phi)\,\cos \phi+\mathcal{O}(\epsilon^3)\;,
	\end{equation}
	where $\alpha$ is the angle between $\textbf{k}_2$ and $\textbf{k}_3$. Then our $\lambda_1$ and $\lambda_2$ simplify to 
	\begin{equation}\label{lambda_1_2_leading}
		\lambda_1=\frac{\tau_1}{3\tau^2}\,\textbf{k}_2\cdot \textbf{k}_3+\mathcal{O}(\epsilon^3)\,,\,\,\,\,\,\,\,\,\,\,\,\lambda_2=\,-i T\,\chi\frac{\tau_1}{3\tau^2}\,\textbf{k}_2\cdot \textbf{k}_3+\mathcal{O}(\epsilon^3)\;.
	\end{equation}
	To compare with Ref.~\cite{Chen-Lin:2018kfl} it should be noted that that reference studies the  diffusion of heat with $\kappa=c D$ being the thermal conductivity, and $c$ the specific heat. 
	This equation is equivalent to $\sigma = \chi \frac{\tau}{3}$ in the present work. 
	
	Considering this information, it is easy to show that $\lambda_1$ and $\lambda_2$ above (with $\textbf{k}_2\cdot \textbf{k}_3$ dropped) are exactly $\lambda$ and $\tilde{\lambda}$ in the cubic Lagrangian of Ref.~\cite{Chen-Lin:2018kfl}. 
	The momentum-dependent factor $-\textbf{k}_2\cdot \textbf{k}_3$ in Fourier space corresponds to $\nabla \cdot \nabla$ in real space. 
	Applying this correspondence to the vertices shown earlier, we then find the corresponding two terms in real space  as  $-\frac{\tau_1}{3}n\boldsymbol{\nabla} n\cdot \boldsymbol{\nabla} \phi_a \frac{1}{2}\equiv\frac{\tau_1}{3}n^2\boldsymbol{\nabla}^2\phi$  and $i T\,\chi \frac{\tau_1}{3}\,n \boldsymbol{\nabla} \phi\cdot \boldsymbol{\nabla} \phi_a$; in complete agreement with Ref.~\cite{Chen-Lin:2018kfl}.

	%


\begin{thebibliography}{10}
	
		
		
		\bibitem{An:2021wof}
		X.~An, M.~Bluhm, L.~Du, G.~V.~Dunne, H.~Elfner, C.~Gale, J.~Grefa, U.~Heinz, A.~Huang and J.~M.~Karthein, \textit{et al.}
		``The BEST framework for the search for the QCD critical point and the chiral magnetic effect,''
		Nucl. Phys. A \textbf{1017} (2022), 122343
		[arXiv:2108.13867 [nucl-th]].
		
		\bibitem{Du:2024wjm}
		L.~Du, A.~Sorensen and M.~Stephanov,
		``The QCD phase diagram and Beam Energy Scan physics: a theory overview,''
		[arXiv:2402.10183 [nucl-th]].
		
		
		\bibitem{Stephanov:2024xkn}
		M.~Stephanov,
		``QCD critical point: recent developments,''
		[arXiv:2410.02861 [nucl-th]].
		
		
		
		\bibitem{Stephanov:2008qz}
		M.~A.~Stephanov,
		``Non-Gaussian fluctuations near the QCD critical point,''
		Phys. Rev. Lett. \textbf{102} (2009), 032301
		[arXiv:0809.3450 [hep-ph]].
		
		\bibitem{An:2020vri}
		X.~An, G.~Ba\c{s}ar, M.~Stephanov and H.~U.~Yee,
		``Evolution of Non-Gaussian Hydrodynamic Fluctuations,''
		Phys. Rev. Lett. \textbf{127} (2021) no.7, 072301
		[arXiv:2009.10742 [hep-th]].
		
		
		
		\bibitem{Chattopadhyay:2024jlh}
		C.~Chattopadhyay, J.~Ott, T.~Schaefer and V.~V.~Skokov,
		``Simulations of Stochastic Fluid Dynamics near a Critical Point in the Phase Diagram,''
		Phys. Rev. Lett. \textbf{133} (2024) no.3, 3
		[arXiv:2403.10608 [nucl-th]].
		
		\bibitem{Tang:2023zvj}
		S.~Tang, S.~Wu and H.~Song,
		``Dynamical critical fluctuations near the QCD critical point with hydrodynamic cooling rate,''
		Phys. Rev. C \textbf{108} (2023) no.3, 034901
		[arXiv:2303.15017 [nucl-th]].
		
		
		\bibitem{Nahrgang:2018afz}
		M.~Nahrgang, M.~Bluhm, T.~Schaefer and S.~A.~Bass,
		``Diffusive dynamics of critical fluctuations near the QCD critical point,''
		Phys. Rev. D \textbf{99} (2019) no.11, 116015
		[arXiv:1804.05728 [nucl-th]].
		
		
		\bibitem{An:2022tfk}
		X.~An,
		``Non-Gaussian Fluctuation Dynamics,''
		Acta Phys. Polon. Supp. \textbf{16} (2023) no.1, 1-A47
		[arXiv:2209.15005 [hep-th]].
		
		
		\bibitem{Romatschke:2015gic}
		P.~Romatschke,
		``Retarded correlators in kinetic theory: branch cuts, poles and hydrodynamic onset transitions,''
		Eur. Phys. J. C \textbf{76} (2016) no.6, 352
		[arXiv:1512.02641 [hep-th]].
		
		
		
		\bibitem{Martin}
		L.~P.~Kadanoff and P.~C.~Martin, 
		``Hydrodynamic equations and correlation functions,''
		Ann. Phys. 24, 419–469, 1963.
		
		\bibitem{Kovtun:2012rj}
		P.~Kovtun,
		``Lectures on hydrodynamic fluctuations in relativistic theories,''
		J. Phys. A \textbf{45} (2012), 473001
		doi:10.1088/1751-8113/45/47/473001
		[arXiv:1205.5040 [hep-th]].
		
		
		\bibitem{Akamatsu:2016llw}
		Y.~Akamatsu, A.~Mazeliauskas and D.~Teaney,
		``A kinetic regime of hydrodynamic fluctuations and long time tails for a Bjorken expansion,''
		Phys. Rev. C \textbf{95} (2017) no.1, 014909
		[arXiv:1606.07742 [nucl-th]].
		
		
		\bibitem{Chao:2020kcf}
		J.~Chao and T.~Schaefer,
		``Multiplicative noise and the diffusion of conserved densities,''
		JHEP \textbf{01} (2021), 071
		[arXiv:2008.01269 [hep-th]].
		
		
		\bibitem{Chen-Lin:2018kfl}
		X.~Chen-Lin, L.~V.~Delacr\'etaz and S.~A.~Hartnoll,
		``Theory of diffusive fluctuations,''
		Phys. Rev. Lett. \textbf{122}, no.9, 091602 (2019)
		[arXiv:1811.12540 [hep-th]].
		
		
		\bibitem{Jain:2020zhu}
		A.~Jain and P.~Kovtun,
		``Late Time Correlations in Hydrodynamics: Beyond Constitutive Relations,''
		Phys. Rev. Lett. \textbf{128}, no.7, 7 (2022)
		[arXiv:2009.01356 [hep-th]].
		
		
		\bibitem{Delacretaz:2020nit}
		L.~V.~Delacretaz,
		``Heavy Operators and Hydrodynamic Tails,''
		SciPost Phys. \textbf{9} (2020) no.3, 034
		[arXiv:2006.01139 [hep-th]].
		
		
		
		\bibitem{Basar:2024srd}
		G.~Basar,
		``Recent developments in relativistic hydrodynamic fluctuations,''
		[arXiv:2410.02866 [hep-th]].
		
		
		
		\bibitem{Son:2002sd}
		D.~T.~Son and A.~O.~Starinets,
		``Minkowski space correlators in AdS / CFT correspondence: Recipe and applications,''
		JHEP \textbf{09}, 042 (2002)
		[arXiv:hep-th/0205051 [hep-th]].
		
		
		\bibitem{Bajec:2024jez}
		M.~Bajec, S.~Grozdanov and A.~Soloviev,
		``Spectra of correlators in the relaxation time approximation of kinetic theory,''
		[arXiv:2403.17769 [hep-th]].
		
		
		\bibitem{Kurkela:2017xis}
		A.~Kurkela and U.~A.~Wiedemann,
		``Analytic structure of nonhydrodynamic modes in kinetic theory,''
		Eur. Phys. J. C \textbf{79}, no.9, 776 (2019)
		[arXiv:1712.04376 [hep-ph]].
		
		
		\bibitem{Dash:2023ppc}
		D.~Dash, S.~Jaiswal, S.~Bhadury and A.~Jaiswal,
		``Relativistic second-order viscous hydrodynamics from kinetic theory with extended relaxation-time approximation,''
		Phys. Rev. C \textbf{108} (2023) no.6, 064913
		[arXiv:2307.06195 [nucl-th]].
		
		
		
		
		\bibitem{Brants:2024wrx}
		R.~Brants,
		``New insights into the analytic structure of correlation functions via kinetic theory,''
		[arXiv:2409.09022 [hep-th]].
		
\bibitem{Hu:2024tnn}
J.~Hu,
``Relaxation time approximation revisited and pole/cut structure in retarded correlators,''
[arXiv:2409.05131 [hep-ph]].


		\bibitem{Maldacena:2016hyu}
		J.~Maldacena and D.~Stanford,
		``Remarks on the Sachdev-Ye-Kitaev model,''
		Phys. Rev. D \textbf{94}, no.10, 106002 (2016)
		[arXiv:1604.07818 [hep-th]].
		
		
		\bibitem{Sachdev:2023try}
		S.~Sachdev,
		``Quantum statistical mechanics of the Sachdev-Ye-Kitaev model and charged black holes,''
		[arXiv:2304.13744 [cond-mat.str-el]].
		
		\bibitem{Delacretaz:2023ypv}
		L.~V.~Delacretaz and R.~Mishra,
		``Nonlinear response in diffusive systems,''
		SciPost Phys. \textbf{16}, no.2, 047 (2024)
		[arXiv:2304.03236 [cond-mat.str-el]].
		
		\bibitem{Sogabe:2021svv}
		N.~Sogabe and Y.~Yin,
		``Off-equilibrium non-Gaussian fluctuations near the QCD critical point: an effective field theory perspective,''
		JHEP \textbf{03} (2022), 124
		[arXiv:2111.14667 [nucl-th]].
		
		\bibitem{Pantelidou:2022ftm}
		C.~Pantelidou and B.~Withers,
		`Thermal three-point functions from holographic Schwinger-Keldysh contours,''
		JHEP \textbf{04}, 050 (2023)
		[arXiv:2211.09140 [hep-th]].
		
		
		\bibitem{Jana:2020vyx}
		C.~Jana, R.~Loganayagam and M.~Rangamani,
		``Open quantum systems and Schwinger-Keldysh holograms,''
		JHEP \textbf{07} (2020), 242
		[arXiv:2004.02888 [hep-th]].
		
		
		
		
		\bibitem{Saremi:2011nh}
		O.~Saremi and K.~A.~Sohrabi,
		``Causal three-point functions and nonlinear second-order hydrodynamic coefficients in AdS/CFT,''
		JHEP \textbf{11} (2011), 147
		[arXiv:1105.4870 [hep-th]].
		
		
		\bibitem{Grozdanov:2016fkt}
		S.~Grozdanov and A.~O.~Starinets,
		``Second-order transport, quasinormal modes and zero-viscosity limit in the Gauss-Bonnet holographic fluid,''
		JHEP \textbf{03} (2017), 166
		[arXiv:1611.07053 [hep-th]].
		
		\bibitem{Moore:2010bu}
		G.~D.~Moore and K.~A.~Sohrabi,
		``Kubo Formulae for Second-Order Hydrodynamic Coefficients,''
		Phys. Rev. Lett. \textbf{106} (2011), 122302
		[arXiv:1007.5333 [hep-ph]].
		
		\bibitem{Erdmenger:2008rm}
		J.~Erdmenger, M.~Haack, M.~Kaminski and A.~Yarom,
		``Fluid dynamics of R-charged black holes,''
		JHEP \textbf{01} (2009), 055
		[arXiv:0809.2488 [hep-th]].
		
		\bibitem{Haack:2008xx}
		M.~Haack and A.~Yarom,
		``Universality of second order transport coefficients from the gauge-string duality,''
		Nucl. Phys. B \textbf{813} (2009), 140-155
		[arXiv:0811.1794 [hep-th]].
		
		\bibitem{Grozdanov:2014kva}
		S.~Grozdanov and A.~O.~Starinets,
		``On the universal identity in second order hydrodynamics,''
		JHEP \textbf{03} (2015), 007
		[arXiv:1412.5685 [hep-th]].


		\bibitem{Becker:2014jla}
		M.~Becker, Y.~Cabrera and N.~Su,
		``Finite-temperature three-point function in 2D CFT,''
		JHEP \textbf{09} (2014), 157
		[arXiv:1407.3415 [hep-th]].
		
		
		\bibitem{Rodriguez-Gomez:2021mkk}
		D.~Rodriguez-Gomez and J.~G.~Russo,
		``Thermal correlation functions in CFT and factorization,''
		JHEP \textbf{11} (2021), 049
		[arXiv:2105.13909 [hep-th]].
		
		
		
		
		\bibitem{landau1981kinetic}
		L. D.~ Landau, E. M.~ Lifshitz, 
		``Kinetic Theory: Volume 10 of A Course of Theoretical Physics",
		{Butterworth-Heinemann, 1981}
		
		
		
		\bibitem{SoaresRocha:2024afv}
		G.~Soares Rocha, L.~Gavassino and N.~Mullins,
		``Modeling stochastic fluctuations in relativistic kinetic theory,''
		Phys. Rev. D \textbf{110} (2024) no.1, 016020
		[arXiv:2405.10878 [nucl-th]].
		
		\bibitem{Abbasi:2021fcz}
		N.~Abbasi,
		``Long-time tails in the SYK chain from the effective field theory with a large number of derivatives,''
		JHEP \textbf{04} (2022), 181
		[arXiv:2112.12751 [hep-th]].
		
		\bibitem{Abbasi:2022aao}
		N.~Abbasi, M.~Kaminski and O.~Tavakol,
		``Theory of Nonlinear Diffusion with a Physical Gapped Mode,''
		Phys. Rev. Lett. \textbf{132} (2024) no.13, 131602
		[arXiv:2212.11499 [hep-th]].
		
		
		\bibitem{Grozdanov:2024fle}
		S.~Grozdanov, T.~Lemut, J.~Pelai\v{c} and A.~Soloviev,
		``Analytic structure of diffusive correlation functions,''
		[arXiv:2407.13550 [hep-th]].
		
		\bibitem{Cercignani:2002}
		C. ~Cercignani and G.M. ~Kremer, 
		``The Relativistic Boltzmann Equation: Theory and Applications,''
		Springer (2002).
		
		
		
		\bibitem{Romatschke:2017ejr}
		P.~Romatschke and U.~Romatschke,
		``Relativistic Fluid Dynamics In and Out of Equilibrium,''
		Cambridge University Press, 2019,
		ISBN 978-1-108-48368-1, 978-1-108-75002-8
		[arXiv:1712.05815 [nucl-th]].


      \bibitem{Withers:2018srf}
      B.~Withers,
     ``Short-lived modes from hydrodynamic dispersion relations,''
     JHEP \textbf{06} (2018), 059
   [arXiv:1803.08058 [hep-th]].


        \bibitem{Grozdanov:2019kge}
        S.~Grozdanov, P.~K.~Kovtun, A.~O.~Starinets and P.~Tadi\'c,
        ``Convergence of the Gradient Expansion in Hydrodynamics,''
        Phys. Rev. Lett. \textbf{122} (2019) no.25, 251601
        [arXiv:1904.01018 [hep-th]].

		\bibitem{Heller:2020uuy}
		M.~P.~Heller, A.~Serantes, M.~Spali\'nski, V.~Svensson and B.~Withers,
		``Hydrodynamic gradient expansion in linear response theory,''
		Phys. Rev. D \textbf{104} (2021) no.6, 066002
		[arXiv:2007.05524 [hep-th]].
		
		\bibitem{Heller:2020hnq}
		M.~P.~Heller, A.~Serantes, M.~Spali\'nski, V.~Svensson and B.~Withers,
		``Convergence of hydrodynamic modes: insights from kinetic theory and holography,''
		SciPost Phys. \textbf{10}, no.6, 123 (2021)
		[arXiv:2012.15393 [hep-th]].
		
		
		
		\bibitem{Gangadharan:2024ovs}
		R.~Gangadharan and V.~Roy,
		``The Convergence Problem Of Gradient Expansion In The Relaxation Time Approximation,''
		[arXiv:2405.10846 [nucl-th]].
		
		\bibitem{Davison:2024msq}
		R.~A.~Davison and L.~V.~Delacretaz,
		``Universal thermalization dynamics in (1+1)d QFTs,''
		[arXiv:2409.09112 [hep-th]].
		
		
		
		
		
		
		
		
		\bibitem{Crossley:2015evo}
		M.~Crossley, P.~Glorioso and H.~Liu,
		``Effective field theory of dissipative fluids,''
		JHEP \textbf{09} (2017), 095
		[arXiv:1511.03646 [hep-th]].
		
		
		\bibitem{Liu:2018kfw}
		H.~Liu and P.~Glorioso,
		``Lectures on non-equilibrium effective field theories and fluctuating hydrodynamics,''
		PoS \textbf{TASI2017}, 008 (2018)
		[arXiv:1805.09331 [hep-th]].
		
		\bibitem{Glorioso:2017fpd}
		P.~Glorioso, M.~Crossley and H.~Liu,
		``Effective field theory of dissipative fluids (II): classical limit, dynamical KMS symmetry and entropy current,''
		JHEP \textbf{09} (2017), 096
		[arXiv:1701.07817 [hep-th]].
		
		\bibitem{Glorioso:2017lcn}
		P.~Glorioso, H.~Liu and S.~Rajagopal,
		``Global Anomalies, Discrete Symmetries, and Hydrodynamic Effective Actions,''
		JHEP \textbf{01} (2019), 043
		[arXiv:1710.03768 [hep-th]].
		
		\bibitem{Glorioso:2016gsa}
		P.~Glorioso and H.~Liu,
		``The second law of thermodynamics from symmetry and unitarity,''
		[arXiv:1612.07705 [hep-th]].
		
		
		\bibitem{Jensen:2017kzi}
		K.~Jensen, N.~Pinzani-Fokeeva and A.~Yarom,
		``Dissipative hydrodynamics in superspace,''
		JHEP \textbf{09} (2018), 127
		[arXiv:1701.07436 [hep-th]].
		
		
		\bibitem{Jensen:2018hhx}
		K.~Jensen, R.~Marjieh, N.~Pinzani-Fokeeva and A.~Yarom,
		``An entropy current in superspace,''
		JHEP \textbf{01} (2019), 061
		[arXiv:1803.07070 [hep-th]].
		
		\bibitem{Jensen:2018hse}
		K.~Jensen, R.~Marjieh, N.~Pinzani-Fokeeva and A.~Yarom,
		``A panoply of Schwinger-Keldysh transport,''
		SciPost Phys. \textbf{5} (2018) no.5, 053
		[arXiv:1804.04654 [hep-th]].
		
		
		\bibitem{Haehl:2018lcu}
		F.~M.~Haehl, R.~Loganayagam and M.~Rangamani,
		``Effective Action for Relativistic Hydrodynamics: Fluctuations, Dissipation, and Entropy Inflow,''
		JHEP \textbf{10} (2018), 194
		[arXiv:1803.11155 [hep-th]].
		
		\bibitem{Grozdanov:2013dba}
		S.~Grozdanov and J.~Polonyi,
		``Viscosity and dissipative hydrodynamics from effective field theory,''
		Phys. Rev. D \textbf{91} (2015) no.10, 105031
		[arXiv:1305.3670 [hep-th]].
		
		
		\bibitem{Kovtun:2014hpa}
		P.~Kovtun, G.~D.~Moore and P.~Romatschke,
		``Towards an effective action for relativistic dissipative hydrodynamics,''
		JHEP \textbf{07} (2014), 123
		[arXiv:1405.3967 [hep-ph]].
		
		
		\bibitem{Bu:2015ame}
		Y.~Bu, M.~Lublinsky and A.~Sharon,
		``$U(1)$ current from the AdS/CFT: diffusion, conductivity and causality,''
		JHEP \textbf{04}, 136 (2016)
		[arXiv:1511.08789 [hep-th]].


        \bibitem{Bu:2020jfo}
        Y.~Bu, T.~Demircik and M.~Lublinsky,
        ``All order effective action for charge diffusion from Schwinger-Keldysh holography,''
        HEP \textbf{05} (2021), 187
       [arXiv:2012.08362 [hep-th]].


		\bibitem{Abbasi:2022rum}
		N.~Abbasi, A.~Davody and S.~Tahery,
		``Correlation functions in stable first-order relativistic hydrodynamics,''
		Phys. Rev. D \textbf{109}, no.3, 036006 (2024)
		[arXiv:2212.14619 [hep-th]].
		
		\bibitem{Abbasi:2025}
		N.~Abbasi and D.~H.~Rischke,
		``To appear.''
		
		
		\bibitem{Wang:1998wg}
		E.~Wang and U.~W.~Heinz,
		``A Generalized fluctuation dissipation theorem for nonlinear response functions,''
		Phys. Rev. D \textbf{66}, 025008 (2002)
		[arXiv:hep-th/9809016 [hep-th]].
		
		
		\bibitem{Huber:2023uzd}
		M.~Q.~Huber, W.~J.~Kern and R.~Alkofer,
		``How to Determine the Branch Points of Correlation Functions in Euclidean Space II: Three-Point Functions,''
		Symmetry \textbf{15} (2023) no.2, 414
		[arXiv:2302.01350 [hep-ph]].
		
		
		\bibitem{Grozdanov:2015kqa}
		S.~Grozdanov and N.~Kaplis,
		``Constructing higher-order hydrodynamics: The third order,''
		Phys. Rev. D \textbf{93} (2016) no.6, 066012
		[arXiv:1507.02461 [hep-th]].
        
        \bibitem{Diles:2023tau}
        S.~M.~Diles, A.~S.~Miranda, L.~A.~H.~Mamani, A.~M.~Echemendia and V.~T.~Zanchin,
        ``Third-order relativistic fluid dynamics at finite density in a general hydrodynamic frame,''
       Eur. Phys. J. C \textbf{84} (2024) no.5, 516
      [arXiv:2311.01232 [hep-th]].


		\bibitem{Calzetta:1999xh}
		E.~Calzetta and B.~L.~Hu,
		``Stochastic dynamics of correlations in quantum field theory: From Schwinger-Dyson to Boltzmann-Langevin equation,''
		Phys. Rev. D \textbf{61}, 025012 (2000)
		[arXiv:hep-ph/9903291 [hep-ph]].
		
		
		
		\bibitem{martin1973statistical}
		P.~C.~Martin, E.~D.~Siggia, and H.~A.~Rose,
		``Statistical Dynamics of Classical Systems,''
		Phys. Rev. A \textbf{8}, 423--437 (1973).
		
		
		
		
		
		
		\bibitem{Michailidis:2023mkd}
		A.~A.~Michailidis, D.~A.~Abanin and L.~V.~Delacr\'etaz,
		``Corrections to Diffusion in Interacting Quantum Systems,''
		Phys. Rev. X \textbf{14}, no.3, 031020 (2024)
		[arXiv:2310.10564 [cond-mat.stat-mech]].
		
		\bibitem{Delacretaz:2023pxm}
		L.~V.~Delacretaz,
		``Bound on Thermalization from Diffusive Fluctuations,''
		[arXiv:2310.16948 [cond-mat.str-el]].


        \bibitem{Vardhan:2024qdi}
        S.~Vardhan, S.~Grozdanov, S.~Leutheusser and H.~Liu,
        ``Effective field theories of dissipative fluids with one-form symmetries,''
        [arXiv:2408.12868 [hep-th]].


		\bibitem{Cartwright:2019opv}
		C.~Cartwright and M.~Kaminski,
		``Correlations far from equilibrium in charged strongly coupled fluids subjected to a strong magnetic field,''
		JHEP \textbf{09} (2019), 072
		[arXiv:1904.11507 [hep-th]].
		
		\bibitem{Wondrak:2020tzt}
		M.~F.~Wondrak, M.~Kaminski and M.~Bleicher,
		``Shear transport far from equilibrium via holography,''
		Phys. Lett. B \textbf{811} (2020), 135973
		[arXiv:2002.11730 [hep-ph]].
		
		\bibitem{Erdmenger:2012xu}
		J.~Erdmenger and S.~Lin,
		``Thermalization from gauge/gravity duality: Evolution of singularities in unequal time correlators,''
		JHEP \textbf{10} (2012), 028
		[arXiv:1205.6873 [hep-th]].
		
		\bibitem{Ammon:2016fru}
		M.~Ammon, S.~Grieninger, A.~Jimenez-Alba, R.~P.~Macedo and L.~Melgar,
		``Holographic quenches and anomalous transport,''
		JHEP \textbf{09} (2016), 131
		[arXiv:1607.06817 [hep-th]].
		
		
		\bibitem{Landsteiner:2017lwm}
		K.~Landsteiner, E.~Lopez and G.~Milans del Bosch,
		``Quenching the Chiral Magnetic Effect via the Gravitational Anomaly and Holography,''
		Phys. Rev. Lett. \textbf{120} (2018) no.7, 071602
		[arXiv:1709.08384 [hep-th]].
		
		
		\bibitem{Florkowski:2017olj}
		W.~Florkowski, M.~P.~Heller and M.~Spalinski,
		``New theories of relativistic hydrodynamics in the LHC era,''
		Rept. Prog. Phys. \textbf{81} (2018) no.4, 046001
		[arXiv:1707.02282 [hep-ph]].
		
		
		\bibitem{Jain:2023obu}
		A.~Jain and P.~Kovtun,
		``Schwinger-Keldysh effective field theory for stable and causal relativistic hydrodynamics,''
		JHEP \textbf{01} (2024), 162
		[arXiv:2309.00511 [hep-th]].
		
		
		\bibitem{Hoult:2021gnb}
		R.~E.~Hoult and P.~Kovtun,
		``Causal first-order hydrodynamics from kinetic theory and holography,''
		Phys. Rev. D \textbf{106} (2022) no.6, 066023
		[arXiv:2112.14042 [hep-th]].
		Copy to ClipboardDownload
		
		\bibitem{Moore:2018mma}
		G.~D.~Moore,
		``Stress-stress correlator in $\phi^{4}$ theory: poles or a cut?,''
		JHEP \textbf{05} (2018), 084
		[arXiv:1803.00736 [hep-ph]].
		
		\bibitem{Grozdanov:2018atb}
		S.~Grozdanov, K.~Schalm and V.~Scopelliti,
		``Kinetic theory for classical and quantum many-body chaos,''
		Phys. Rev. E \textbf{99} (2019) no.1, 012206
		[arXiv:1804.09182 [hep-th]].
		
		\bibitem{Denicol:2022bsq}
		G.~S.~Denicol and J.~Noronha,
		``Spectrum of the Boltzmann collision operator for \ensuremath{\lambda}\ensuremath{\phi}4 theory in the classical regime,''
		Phys. Lett. B \textbf{850} (2024), 138487
		[arXiv:2209.10370 [nucl-th]].
		
		
		\bibitem{Ochsenfeld:2023wxz}
		S.~Ochsenfeld and S.~Schlichting,
		``Hydrodynamic and non-hydrodynamic excitations in kinetic theory \textemdash{} a numerical analysis in scalar field theory,''
		JHEP \textbf{09} (2023), 186
		[arXiv:2308.04491 [hep-th]].
		
		
		\bibitem{Rocha:2024cge}
		G.~S.~Rocha, I.~Danhoni, K.~Ingles, G.~S.~Denicol and J.~Noronha,
		``Branch-cut in the shear-stress response function of massless \ensuremath{\lambda}\ensuremath{\varphi}4 with Boltzmann statistics,''
		Phys. Rev. D \textbf{110} (2024) no.7, 076003
		[arXiv:2404.04679 [nucl-th]].
		
		
		\bibitem{Pu:2009fj}
		S.~Pu, T.~Koide and D.~H.~Rischke,
		``Does stability of relativistic dissipative fluid dynamics imply causality?,''
		Phys. Rev. D \textbf{81} (2010), 114039
		[arXiv:0907.3906 [hep-ph]].



		
	\end{thebibliography}
\end{document}